\input harvmac
\input epsf
\input color
\font\small =cmr10 scaled 850
\def\vev#1{\langle#1\rangle}
\def\frac#1#2{{{#1}\over{#2}}}
\def\half{\frac{1}{2}}
\def\centertable#1{ \hbox to \hsize {\hfill\vbox{
                    \offinterlineskip \tabskip=0pt \halign{#1} }\hfill} }

\def\wt{\widetilde}
\def\Box{\hbox{$\rlap{$\sqcap$}\sqcup$}}

\def\eq{\eqn\nolabel}

\def\O{{\cal O}}
\def\J{{\cal J}}

\def\a{\alpha}
\def\b{\beta}

\def\d{\delta}
\def\ad{{\dot\a}}
\def\bd{{\dot\b}}

\def\e{\epsilon}
\def\l{\lambda}

\def\p{\partial}
\def\t{\theta}
\def\tb{\bar\t}
\def\phiT{\Upsilon}

\def\lb{\bar\l}

\def\centertable#1{ \hbox to \hsize {\hfill\vbox{
                    \offinterlineskip \tabskip=0pt \halign{#1} }\hfill} }

\def\date{\number\month/\number\day/\number\yearltd}

\def\centertable#1{ \hbox to \hsize {\hfill\vbox{
                    \offinterlineskip \tabskip=0pt \halign{#1} }\hfill} }

\def\ev#1{\langle#1\rangle}
\def\p{\partial}

\noblackbox
\def\IL{\relax{\rm I\kern-.18em L}}
\def\IH{\relax{\rm I\kern-.18em H}}
\def\IR{\relax{\rm I\kern-.18em R}}
\def\IC{\relax\hbox{$\inbar\kern-.3em{\rm C}$}}
\def\IZ{\relax\ifmmode\mathchoice
{\hbox{\cmss Z\kern-.4em Z}}{\hbox{\cmss Z\kern-.4em Z}}
{\lower.9pt\hbox{\cmsss Z\kern-.4em Z}} {\lower1.2pt\hbox{\cmsss
Z\kern-.4em Z}}\else{\cmss Z\kern-.4em Z}\fi}
\def\CM {{\cal M}}

\def\CJ {{\cal J}}

\def\CL {{\cal L}}
\def\CV {{\cal V}}
\def\CO {{\cal O}}

\def\O{{\cal O}}
\def\J{{\cal J}}
\def\M{{\cal M}}
\def\F{{\cal F}}

\def\a{\alpha}
\def\ad{{\dot\alpha}}
\def\b{\beta}

\def\d{\delta}
\def\e{\epsilon}
\def\l{\lambda}
\def\lb{\bar\lambda}

\def\p{\partial}
\def\t{\theta}
\def\tb{\bar\t}


\def\D{{\cal D}}

\def\det{{\rm det}}
\def\Tr{{\rm Tr}}

\font\manual=manfnt \def\dbend{\lower3.5pt\hbox{\manual\char127}}

\def\IZ{\relax\ifmmode\mathchoice
{\hbox{\cmss Z\kern-.4em Z}}{\hbox{\cmss Z\kern-.4em Z}}
{\lower.9pt\hbox{\cmsss Z\kern-.4em Z}} {\lower1.2pt\hbox{\cmsss
Z\kern-.4em Z}}\else{\cmss Z\kern-.4em Z}\fi}
\def\half {{1\over 2}}

\def\lfm#1{\medskip\noindent\item{#1}}
\def\p{\partial}

\font\small =cmr10 scaled 850
\def\vev#1{\langle#1\rangle}
\def\frac#1#2{{{#1}\over{#2}}}
\def\half{\frac{1}{2}}
\def\centertable#1{ \hbox to \hsize {\hfill\vbox{
                    \offinterlineskip \tabskip=0pt \halign{#1} }\hfill} }

\def\wt{\widetilde}
\def\Box{\hbox{$\rlap{$\sqcap$}\sqcup$}}

\def\J{{\cal J}}
\vskip-.2in
\lref\aglr{
  N.~Arkani-Hamed, G.~F.~Giudice, M.~A.~Luty and R.~Rattazzi,
  ``Supersymmetry-breaking loops from analytic continuation into  superspace,''
  Phys.\ Rev.\  D {\bf 58}, 115005 (1998)
  [arXiv:hep-ph/9803290].
}
\lref\GiudiceCA{
  G.~F.~Giudice, H.~D.~Kim and R.~Rattazzi,
  ``Natural mu and Bmu in gauge mediation,''
  Phys.\ Lett.\  B {\bf 660}, 545 (2008)
  [arXiv:0711.4448 [hep-ph]].
}
\lref\DineXT{
  M.~Dine and J.~Mason,
  ``Gauge mediation in metastable vacua,''
  Phys.\ Rev.\  D {\bf 77}, 016005 (2008)
  [arXiv:hep-ph/0611312].
}
\lref\FortinGX{
  J.~F.~Fortin,
  ``Spontaneously Broken Gauge Symmetry in SUSY Yang-Mills Theories with
  Matter,''
  arXiv:0710.2131 [hep-th].
}
\lref\PolchinskiAJ{
  J.~Polchinski,
  ``Gauge Fermion Masses In Supersymmetric Hierarchy Models,''
  Phys.\ Rev.\  D {\bf 26}, 3674 (1982).
}
\lref\IntriligatorPY{
  K.~A.~Intriligator, N.~Seiberg and D.~Shih,
  ``Supersymmetry Breaking, R-Symmetry Breaking and Metastable Vacua,''
  JHEP {\bf 0707}, 017 (2007)
  [arXiv:hep-th/0703281].
}
\lref\BuicanWS{
  M.~Buican, P.~Meade, N.~Seiberg and D.~Shih,
  ``Exploring General Gauge Mediation,''
  JHEP {\bf 0903}, 016 (2009)
  [arXiv:0812.3668 [hep-ph]].
}
\lref\LutySN{
  M.~A.~Luty,
  ``2004 TASI lectures on supersymmetry breaking,''
  arXiv:hep-th/0509029.
}
\lref\ColemanJX{
  S.~R.~Coleman and E.~J.~Weinberg,
  ``Radiative Corrections As The Origin Of Spontaneous Symmetry Breaking,''
  Phys.\ Rev.\  D {\bf 7}, 1888 (1973).
}
\lref\mss{
  P.~Meade, N.~Seiberg and D.~Shih,
  ``General Gauge Mediation,''
  Prog.\ Theor.\ Phys.\ Suppl.\  {\bf 177}, 143 (2009)
  [arXiv:0801.3278 [hep-ph]].
}
\lref\ADS{
  I.~Affleck, M.~Dine and N.~Seiberg,
  ``Dynamical Supersymmetry Breaking In Four-Dimensions And Its
  Phenomenological Implications,''
  Nucl.\ Phys.\  B {\bf 256}, 557 (1985).
}
\lref\FayetJB{
  P.~Fayet and J.~Iliopoulos,
  ``Spontaneously Broken Supergauge Symmetries and Goldstone Spinors,''
  Phys.\ Lett.\  B {\bf 51}, 461 (1974).
}
\lref\GiudiceBP{
  G.~F.~Giudice and R.~Rattazzi,
  ``Theories with gauge-mediated supersymmetry breaking,''
  Phys.\ Rept.\  {\bf 322}, 419 (1999)
  [arXiv:hep-ph/9801271].
}
\lref\ShoreBH{
  G.~M.~Shore,
  ``Supersymmetric Higgs Mechanism With Nondoubled Goldstone Bosons,''
  Nucl.\ Phys.\  B {\bf 248}, 123 (1984).
}
\lref\ShoreTF{
  G.~M.~Shore,
  ``The Supersymmetric Higgs Mechanism: Quartet Decoupling And Nondoubled
  Goldstone Bosons,''
  Annals Phys.\  {\bf 168}, 46 (1986).
}
\lref\ShoreTG{
  G.~M.~Shore,
  ``Dynamical Gauge Symmetry Breaking In Supersymmetry Theories,''
  Nucl.\ Phys.\  B {\bf 261}, 522 (1985).
}
\lref\GrisaruWC{
  M.~T.~Grisaru, W.~Siegel and M.~Rocek,
  ``Improved Methods For Supergraphs,''
  Nucl.\ Phys.\  B {\bf 159}, 429 (1979).
}
\lref\RayWK{
  S.~Ray,
  ``Some properties of meta-stable supersymmetry-breaking vacua in Wess-Zumino
  models,''
  Phys.\ Lett.\  B {\bf 642}, 137 (2006)
  [arXiv:hep-th/0607172].
}
\lref\ShoreRJ{
  G.~M.~Shore,
  ``Vacuum Alignment And Dynamical Mass Generation In Supersymmetric Gauge
  Theories,''
  Nucl.\ Phys.\  B {\bf 245}, 399 (1984).
}
\lref\NardecchiaNH{
  M.~Nardecchia, A.~Romanino and R.~Ziegler,
  ``General Aspects of Tree Level Gauge Mediation,''
  arXiv:0912.5482 [hep-ph].
}
\lref\OvrutWA{
  B.~A.~Ovrut and J.~Wess,
  ``Supersymmetric R(Xi) Gauge And Radiative Symmetry Breaking,''
  Phys.\ Rev.\  D {\bf 25}, 409 (1982).
}
\lref\superspace{
  S.~J.~Gates, M.~T.~Grisaru, M.~Rocek and W.~Siegel,
  ``Superspace, or one thousand and one lessons in supersymmetry,''
  Front.\ Phys.\  {\bf 58}, 1 (1983)
  [arXiv:hep-th/0108200].
}
\lref\FortinGX{
  J.~F.~Fortin,
  ``Spontaneously Broken Gauge Symmetry in SUSY Yang-Mills Theories with
  Matter,''
  arXiv:0710.2131 [hep-th].
}
\lref\PolchinskiQK{
  J.~Polchinski,
  ``Effective Potentials For Supersymmetric Three Scale Hierarchies,''
  Phys.\ Rev.\  D {\bf 27}, 1320 (1983); B. Zumino, unpublished, as cited therein.
}
\lref\ShoreRI{
  G.~M.~Shore and G.~Veneziano,
  ``CURRENT ALGEBRA AND SUPERSYMMETRY,''
  Int.\ J.\ Mod.\ Phys.\  A {\bf 1}, 499 (1986).
}
\lref\sv{
  M.~A.~Shifman and A.~I.~Vainshtein,
  ``Solution of the Anomaly Puzzle in SUSY Gauge Theories and the Wilson
  Operator Expansion,''
  Nucl.\ Phys.\  B {\bf 277}, 456 (1986)
  [Sov.\ Phys.\ JETP {\bf 64}, 428 (1986\ ZETFA,91,723-744.1986)].
}
\lref\DimopoulosGM{
  S.~Dimopoulos and S.~Raby,
  ``Geometric Hierarchy,''
  Nucl.\ Phys.\  B {\bf 219}, 479 (1983).
}
\lref\LercheQA{
  W.~Lerche,
  ``On Goldstone Fields In Supersymmetric Theories,''
  Nucl.\ Phys.\  B {\bf 238}, 582 (1984).
}
\lref\AGLR{
  N.~Arkani-Hamed, G.~F.~Giudice, M.~A.~Luty and R.~Rattazzi,
  ``Supersymmetry-breaking loops from analytic continuation into  superspace,''
  Phys.\ Rev.\  D {\bf 58}, 115005 (1998)
  [arXiv:hep-ph/9803290].
}

\lref\DistlerBT{
  J.~Distler and D.~Robbins,
  ``General F-Term Gauge Mediation,''
  arXiv:0807.2006 [hep-ph].
}
\lref\OoguriEZ{
  H.~Ooguri, Y.~Ookouchi, C.~S.~Park and J.~Song,
  ``Current Correlators for General Gauge Mediation,''
  arXiv:0806.4733 [hep-th].
}
\lref\PolchinskiAN{
  J.~Polchinski and L.~Susskind,
  ``Breaking Of Supersymmetry At Intermediate-Energy,''
  Phys.\ Rev.\  D {\bf 26}, 3661 (1982).
}
\lref\CarpenterWI{
  L.~M.~Carpenter, M.~Dine, G.~Festuccia and J.~D.~Mason,
  ``Implementing General Gauge Mediation,''
  arXiv:0805.2944 [hep-ph].
}
\lref\EndoGI{
  M.~Endo and K.~Yoshioka,
  ``Low-scale Gaugino Mass Unification,''
  arXiv:0804.4192 [hep-ph].
}
\lref\MartinNS{
  S.~P.~Martin,
  ``A Supersymmetry Primer,''
  arXiv:hep-ph/9709356.
}
\lref\BanksMG{
  T.~Banks and V.~Kaplunovsky,
  ``Nosonomy Of An Upside Down Hierarchy Model. 1,''
  Nucl.\ Phys.\  B {\bf 211}, 529 (1983).
}
\lref\Vadim{
  V.~Kaplunovsky,
  ``Nosonomy Of An Upside Down Hierarchy Model. 2,''
  Nucl.\ Phys.\  B {\bf 233}, 336 (1984).
}
\lref\IbeSI{
  M.~Ibe, Y.~Nakayama and T.~T.~Yanagida,
  ``Conformal Gauge Mediation and Light Gravitino of Mass $m_{3/2} <O(10eV)$,"
arXiv:0804.0636 [hep-ph].
}
\lref\IbeAB{
  M.~Ibe and R.~Kitano,
  ``Minimal Direct Gauge Mediation,''
  Phys.\ Rev.\  D {\bf 77}, 075003 (2008)
  [arXiv:0711.0416 [hep-ph]].
}
\lref\MartinVX{
  S.~P.~Martin,
  ``Two-loop effective potential for a general renormalizable theory and
  softly broken supersymmetry,''
  Phys.\ Rev.\  D {\bf 65}, 116003 (2002)
  [arXiv:hep-ph/0111209].
}
\lref\BrignoleKG{
  A.~Brignole,
  ``One-loop Kaehler potential in non-renormalizable theories,''
  Nucl.\ Phys.\  B {\bf 579}, 101 (2000)
  [arXiv:hep-th/0001121].
}
\lref\WessCP{
  J.~Wess and J.~Bagger,
  ``Supersymmetry and supergravity,''
{\it  Princeton, USA: Univ. Pr. (1992) 259 p}
}
\lref\LangackerIP{
  P.~Langacker, G.~Paz, L.~T.~Wang and I.~Yavin,
  ``Aspects of Z'-mediated Supersymmetry Breaking,''
  Phys.\ Rev.\  D {\bf 77}, 085033 (2008)
  [arXiv:0801.3693 [hep-ph]].
}
\lref\LuoKF{
  M.~Luo and S.~Zheng,
  JHEP {\bf 0904}, 122 (2009)
  [arXiv:0901.2613 [hep-ph]].
}

\lref\GrisaruSR{
  M.~T.~Grisaru, M.~Rocek and A.~Karlhede,
  ``The Superhiggs Effect In Superspace,''
  Phys.\ Lett.\  B {\bf 120}, 110 (1983).
}
\lref\os{
  H.~Osborn,
  ``N = 1 superconformal symmetry in four-dimensional quantum field theory,''
  Annals Phys.\  {\bf 272}, 243 (1999)
  [arXiv:hep-th/9808041].
}
\lref\gr{
  G.~F.~Giudice and R.~Rattazzi,
  ``Extracting supersymmetry-breaking effects from wave-function
  renormalization,''
  Nucl.\ Phys.\  B {\bf 511}, 25 (1998)
  [arXiv:hep-ph/9706540].
}
\lref\DermisekQJ{
  R.~Dermisek, H.~D.~Kim and I.~W.~Kim,
  ``Mediation of supersymmetry breaking in gauge messenger models,''
  JHEP {\bf 0610}, 001 (2006)
  [arXiv:hep-ph/0607169].
}
\lref\dMM{
  A.~de Gouvea, T.~Moroi and H.~Murayama,
  ``Cosmology of supersymmetric models with low-energy gauge mediation,''
  Phys.\ Rev.\  D {\bf 56}, 1281 (1997)
  [arXiv:hep-ph/9701244].
}
\lref\semidirect{
  N.~Seiberg, T.~Volansky and B.~Wecht,
  ``Semi-direct Gauge Mediation,''
  JHEP {\bf 0811}, 004 (2008)
  [arXiv:0809.4437 [hep-ph]].
}
\lref\egms{
  E.~Gorbatov and M.~Sudano,
  ``Sparticle Masses in Higgsed Gauge Mediation,''
  arXiv:0802.0555 [hep-ph].
}
\lref\DistlerBT{
  J.~Distler and D.~Robbins,
  ``General F-Term Gauge Mediation,''
  arXiv:0807.2006 [hep-ph].
}
\lref\DineVC{
  M.~Dine, A.~E.~Nelson and Y.~Shirman,
  ``Low-Energy Dynamical Supersymmetry Breaking Simplified,''
  Phys.\ Rev.\  D {\bf 51}, 1362 (1995)
  [arXiv:hep-ph/9408384].
}
\lref\IntriligatorFE{
  K.~Intriligator, D.~Shih and M.~Sudano,
  ``Surveying Pseudomoduli: the Good, the Bad and the Incalculable,''
  JHEP {\bf 0903}, 106 (2009)
  [arXiv:0809.3981 [hep-th]].
}
\lref\IntriligatorFR{
  K.~A.~Intriligator and M.~Sudano,
  ``Comments on General Gauge Mediation,''
  JHEP {\bf 0811}, 008 (2008)
  [arXiv:0807.3942 [hep-ph]].
}
\lref\ElvangGK{
  H.~Elvang and B.~Wecht,
  ``Semi-Direct Gauge Mediation with the 4-1 Model,''
  JHEP {\bf 0906}, 026 (2009)
  [arXiv:0904.4431 [hep-ph]].
}
\lref\DineAG{
  M.~Dine, A.~E.~Nelson, Y.~Nir and Y.~Shirman,
  ``New tools for low-energy dynamical supersymmetry breaking,''
  Phys.\ Rev.\  D {\bf 53}, 2658 (1996)
  [arXiv:hep-ph/9507378].
}
\lref\CohenQC{
  A.~G.~Cohen, T.~S.~Roy and M.~Schmaltz,
  ``Hidden sector renormalization of MSSM scalar masses,''
  JHEP {\bf 0702}, 027 (2007)
  [arXiv:hep-ph/0612100].
}
\lref\IzawaGS{
  K.~I.~Izawa, Y.~Nomura, K.~Tobe and T.~Yanagida,
  ``Direct-transmission models of dynamical supersymmetry breaking,''
  Phys.\ Rev.\  D {\bf 56}, 2886 (1997)
  [arXiv:hep-ph/9705228].
}
\lref\SchwingerTN{
  J.~S.~Schwinger,
  ``GAUGE INVARIANCE AND MASS,''
  Phys.\ Rev.\  {\bf 125}, 397 (1962);
J.~S.~Schwinger,
  ``Gauge Invariance And Mass. 2,''
  Phys.\ Rev.\  {\bf 128}, 2425 (1962).
}
\lref\WessCP{
  J.~Wess and J.~Bagger,
  ``Supersymmetry and supergravity,''
{\it  Princeton, USA: Univ. Pr. (1992) 259 p}
}
\lref\GiudiceCA{
  G.~F.~Giudice, H.~D.~Kim and R.~Rattazzi,
  ``Natural mu and Bmu in gauge mediation,''
  Phys.\ Lett.\  B {\bf 660}, 545 (2008)
  [arXiv:0711.4448 [hep-ph]].
}
\lref\KomargodskiAX{
  Z.~Komargodski and N.~Seiberg,
  ``mu and General Gauge Mediation,''
  JHEP {\bf 0903}, 072 (2009)
  [arXiv:0812.3900 [hep-ph]].
}
\lref\MurayamaGE{
  H.~Murayama, Y.~Nomura and D.~Poland,
  ``More Visible Effects of the Hidden Sector,''
  Phys.\ Rev.\  D {\bf 77}, 015005 (2008)
  [arXiv:0709.0775 [hep-ph]].
}

\lref\DineYW{
  M.~Dine and A.~E.~Nelson,
  ``Dynamical supersymmetry breaking at low-energies,''
  Phys.\ Rev.\  D {\bf 48}, 1277 (1993)
  [arXiv:hep-ph/9303230].
}
\lref\RoyNZ{
  T.~S.~Roy and M.~Schmaltz,
  ``A hidden solution to the $\mu/B_\mu$ problem in gauge mediation,''
  Phys.\ Rev.\  D {\bf 77}, 095008 (2008)
  [arXiv:0708.3593 [hep-ph]].
}
\lref\GrisaruVE{
  M.~T.~Grisaru, M.~Rocek and R.~von Unge,
  ``Effective K\"ahler Potentials,''
  Phys.\ Lett.\  B {\bf 383}, 415 (1996)
  [arXiv:hep-th/9605149].
}

\lref\oops{
  H.~Ooguri, Y.~Ookouchi, C.~S.~Park and J.~Song,
  ``Current Correlators for General Gauge Mediation,''
  arXiv:0806.4733 [hep-th].
}
\lref\LutyFK{
  M.~A.~Luty,
  ``Naive dimensional analysis and supersymmetry,''
  Phys.\ Rev.\  D {\bf 57}, 1531 (1998)
  [arXiv:hep-ph/9706235].
}
\lref\DineXK{
  M.~Dine, Y.~Nir and Y.~Shirman,
  ``Variations on minimal gauge mediated supersymmetry breaking,''
  Phys.\ Rev.\  D {\bf 55}, 1501 (1997)
  [arXiv:hep-ph/9607397].
}
\lref\MartinZK{
  S.~P.~Martin and M.~T.~Vaughn,
  ``Two Loop Renormalization Group Equations For Soft Supersymmetry Breaking
  Couplings,''
  Phys.\ Rev.\  D {\bf 50}, 2282 (1994)
  [Erratum-ibid.\  D {\bf 78}, 039903 (2008)]
  [arXiv:hep-ph/9311340].
}
\lref\KomargodskiJF{
  Z.~Komargodski and D.~Shih,
  ``Notes on SUSY and R-Symmetry Breaking in Wess-Zumino Models,''
  JHEP {\bf 0904}, 093 (2009)
  [arXiv:0902.0030 [hep-th]].
}
\lref\ArgurioGE{
  R.~Argurio, M.~Bertolini, G.~Ferretti and A.~Mariotti,
  ``Patterns of Soft Masses from General Semi-Direct Gauge Mediation,''
  arXiv:0912.0743 [hep-ph].
}

\lref\DineZA{
  M.~Dine, W.~Fischler and M.~Srednicki,
  ``Supersymmetric Technicolor,''
  Nucl.\ Phys.\  B {\bf 189}, 575 (1981).
}
\lref\DimAU{
  S.~Dimopoulos and S.~Raby,
  ``Supercolor,''
  Nucl.\ Phys.\  B {\bf 192}, 353 (1981).
}
\lref\DimopoulosIG{
  S.~Dimopoulos and G.~F.~Giudice,
  ``Multi-messenger theories of gauge-mediated supersymmetry breaking,''
  Phys.\ Lett.\  B {\bf 393}, 72 (1997)
  [arXiv:hep-ph/9609344].
}

\lref\DineGU{
  M.~Dine and W.~Fischler,
  ``A Phenomenological Model Of Particle Physics Based On Supersymmetry,''
  Phys.\ Lett.\  B {\bf 110}, 227 (1982).
}
\lref\ISS{
  K.~Intriligator, N.~Seiberg and D.~Shih,
  ``Dynamical SUSY breaking in meta-stable vacua,''
  JHEP {\bf 0604}, 021 (2006)
  [arXiv:hep-th/0602239].
}

\lref\NappiHM{
  C.~R.~Nappi and B.~A.~Ovrut,
  ``Supersymmetric Extension Of The SU(3) X SU(2) X U(1) Model,''
  Phys.\ Lett.\  B {\bf 113}, 175 (1982).
}
\lref\WittenKV{
  E.~Witten,
  ``Mass Hierarchies In Supersymmetric Theories,''
  Phys.\ Lett.\  B {\bf 105}, 267 (1981).
}

\lref\Alvarez{
  L.~Alvarez-Gaume, M.~Claudson and M.~B.~Wise,
  ``Low-Energy Supersymmetry,''
  Nucl.\ Phys.\  B {\bf 207}, 96 (1982).
}
\lref\DimGM{
  S.~Dimopoulos and S.~Raby,
  ``Geometric Hierarchy,''
  Nucl.\ Phys.\  B {\bf 219}, 479 (1983).
}
\lref\NibbelinkSI{
  S.~G.~Nibbelink and T.~S.~Nyawelo,
  ``Effective action of softly broken supersymmetric theories,''
  Phys.\ Rev.\  D {\bf 75}, 045002 (2007)
  [arXiv:hep-th/0612092].
}
\lref\MartinZK{
  S.~P.~Martin and M.~T.~Vaughn,
  ``Two Loop Renormalization Group Equations For Soft Supersymmetry Breaking
  Couplings,''
  Phys.\ Rev.\  D {\bf 50}, 2282 (1994)
  [Erratum-ibid.\  D {\bf 78}, 039903 (2008)]
  [arXiv:hep-ph/9311340].
}

\lref\YamadaID{
  Y.~Yamada,
  ``Two loop renormalization group equations for soft SUSY breaking scalar
  interactions: Supergraph method,''
  Phys.\ Rev.\  D {\bf 50}, 3537 (1994)
  [arXiv:hep-ph/9401241].
}
\lref\OoguriEZ{
  H.~Ooguri, Y.~Ookouchi, C.~S.~Park and J.~Song,
  ``Current Correlators for General Gauge Mediation,''
  Nucl.\ Phys.\  B {\bf 808}, 121 (2009)
  [arXiv:0806.4733 [hep-th]].
}
\lref\DistlerBT{
  J.~Distler and D.~Robbins,
  ``General F-Term Gauge Mediation,''
  arXiv:0807.2006 [hep-ph].
}
\lref\IntriligatorFR{
  K.~A.~Intriligator and M.~Sudano,
  ``Comments on General Gauge Mediation,''
  JHEP {\bf 0811}, 008 (2008)
  [arXiv:0807.3942 [hep-ph]].
}
\lref\BenakliPG{
  K.~Benakli and M.~D.~Goodsell,
  ``Dirac Gauginos in General Gauge Mediation,''
  Nucl.\ Phys.\  B {\bf 816}, 185 (2009)
  [arXiv:0811.4409 [hep-ph]].
}
\lref\CarpenterHE{
  L.~M.~Carpenter,
  ``Surveying the Phenomenology of General Gauge Mediation,''
  arXiv:0812.2051 [hep-ph].
}
\lref\BuicanWS{
  M.~Buican, P.~Meade, N.~Seiberg and D.~Shih,
  ``Exploring General Gauge Mediation,''
  JHEP {\bf 0903}, 016 (2009)
  [arXiv:0812.3668 [hep-ph]].
}

\lref\RajaramanGA{
  A.~Rajaraman, Y.~Shirman, J.~Smidt and F.~Yu,
  ``Parameter Space of General Gauge Mediation,''
  Phys.\ Lett.\  B {\bf 678}, 367 (2009)
  [arXiv:0903.0668 [hep-ph]].
} 
\lref\GinspargTQ{
  P.~H.~Ginsparg,
  ``Finite Temperature Behavior Of Mass Hierarchies In Supersymmetric
  Theories,''
  Phys.\ Lett.\  B {\bf 112}, 45 (1982).
}
\def\gmed{\refs{\DineZA\DimAU\WittenKV\NappiHM\Alvarez\DimGM\ADS\DineYW\DineVC\DineAG\DineXK-\GiudiceBP}}
\Title{\vbox{\baselineskip12pt \hbox{UCSD-PTH-09-09}\hbox{IPMU 09-0141}}}
{\vbox{\centerline{General Gauge Mediation with Gauge Messengers}}}
\smallskip
\centerline{Kenneth Intriligator$^1$ and Matthew Sudano$^{1,2}$}
\smallskip
\centerline{$^1${\it Department of Physics, University of
California, San Diego, La Jolla, CA 92093 USA}}
\centerline{$^2${\it Institute for the Physics and Mathematics of the Universe}}
\centerline{{\it University of Tokyo, Kashiwa, Chiba 277-8568, Japan}}
\bigskip
\vskip 1cm

\noindent 
We generalize the General Gauge Mediation formalism \mss\ to allow for the possibility of gauge messengers.  Gauge messengers occur when charged matter fields of the susy-breaking sector have non-zero F-terms, which leads to tree-level, susy-breaking mass splittings in the gauge fields.   A classic example is that $SU(5)_{GUT}/SU(3)\times SU(2)\times U(1)$ gauge fields could be gauge messengers.  
We give a completely general, model independent, current-algebra based analysis of gauge messenger mediation of susy-breaking to the visible sector.  Characteristic aspects of gauge messengers include enhanced contributions to gaugino masses, (tachyonic) sfermion mass-squareds generated already at one loop, and also at two loops, and significant  one-loop A-terms, already at the messenger scale.

\bigskip

\Date{January 2010}

\newsec{Introduction}

In gauge mediation of supersymmetry breaking (see e.g.~\gmed), a susy-breaking hidden sector is coupled to a supersymmetric extension of the standard model (SSM) only through gauge interactions.  In direct gauge mediation, the SSM interacts with the hidden sector through the $SU(3)\times SU(2)\times U(1)$ gauge interactions of the standard model.  Another model-building option is indirect gauge mediation, where susy breaking is communicated from the hidden sector to an intermediate, messenger sector through some new gauge interactions, and then from the messenger sector to the MSSM via the SM gauge interactions.  In what follows, we'll consider general aspects of susy breaking mediation from a susy-breaking ``hidden sector" to an otherwise supersymmetric ``visible sector,'' via gauge interactions\foot{ One could also consider including direct superpotential coupling of the sectors, as is sometimes useful in considering the Higgs sector (see e.g. \GiudiceCA, \KomargodskiAX), but we will not do so here.}  
\eqn\gaugev{\CL_{int}\supset g\int d^4 \theta (\CJ _{hidden}+\CJ_{vis})\CV,}
where $\CJ$ are the currents in the two sectors and $\CV$ is the gauge vector multiplet.  We refer to the sectors as ``hidden" and ``visible," but our discussion will be completely general and will not assume that the ``visible" sector is actually the SSM.  Thus our general analysis can also be applied in models where the ``visible sector" is instead replaced with an intermediate messenger sector.

We will here consider general aspects of gauge mediation of susy breaking {\it with gauge messengers}, i.e. when some of the gauge fields $\CV$ in \gaugev\ themselves have a tree-level susy-split spectrum.  The majority of the literature on gauge mediation focuses on the case {\it without} gauge messengers, where the gauge multiplet spectrum is susy-split only at the loop-level.  Indeed, even the general gauge mediation framework of \mss\ focuses on that case.  See also e.g.\ \refs{\OoguriEZ\DistlerBT\IntriligatorFR\BenakliPG\CarpenterHE\BuicanWS\RajaramanGA-\ArgurioGE} for further work on non-gauge-messenger, general gauge mediation.   Although gauge messengers are less studied, it is not such an exotic pheonomenon: 
gauge messengers occur whenever {\it charged} hidden sector fields get non-zero F-components\foot{Gauge messengers can also occur with D-term breaking.  As we will discuss later, D-term breaking is already covered by our F-term based analysis whenever the gauge group is Higgsed: when $m_V^2\neq 0$, $\ev{D}$ is not an independent variable.   }.  

Gauge messengers have a long history. The Fayet-Iliopoulos model \FayetJB, has gauge messengers (for FI term sufficiently large), taking the Higgsed $U(1)$ gauge group of the model as a  $U(1)'$ to communicate susy breaking to the MSSM.  
Another classic with gauge messengers is Witten's inverted hierarchy model \WittenKV, where the charged fields breaking the GUT group also have non-zero, susy-breaking F-terms.   This scenario was analyzed in many following works long ago e.g.\ \refs{\DimGM, \GinspargTQ \BanksMG \Vadim \PolchinskiAN\PolchinskiAJ- \PolchinskiQK} and also more recently, e.g.\ in \DermisekQJ, where it was noted that gauge messengers can help alleviate the little hierarchy problem and they provide interesting, and somewhat unexplored, new avenues for model building.

With gauge messengers, the  gauge group is Higgsed (at least partially) together with
supersymmetry being broken.
Possible candidates for gauge messengers are
{($i$)} the massive vector bosons of a GUT gauge group; 
{$(ii)$} some new Higgsed gauge sector, such as a $U(1)'$ coupling to the MSSM; 
{$(iii)$}  the $W^\pm$ and $Z^0$ sector of the MSSM;
{$(iv)$} a new gauge sector, coupling to a separate messenger sector, which is then coupled to the MSSM (this possibility includes Semi-direct Gauge Mediation \semidirect). 
We will here generalize the framework of \mss\ to include the possibility of gauge messengers, and the possibility that the susy-mediating gauge group is (partially or fully) Higgsed.  We considered aspects of gauge mediation by susy-preserving Higgsed gauge groups in \refs{\egms, \IntriligatorFR}, and here we generalize that to the case of susy-breaking Higgsed gauge groups.  Because gauge messengers are relatively unexplored, we will also note many basic properties which are not widely known and/or are incompletely treated in some of the literature, and resolve some puzzling issues.

For theories where the messengers' susy-splitting is small relative to their susy mass component, e.g.\ if $F_i\sim \ev{Q^2X_i}$ and $M_i\sim \ev{X_i}$ have $|F_i|\ll |M_i|^2$, the leading visible sector soft masses can be obtained by using the technique of analytic continuation in superspace \refs{\Vadim, \gr, \AGLR}. This method works equally well whether or not there are gauge messengers.  
 Indeed, this technique was first developed \Vadim\ to study gauge messenger models, and 
some key differences between gauge-messenger and non-gauge-messenger models were discussed in \gr.   Our general analysis will include the small-$F$-term results as a limiting case.  We will note and explain, however, some differences from the literature.   For example, we note that the vector multiplet coupling to the visible sector generally has $Str M_V^2\neq 0$.  Also, much of the literature has focused on the case where the same, single chiral superfield $\Sigma = M+\theta ^2 F$ breaks both the gauge symmetry and supersymmetry.  We note that this is an oversimplification in actual models of spontaneous susy-breaking, such as the O'Raifearteigh model.  We note that in the more general case, gauge messengers also lead to non-vanishing one-loop contributions to the sfermion $m^2$s.

Gauge mediation without gauge messengers has some generic properties, e.g.

\lfm{1.} Vanishing one-loop contributions to visible-sector sfermion soft-breaking masses.  

\lfm{2.} Non-zero two-loop $m_{sfermion}^2$, which are typically non-tachyonic at the messenger scale (aside from contributions from fields with $D$-type masses).

\lfm{3.} Though one-loop gaugino masses $m_{gaugino}$ are generated, they vanish to leading order in $F/M^2$ if susy is spontaneously broken \PolchinskiAN.  Non-zero one-loop gaugino masses are generated at order $F^3/M^5$, as seen e.g.\ in  \IzawaGS, but they tend to be small and so $m_{gaugino}$ tends to be anomalously small compared with $m_{sfermion}$.  See \KomargodskiJF\ for a recent discussion and how this could be evaded by a certain type of metastability.  

\lfm{4.} Insignificant $A$ terms.  $A$ terms vanish to one-loop at the messenger scale, and are only generated at lower scales with two-loop factors, from one-loop RG running induced by the (already small) one-loop gaugino masses.  

Gauge messenger models, on the other hand, have qualitatively different properties:

\lfm{1$'$.} Non-zero (generally tachyonic) one-loop\foot{This should not be confused with the well-known (see e.g. \refs{\DineGU, \DimopoulosIG})
one-loop contributions to $m_Q^2$, proportional to the hypercharge,  coming from a one-loop induced $\ev{D^A}\neq 0$.  Messenger parity  \DineGU, a  $J^A\to -J^A$ symmetry to keep $\ev{D^A}=0$, is introduced to eliminate those problematic one-loop $m_Q^2$s.      The one-loop $m_Q^2s$ we're discussing have nothing to do with $\ev{D^A}$ and can occur for non-abeliang groups, regardless of whether or not messenger parity is imposed.}  contributions to  soft masses $m_{sfermion}^2$.  

\lfm{2$'$.}   Tachyonic contributions to the two-loop sfermion $m^2_{sfermion}$ at the messenger scale.

\lfm{3$'$.} Non-zero, one-loop gaugino masses, already at $\CO(F/M)$, included for the gauginos associated with massless (unhiggsed) gauge field subgroups. 

\lfm{4$'$.} Significant $A$-terms, $V\supset A_Q Q \partial _{Q }W+h.c.$ with $A_Q$ generated at one-loop and non-zero at the messenger scale.    

These observations have a long history.  In the context of the inverse-hierarchy \WittenKV\ type models, a diagram leading to (1$'$) one-loop $m^2_{sfermion}$  was mentioned in \DimGM, but not explicitly computed as it was observed to be insignificant in their model because of an additional suppression by $(m_{susy}/M_{GUT})^2\ll 1$ as compared with the two-loop $m^2_{sfermion}$.  In the context of analytic continuation in superspace \refs{\Vadim, \gr}, a proof was given that there cannot be one-loop contributions to $m^2_{sfermion}$ (as we will discuss, the loophole in the proof is simply that $ \int d^4 \theta \ln (\sum _{i=1}^N\bar X_i X_i) \neq 0$ if $N>1$).   The tachyonic two-loop $m^2_{sfermion}$ contributions (2$'$) were discussed using analytic continuation in superspace in \gr, and it was noted that the net result is typically tachyonic.  The fact (3$'$) that gauge messengers evade the issue of vanishing gaugino mass contribution was  noted in \PolchinskiAN, and an explicit computation of the non-zero, one-loop, $\CO(F/M)$ gaugino mass was given in \PolchinskiAJ\ for a model with gauge messengers.  The fact (4$'$) that there can be significant $A$ terms with gauge messengers was noted using analytic continuation in superspace \refs{\Vadim, \gr}.  

We aim to clarify these issues, and to generalize them in a model-independent, current-algebra based framework, much as in \mss, which does not rely on a weakly coupled hidden sector.  The gauginos and visible sector sfermions get susy-breaking soft masses from diagrams similar to those of \mss, see Fig. 1, except that here the blobs denote the full propagators (the sum of the series of 1PI propagators) of the vector-multiplet fields. 
\def\size{2}
\vskip.2in
\centertable{
\vrule height.3ex depth0.25ex width 0pt \tabskip=1em  \hfil#\hfil\cr
\epsfxsize=\size truein\epsfbox{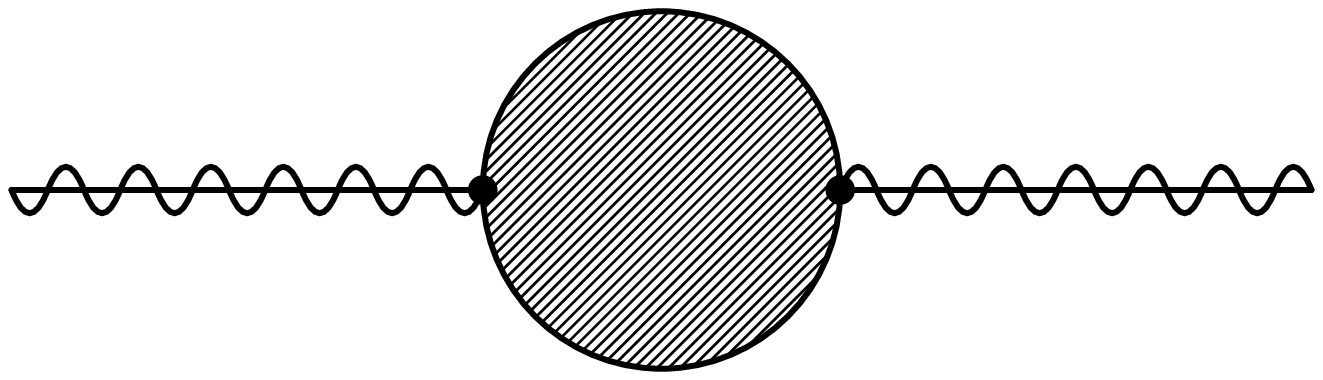}\cr
\cr
$D1$\cr
}
\vskip.2in
\centertable{
\vrule height.3ex depth0.25ex width 0pt \tabskip=1em  \hfil#\hfil\qquad&\qquad\hfil#\hfil\cr
\epsfxsize=\size truein\epsfbox{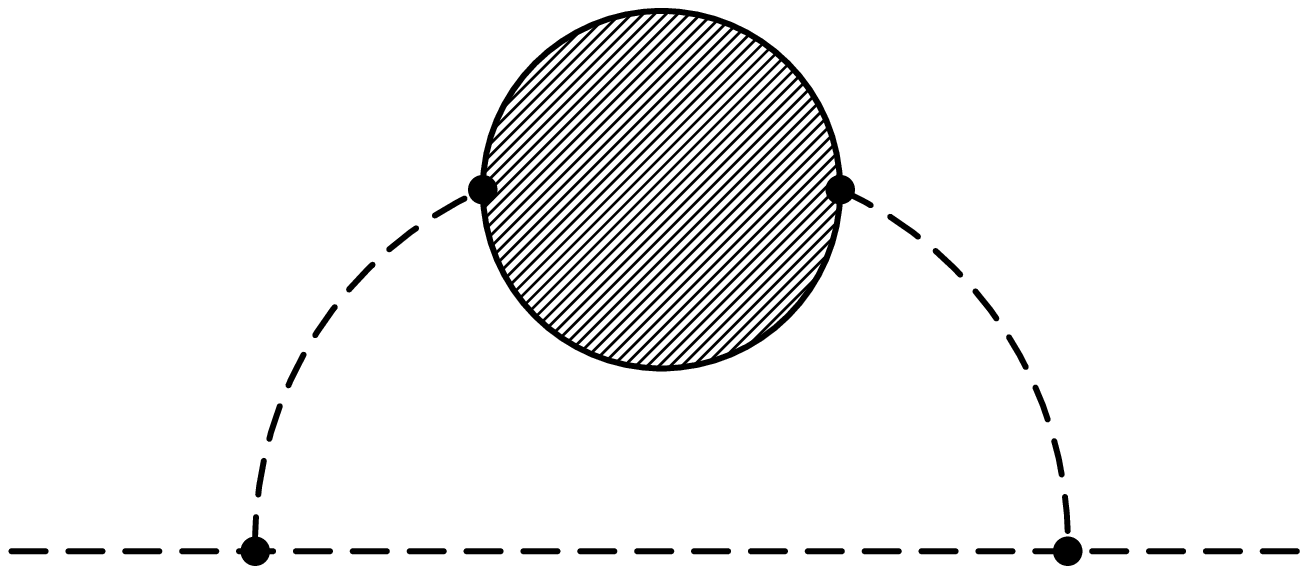}&\epsfxsize=\size truein\epsfbox{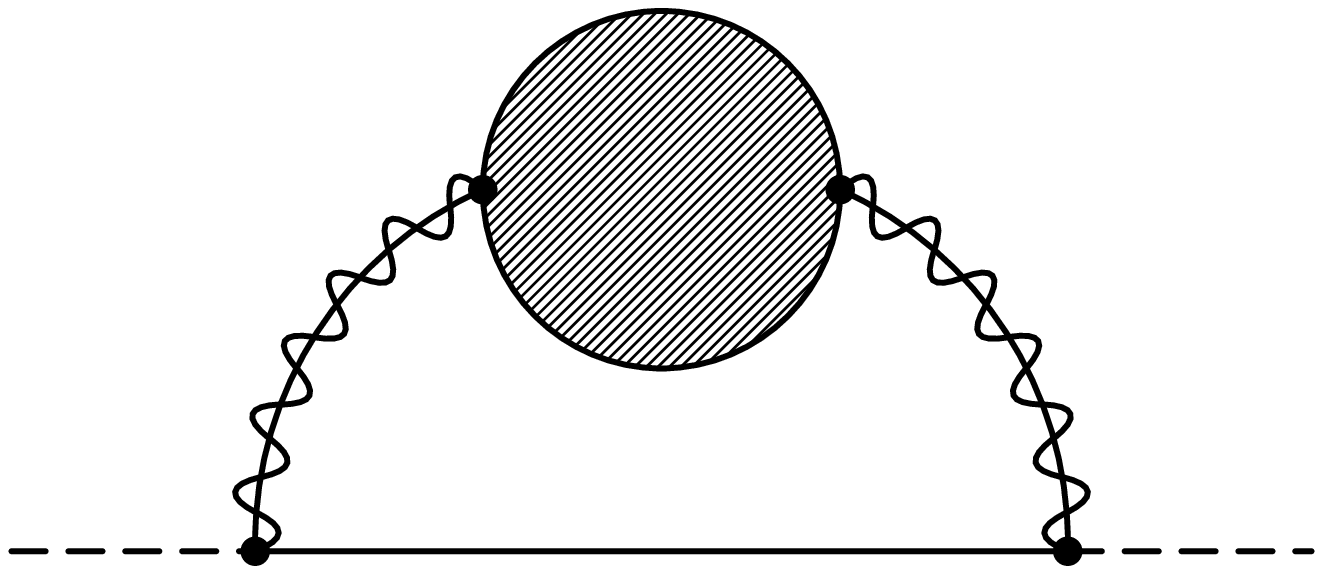}\cr
&\cr
$D2$&$D3$\cr
}
\vskip.2in
\centertable{
\vrule height.3ex depth0.25ex width 0pt \tabskip=1em  \hfil#\hfil\qquad&\qquad\hfil#\hfil\cr
\epsfxsize=\size truein\epsfbox{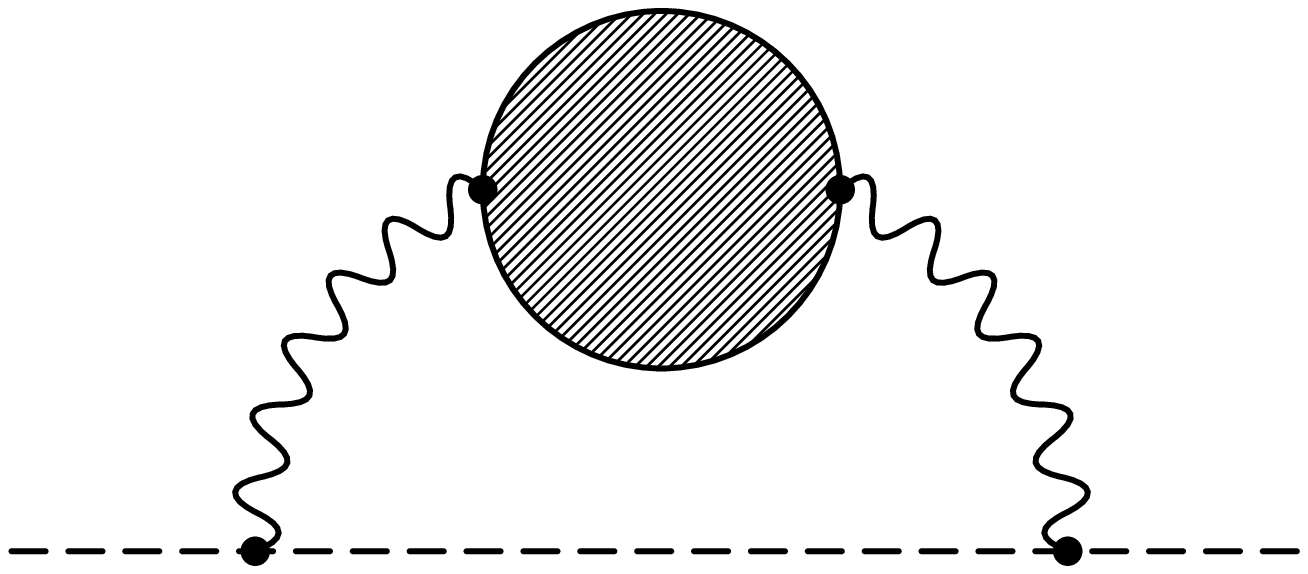}&\epsfxsize=\size truein\epsfbox{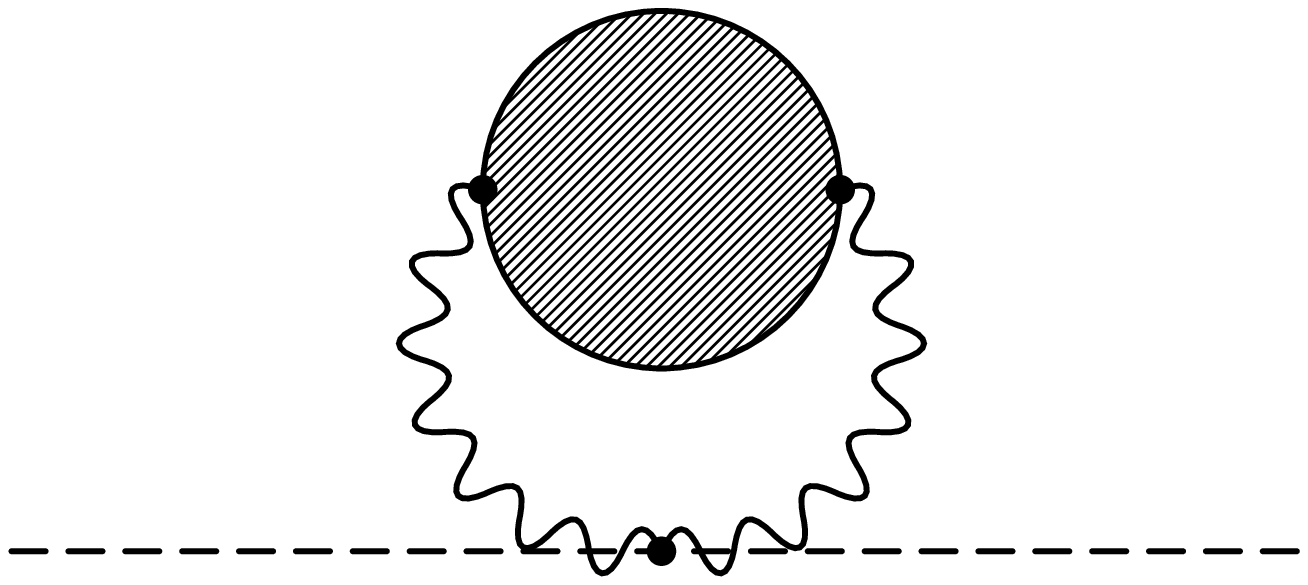}\cr
&\cr
$D4$&$D5$\cr
}
{\small\noindent{\bf Figure 1.}  Diagram $D1$ is the chirality-flipped propagator $\Sigma (p^2)$.  Diagrams $D2$-$D5$ contribute to the masses of sfermions and involve the spin $0$, $1/2$, and $1$ gauge superfield propagators, $\Delta (p^2)$, $\Delta _{\alpha \beta}(p^2)$, and $\Delta _{\mu \nu}(p^2)$, respectively.} 
\vskip.2in

Consider a general situation with gauge group $G'$ in the UV\foot{We use primes to denote UV extensions or versions of quantities.}, spontaneously broken to subgroup $G\subset G'$ at some energy scale $m_V=m_{G'/G}$.
We separate our discussion into effects associated with the propagators of massless gauge fields (denoted by $A$), and those associated with the propagators of massive gauge fields (denoted by $A'$).  In the general case where the gauge group is partially broken, with some gauge fields remaining massless, both effects are present, e.g.\ sfermions get masses from both the massless gauge fields and massive gauge fields, as in Fig. 2. 
$$\epsfxsize=2in\epsfbox{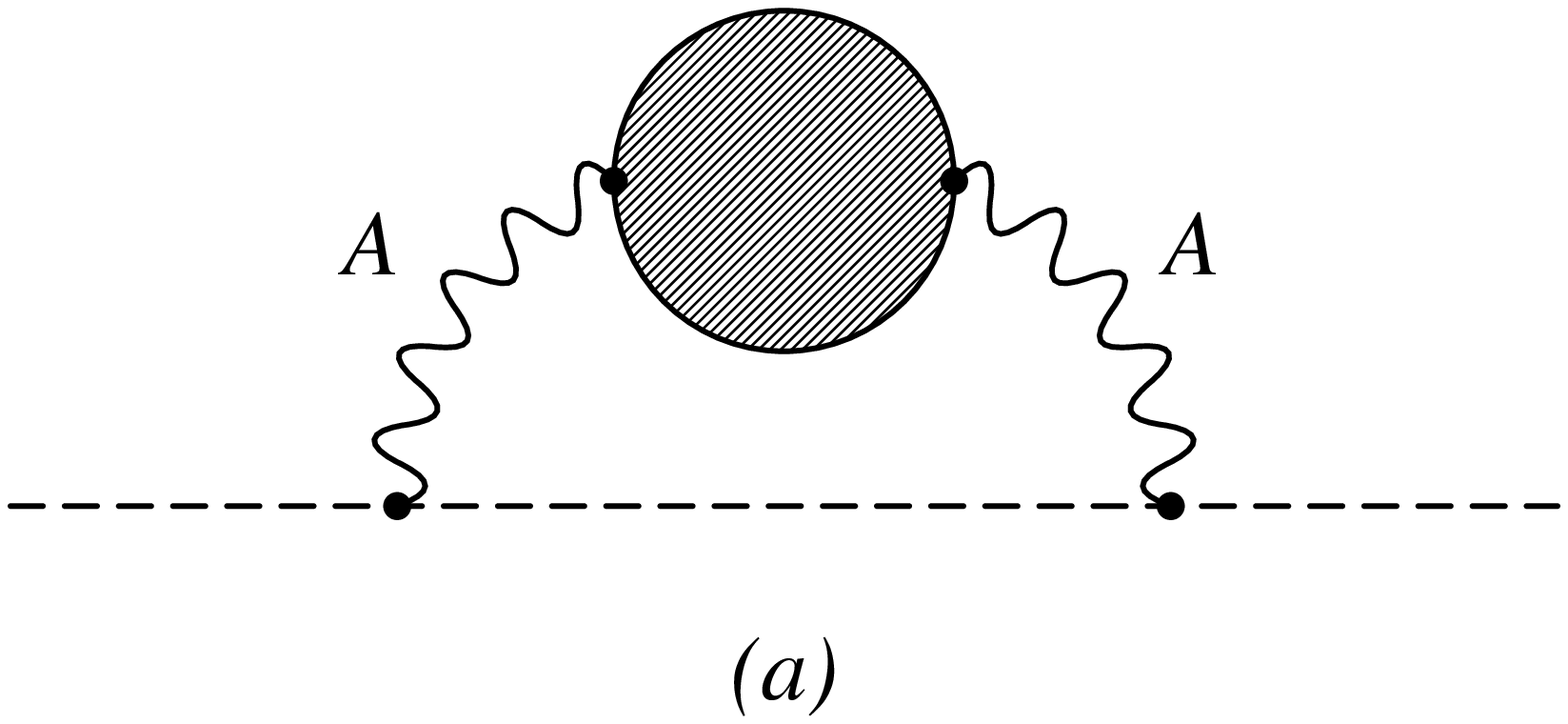}\qquad\epsfxsize=2in\epsfbox{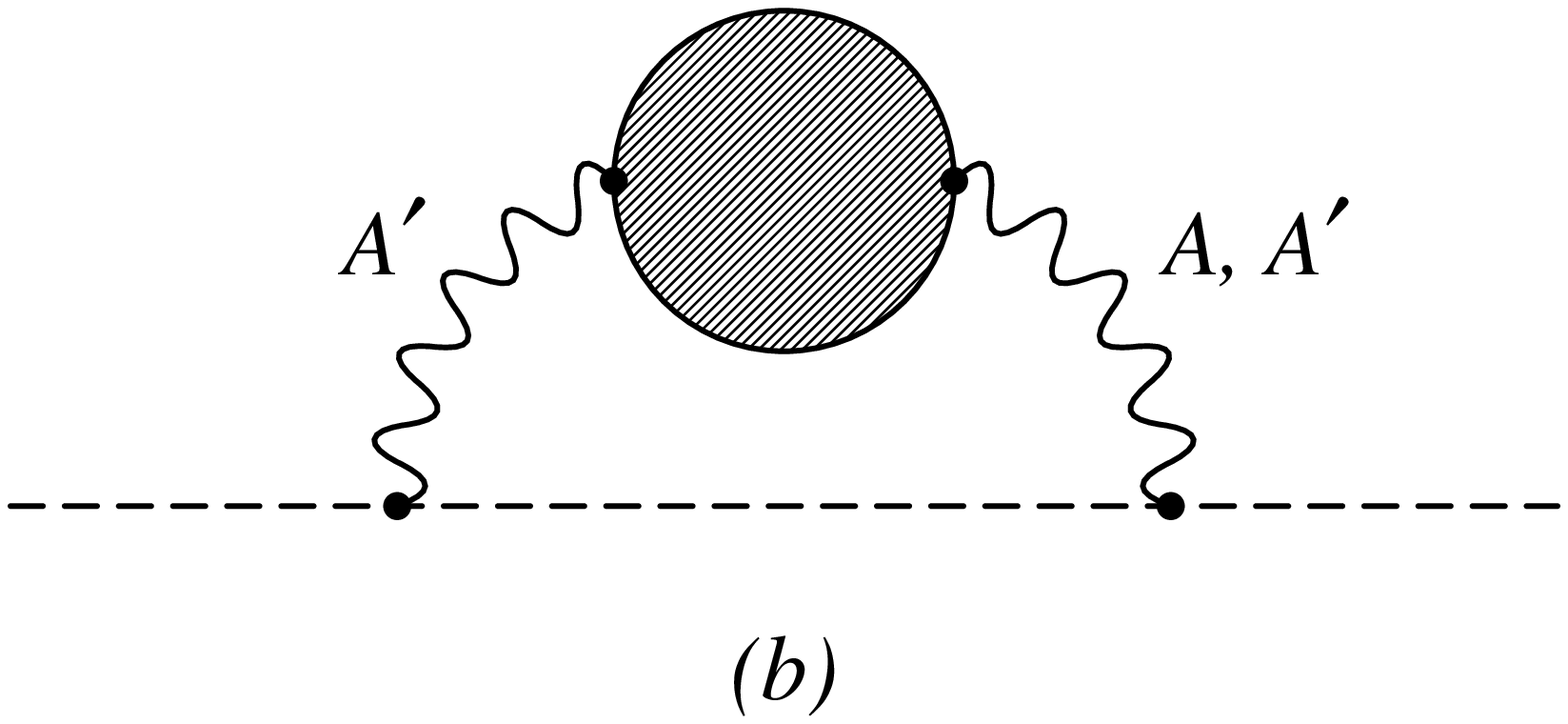}$$
\vskip- .1in
{\small\noindent{\bf Figure 2.}   The gauge fields in $(a)$ are massless, while in $(b)$, they are massive.}
\vskip.2in

\noindent
For example, with $SU(5)\to SU(3)\times SU(2)\times U(1)$, the first diagram (a) includes gauge mediation by the Standard Model gauge fields (including massive $SU(5)/[SU(3)\times SU(2)\times U(1)]$ in the blob), while the second diagram (b) includes new effects from the additional massive gauge fields of $SU(5)$.    

The unbroken gauge fields have supersymmetric tree-level spectrum.  (Clearly, $\ev{F^{\dagger i}}=-K^{ij}\partial _{\phi ^j}W$ only breaks those generators which are already broken by $\ev{\phi ^i}$.)  So the unbroken gauge fields in Fig. 2a yield a contribution to  the sfermion soft-masses which is two-loop and similar to \mss, except that the massive gauge fields contribute to the propagators of the unbroken gauge fields, and lead to the new effects (2$'$) and (3$'$) mentioned above.  
  The new effects (1$'$) and (4$'$) mentioned above comes from direct coupling of visible sector to the broken gauge fields, Fig. 2b, which can have susy-splittings already at tree-level.   

The outline of this paper is as follows.  In sect. 2, we will provide a technical outline and summary of our results.   Section 3 is devoted to a general discussion of gauge field propagators and current algebra, including discussion of spontaneous symmetry breaking, NG boson supermultiplets, and how the super-gauge field multiplet components get their masses.  In sect. 4, we discuss gauge mediation by any remaining massless gauge fields, as in Fig. 2a, and  in particular the new qualitative effects from the contributions of the massive gauge messenger fields to the correlators of the unbroken currents, as in Fig. 3.   These lead to the characteristic properties (2$'$) and (3$'$) of gauge messengers mentioned above.  

$$\epsfxsize=2in\epsfbox{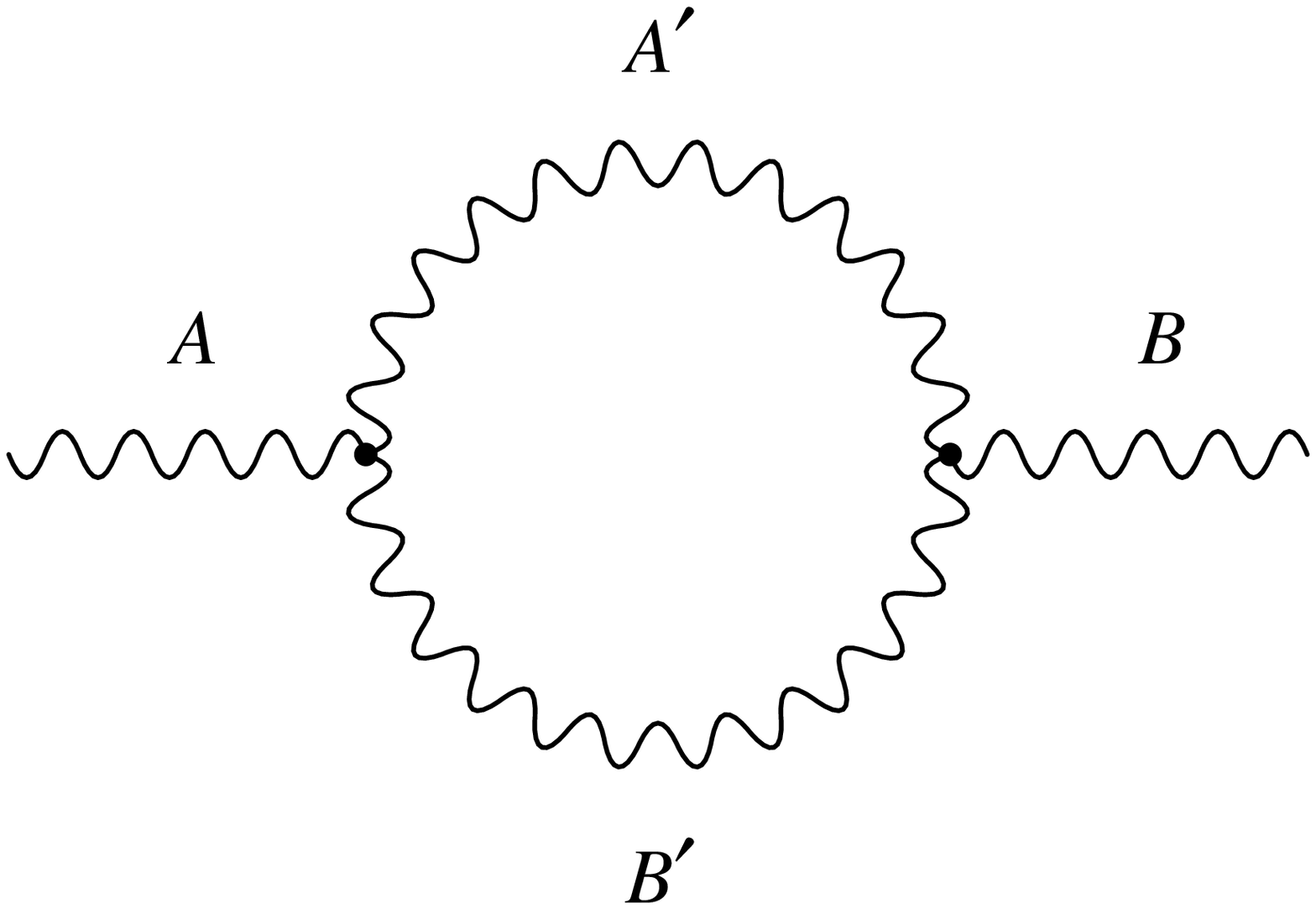}$$
\vskip-.1in
{\small\noindent{\bf Figure 3.}    The external gauge fields are massless, and the internal ones are massive.}
\vskip.2in
\noindent
 In sect. 5, we discuss the effects of coupling the massive gauge messenger gauge fields directly to the visible sector matter, as in Fig.~2b, and how this leads to the properties (1$'$) and (4$'$) above.  In sect. 6, the susy algebra structure of the soft terms, and the small susy-breaking limit are considered.  Section 7 is devoted to examples and, in sect. 8, we conclude.  Appendix A summarizes some basic aspects of susy and notation, susy breaking, spurions, and analytic continuation in superspace.  In appendix B we demonstrate the equivalence of three methods of computing the effective potential,  which generally have different regimes of validity, namely our current-correlator expression vs the Coleman-Weinberg potential vs the effective K\"ahler potential, where the methods are simultaneously valid.

\lref\BuicanVV{
  M.~Buican and Z.~Komargodski,
  ``Soft Terms from Broken Symmetries,''
  arXiv:0909.4824 [hep-ph].
}
\lref\LeeKB{
  J.~Y.~Lee,
  ``Renormalization in General Gauge Mediation,''
  arXiv:1001.1940 [hep-ph].
}
{\it Note added:} After this work was substantially completed, and it was presented at the Simons Summer Workshop  \ref\kisem{K.I., online seminar at the Simons Workshop, August 5,  2009.}, we learned of  M. Buican and Z. Komargodski's  independent work on a similar topic \BuicanVV; we thank them for bringing their work to our attention prior to its completion, and for communications and comments on our paper.   There is also a small amount of overlap (e.g. in parts of our sect. 3.1) with parts of the otherwise complementary recent work \NardecchiaNH, and also with \LeeKB.

\newsec{A technical outline and summary of results}

In this paper, we suppose that the the hidden, susy-breaking sector couples to the visible sector via gauge interactions \gaugev, and work in small $g$ perturbation theory.  The hidden sector can (fully or partially) break the gauge symmetry, in addition to supersymmetry.  The broken gauge fields have mass $m_V$, which we treat  as order $m_V=\CO(g^0)$ in our perturbative expansion.   Much as in \mss, we do not need to know every detail of the hidden sector -- the relevant information is how the hidden sector contributes to the current correlators.

\subsec{Tree level contact interactions}

One effect, which we mention for completeness but which will not be so important for the considerations of this paper, is present already at tree-level, and regardless of any susy breaking.   Integrating out massive vector multiplets in \gaugev\ induces a well-known contact interaction in the low-energy theory
\eqn\leffvis{\CL _{low}\supset  -\sum _{AB}{1\over 2(m_V^2)^{AB}}\int d^4\theta \CJ^A \CJ^B \supset  -\sum _{AB}{1\over (m_V^2)^{AB}}\int d^4\theta \CJ^A _{vis}\CJ^B _{hidden},}
which is a correction to the low-energy K\"ahler potential (see \BuicanWS\ for a recent discussion of its curvature).   Here $A$ and $B$ are gauge adjoint indices (which we'll often suppress).  This interaction accounts for the coupling of $\CJ^A$ to $\ev{\CV ^A}=-(m_V^{-2})^{AB}\CJ^B$,  and illustrates that the visible sector notices the hidden sector just via its effect on the gauge fields.  

The interaction \leffvis\ can yield tree-level gauge meditation to the visible sector, namely visible sector susy-breaking  D-type soft sfermion masses, if $\ev{\CJ ^B_{hidden}}|_{\theta ^2\bar\theta ^2}\neq 0$, i.e. if $F_0 T^B F_0\neq 0$.  In this case, as noted long ago  \PolchinskiAN, the EOM yield
\eqn\dfreln{\ev{D^A}(m_V^2)^{AB}=-2g\sum_r\ev{\bar F_r T_r^B F_r}\equiv -2g\bar F_0T^BF_0.}
Current conservation (gauge invariance) ensures that $\ev{\CJ ^B_{hidden}}|_{\theta ^2}=0$.  
The relation \dfreln\  implies  that $\ev{D}$ is not an independent susy-breaking variable when $m_V^2\neq 0$.  So $D$-term breaking is included in a general analysis of $F$-term breaking, except in the very limited class of models (abelian gauge theories with added FI term) where $\ev{D}\neq 0$ and $m_V^2=0$.

If the hidden sector is an O'Raifeartaigh-type model of F-term susy breaking, with gauge interactions included by taking the fields $\Phi$ in reps of a gauge group $G$ (as in e.g. \refs{\WittenKV - \DimGM, \GinspargTQ- \PolchinskiQK, \ISS\DineXT-\IntriligatorPY}\ etc.) then we can take $\bar \phi _0 T^B \phi _0=0$, and then gauge invariance of the superpotential and the EOM (assuming canonical K\"ahler potential) imply that $\bar F_0 T^B F_0=0$ (see e.g. \PolchinskiQK).  So in these cases $\ev{\CJ ^B_{hidden}}|_{\theta ^2\bar\theta ^2}\sim \ev{D^B}=0$ and there is no tree-level susy-breaking mediation to the visible sector.  To have \dfreln\ non-zero, the hidden sector must break susy by a combination of $D$ and $F$-terms (and/or have non-canonical K\"ahler potential) as in the $SU(3)\times SU(2)$ theory \ADS\ discussed in \semidirect; see also  \ElvangGK.

We will here mostly focus on cases without tree-level gauge mediation, where \dfreln\ vanishes, and the leading susy-breaking gauge mediation to the light fields of the visible sector is via loops.  We'll illustrate some aspects of the case where \dfreln\ is non-zero in the FI model \FayetJB\ example, in section 7.  See \NardecchiaNH\ for a recent, thorough discussion of tree-level gauge mediation.

\subsec{Gauge field's full propagators are the fundamental quantities}

The observable, susy-breaking effects in the gauge and visible sectors can be obtained from the propagators for the component fields of the gauge vector supermultiplet\foot{We normalize these fields to have canonical tree-level kinetic term.}
\eqn\gaugeprops{\eqalign{\Delta ^{AB}(p^2)&\equiv i\ev{D^A(p)D^B(-p)}\equiv \Delta _0^{AB}(p^2), \cr  \Delta ^{AB}_{\alpha \dot \alpha}(p^2)&\equiv i\ev{\lambda ^A_\alpha (p)\bar \lambda ^B_{\dot \alpha}(-p)}\equiv p_{\alpha \dot \alpha}\Delta _{1/2}(p^2)/p^2, \cr  \Delta _{\mu \nu}^{AB}(p^2)&\equiv i\ev{V^A_\mu (p)V^B_\nu(-p)}\equiv (\eta _{\mu \nu}-p_\mu p_\nu /p^2)\Delta _1(p^2)/p^2. }}
Here $A$, $B$ are adjoint group indices.  Unbroken susy would imply $\Delta _0=\Delta _{1/2}=\Delta _1=\Delta _{susy}(p^2)$, and $\Delta =1$ for a free, supersymmetric vector multiplet.  
The mass spectrum of any Higgsed gauge fields, in particular,  shows up as poles in these propagators.  Since the masses of the Higgsed gauge field can have susy-breaking mass differences already at tree-level, $\CO(g^0)$, so can the location of the poles of the propagators \gaugeprops.  
The susy-breaking effects for the visible sector matter and gauginos will now  be expressed in terms of the following super-traced combination of the above propagators, 
\eqn\xiis{\eqalign{\Xi ^{AB}(p^2)&\equiv \Delta ^{AB}(p^2)-2p^{\alpha \dot \alpha}\Delta ^{AB}_{\alpha \dot \alpha}(p^2)+g^{\mu \nu}p^2\Delta ^{AB}_{\mu \nu}(p^2),\cr 
&=\Delta _0^{AB}(p^2)-4\Delta _{1/2}^{AB}(p^2)+3\Delta _1^{AB}(p^2),}}
together with the chirality-flipped gaugino propagator,
\eqn\sigmais{\Sigma ^{AB}(p^2)={i\over 2}\epsilon ^{\dot \alpha \dot \beta}\ev{\bar \lambda ^A_{\dot \alpha}(p)\bar \lambda ^B_{\dot \beta}(-p)},}
both of which would vanish if susy were unbroken, $\Xi _{susy}=\Sigma _{susy}=0$.  The propagator \sigmais\ would also vanish if there is an unbroken R-symmetry, as $R(\Sigma )=-2$.  

\subsec{Visible sector soft masses and a-terms, from the propagators}

Our general results can be summarized as follows.  The $\CO(g^2)$ soft susy-breaking gaugino masses (in particular for the gaugino superpartners of any remaining massless gauge field) is 
\eqn\mgaugino{m_{gaugino}^{AB}=\lim _{p^2\to 0}\left(p^2 \Sigma ^{AB} (p^2)\right).}
The visible sector sfermion soft masses are computed from the gauge-field propagators as
\eqn\sfermint{m^2_Q=g^2 \ev{J^A}T^A_{r_Q}+g^2 \sum _{A,B} T_{r_Q}^AT^B_{r_Q}\int {d^4 p\over (2\pi )^4} {1\over p^2}\Xi ^{AB}(p^2) +\CO(g^6), }
where $T^A_{r_Q}$ is the group generator in the representation $r_Q$ of the matter field $Q$.   The first term in \sfermint\ generalizes the hypercharge contribution of the MSSM, the $Y_f\xi$ contribution\foot{It can be written in terms of a $D^A$ expectation value using 
\eqn\djvev{\ev{D^A}=-g\ev{J^A}.}} of \mss; one often imposes messenger parity to kill this term.   The second term in \sfermint\ is generated by the diagrams of fig. 1 (diagrams D2, D3, D4 give the terms in \xiis).   

Finally, there are visible sector susy-breaking A-terms.  These 
 are conveniently described by 
keeping the auxiliary components $F_Q$ of the visible sector matter fields $Q$, as then the $A$-terms all arise from the susy-breaking term
\eqn\leffa{\CL _{eff}\supset A_Q \bar F_Q Q\qquad  \hbox{which implies}\qquad V\supset A_QQ\partial _Q W.}
Clearly, such a term is non-zero only if there is R-symmetry breaking, as $R(A_Q)=-2$.  
If the visible sector has $W_{tree}=m_{ij}Q_iQ_j+Y_{ijk}Q_iQ_jQ_k$, then \leffa\ yields ``non-diagonal" soft masses $(m^2_{od})_{ij}\equiv b_{ij}=(A_i+A_j)m_{ij}$, and scalar trilinear couplings $a_{ijk}=(A_i+A_j+A_k)Y_{ijk}$.  By keeping the auxiliary field $F_Q$ in \leffa, these terms have a unified description.  
The coefficient $A_Q$ is generated by a loop diagram similar to diagram D3 in fig 1, but with the external $Q^\dagger$ replaced with $F_Q^\dagger$.  The loop then has a  $Q$ matter propagator\foot{We use the massless $1/p^2$ propagator for the $Q$ matter, as in \sfermint, because  the  momentum integral gets its main contribution at the messenger scale, where we assume that any tree-level $Q$ mass is negligible.  Otherwise, the visible sector mass should be included in the propagator.} and the $R$-symmetry breaking gaugino mass propagator given in \sigmais,  giving our result 
\eqn\aqloopo{A_Q=-2g^2 \sum _{AB} T^A_{r_Q} T^B_{r_Q} \int {d^4p\over (2\pi)^4}{1\over p^2}\Sigma ^{AB}(p^2).}

The fact that \xiis\ and \sigmais\ are related to susy-breaking can be seen by noting that
\eqn\sigmasusy{\Sigma ^{AB}(p^2)={i\over 4p^2}\ev{Q^2(D^A(p)D^B(-p))},}
\eqn\xisusy{\Xi ^{AB}(p^2)={i\over 8p^2}\ev{\bar Q^2 (Q^2(D^A(p)D^B(-p)))},}
where $Q^2(\dots)\equiv \{Q^\alpha,[Q_\alpha,(\dots)]\}$; thus \sigmasusy\ and \xisusy\ would vanish in a susy vacuum.
 Using \sigmasusy\ and \xisusy, we can now rewrite \mgaugino, \sfermint, and \aqloopo\ as
 \eqn\mgauginoq{m_{gaugino}^{AB}=\lim _{p^2\to 0} {i\over 4}\ev{Q^2 D^A(p)D^B(-p)},}
 \eqn\sfermintq{m_Q^2=g^2 \ev{J^A}T^A_{r_Q}+{g^2 \over 8}\sum _{AB}T^A_{r_Q}T^B_{r_Q}i\int {d^4p\over (2\pi)^4 p^4}\ev{\bar Q^2 (Q^2(D^A(p)D^B(-p)))},}
\eqn\aqloopoq{A_Q=-{g^2\over 2}\sum _{AB}T^A_{r_Q}T^B_{r_Q}i\int {d^4p\over (2\pi)^4 p^4}\ev{Q^2(D^A(p)D^B(-p)))},}

We could also consider the diagrams of fig. 1 with non-zero external momentum $k$ and, taking 
$\partial/\partial k^2$, compute the wavefunction renormalization $Z_Q$.  Here we'll simply note that the result in the case of unbroken susy is 
\eqn\zqsusy{Z_Q\supset 1-2g^2\sum _{AB}T^A_{r_Q}T^B_{r_Q}\int {d^4p\over (2\pi)^4}{1\over p^4} \Delta ^{AB}_{susy}(p^2).} 
We'll use \zqsusy\ to connect our general results \sfermintq\ and \aqloopoq\ with spurion-method based results in the limit of small susy-breaking.   

\subsec{A quick check, comparing with RG running, with explicit soft susy breaking}

Though we're here primarily interested in spontaneous susy breaking, we can also apply the above expressions to the case of explicit susy breaking.  We'll use this as a quick check of the normalizations, by connecting with known expressions for RG running.  As a warmup, consider \zqsusy\ and note that, 
 since $\lim _{p^2\to \infty} \Delta ^{AB}_{susy}(p^2)\to \delta ^{AB}+\CO(g^2)$, the $\CO(g^2)$ contribution in \zqsusy\ has a log divergence at large $p^2$, which is cut off using
\eqn\cutoff{\int {d^4p\over (2\pi)^4}{1\over p^4}\to {1\over 16\pi ^2} \ln \Lambda ^2+\hbox{finite}.}
Defining $t\equiv \ln \mu$, and using  $d/d\ln \Lambda =-d/dt$, we thus obtain from \zqsusy\
\eqn\gammaone{{d\ln Z_Q\over d\ln \Lambda }=-{d\ln Z_Q\over dt}=-4{g^2\over 16\pi ^2} c_2(r_Q)+\CO(g^4),}
which is the correct expression for the one-loop anomalous dimension. 

Now consider explicit susy breaking, with the  gauginos given a soft-breaking mass $M_\lambda$, and the gauge group unbroken.    We then have at tree-level $\Delta _0(p^2)=\Delta _1(p^2)=1$,  and 
\eqn\explicit{\Sigma ^{AB}(p^2) ={M_\lambda \over p^2+|M_\lambda|^2}\delta ^{AB}+\CO(g^2), \qquad \Xi ^{AB}(p^2) =4{|M_\lambda|^2\over p^2+|M_\lambda |^2}+\CO(g^2)\qquad\hbox{(explicit breaking)}.}
Using these in the expressions \sfermint\ and \aqloopo, both the $m_Q^2$ and $A_Q$ momentum loop integrals have a log divergence at large $p^2$, which using  \cutoff\ gives 
\eqn\maderivs{{d\over dt} m_Q^2\supset -8{g^2\over 16\pi ^2}c_2(r_Q)|M_\lambda|^2, \qquad {d\over dt} A_Q\supset 4{g^2\over 16\pi ^2} c_2(r_Q)M_\lambda.} These agree with the soft term beta functions (see e.g. \MartinZK), giving a check of the minus signs and factors of 2 in \sfermint\ and \aqloopo.  From here on, we'll only consider spontaneous susy breaking.

\subsec{Computing the propagators, from hidden sector current correlators}

To use the results \mgauginoq - \aqloopoq, we express the above gauge field propagators $\ev{\CV \CV}$ in terms of $\ev{\CJ\CJ}$ current 2-point functions.  The main distinction from the unbroken gauge field case is that we generally need to consider the full (rather than 1PI) propagator, summing the series of 1PI contributions.  
 The full propagator of the spin 1 gauge field (in Landau gauge\foot{The final expression for physical masses is, of course, independent of the gauge choice. The effective potential does depend on the gauge choice.  As illustrated in \GrisaruVE, Landau gauge simplifies e.g. the form of the effective K\"ahler potential.}) is related to the spin 1 current $j_\mu$ 2-point function's function $\wt C_1(p^2)$ by
\eqn\fullgauge{\Delta _{\mu \nu}(p^2)\equiv i\ev{V_\mu (p)V_\nu(-p)}={1\over p^2(1+g^2\widetilde C_1(p^2))}\left(g_{\mu \nu}-{p_\mu p_\nu \over p^2}\right) +\CO(g^4).}
Similarly, the scalar component $D$  of the vector multiplet $V$ has full propagator 
\eqn\Dprop{\Delta (p^2)\equiv i\ev{D(p)D(-p)}={1\over 1+g^2\wt C_0(p^2)}+\CO (g^4).}
The gaugino has full propagator given in terms of the $\ev{j_\alpha \bar j_{\dot \alpha}}$ and $\ev{j_\alpha j_\beta}$ functions by   
\eqn\lamprop{\Delta _{\alpha \dot \alpha}(p^2)\equiv i\ev{\lambda _\alpha (p)\bar\lambda _{\dot \alpha}(-p)}=
\frac{p_{\a\ad}/p^2}{1+g^2\wt C_{1/2}+g^4\wt B_{1/2}^\dagger (1+g^2\wt C_{1/2})^{-1}\wt B_{1/2}/p^2}+\CO(g^4),}
and mass-type, R-symmetry breaking ($R(\Sigma )=-2$)  propagator 
\eqn\altlampropb{\eqalign{
\Sigma (p^2) &={i\over 2}\epsilon ^{\ad \bd}\ev{\bar\lambda _{\ad}(p)\bar\lambda _{\dot\beta} (-p)}\cr
&=g^2\frac1{1+g^2\wt C_{1/2}}\wt B_{1/2}\frac{1}{(1+g^2\wt C_{1/2})p^2+g^4\wt B_{1/2}^\dagger(1+g^2\wt C_{1/2})^{-1}\wt B_{1/2}}+\CO(g^4).}}
Using these propagators in \xiis,  the sfermion mass is given by \sfermint, with 
\eqn\sfermmass{\Xi ^{AB}(p^2)=\left({1\over 1+g^2\wt C_0}-{4\over 1+g^2\wt C_{1/2}+g^4\wt B_{1/2}^\dagger (1+g^2\wt C_{1/2})^{-1}\wt B_{1/2}/p^2}+{3\over 1+g^2\wt C_1}\right)^{AB}.}
These expressions apply in full generality, whether or not the gauge group is broken, and whether or not there are gauge messengers.  

\subsec{Broken vs unbroken symmetries}

Let's now summarize how the current correlators,  and consequently the gauge multiplet propagators, are affected by spontaneous symmetry breaking.  For unbroken generators, the current correlators are regular at low-momentum, and are $\CO(g^0)$: $\wt C_a^{unbr}=\wt C_a^{reg}\sim g^0$ and $\wt B_{1/2}^{unbr}=\wt B_{1/2}^{reg}\sim g^0$.  

For broken generators, on the other hand, the current 2-point functions get pole contributions, where they factorize on the NG boson states:\eqn\cbpolereg{\wt C^{brok}_a(p^2)=\wt C_a^{pole}(p^2)+ \wt C_a^{reg}(p^2), \qquad \wt B^{brok}_{1/2}=\wt B_{1/2}^{pole}(p^2)+\wt B_{1/2}^{reg}(p^2),}
where $a=0,1/2, 1$ and the ``pole" contributions are only present for the Higgsed part of the gauge group.  In particular, 
\eqn\conepole{(\wt C_1^{AB})^{pole}={(m_V^2)^{AB}\over g^2 p^2}, \qquad\hbox{with}\qquad (m_V^2)^{AB}=g^2 (v^2)^{AB},}
and $\wt C^{pole}_0$, $\wt C_{1/2}^{pole}$, and $\wt B^{1/2}_{pole}$ can have similar expressions, though their poles are generally displaced away from $p^2=0$, to  $p^2=\delta m_a^2$, by susy-breaking effects.  
All of the pole contributions in \cbpolereg\ are proportional to $m_V^2/g^2$.  

A key point is that we consider a perturbative expansion, of small $g$, but with  fixed $m_V$.  Then,  while the ``reg" contributions in \cbpolereg\ are $\CO(g^0)$ in our counting, the  pole contributions $\wt C_a^{pole}$ and $\wt B_{1/2}^{pole}$ are counted as $\CO(g^{-2})$, and then the full propagators, with the 1PI contributions resumed, are needed.  The pole contribution \conepole\ then shifts the pole of the full gauge field propagator to be at $m_V^2$, corresponding to the gauge field getting a mass from the Higgs (Schwinger) mechanism.  The current algebra approach does not assume weak coupling.  The other full propagators are similar: the other components of the massive vector multiplet get mass $m_V^2+\delta m_a^2$, where $\delta m_a^2$ are susy-breaking shifts.

\subsec{Special cases and connecting with earlier results}

For unbroken gauge groups, or more generally for unbroken subgroups of partially broken groups, the above resumed propagators are overkill: since their hidden sector current-correlators are $\CO(g^0)$ and we work to $\CO(g^2)$,  we should simply expand out the denominators in \fullgauge, \Dprop, \lamprop, and \altlampropb.  Our expressions then reduce to 
\eqn\ximss{\Xi ^{AB}(p^2)= -g^2\delta ^{AB}\big[\wt C_0(p^2)-4\wt C_{1/2}(p^2)+3\wt C_1(p^2)\big]+\CO (g^4) \qquad\hbox{(unbroken (sub) groups)},}
\eqn\sigmss{\Sigma ^{AB}(p^2)=g^2\delta ^{AB}{\wt B_{1/2}(p^2)\over p^2}+\CO(g^4)   \qquad\hbox{(unbroken (sub) groups)}.}
And then  \mgaugino\ and \sfermint\ properly reduce to the results of \mss\ for the special case of unbroken gauge groups.  

Also, in that case, our general relations \sigmasusy\ and \xisusy\ reduce to the 
expressions were given in \BuicanWS, there directly in terms of the hidden sector current-correlators rather than the vector propagators\foot{The difference is a Legendre transform, since the currents are sources for the gauge fields, e.g. $D\leftrightarrow \delta/\delta J$.}:
 \eqn\bmssx{\eqalign{\ev{Q^2(J(p)J(-p))}&=-4\wt B_{1/2}(p^2),\cr
 \ev{\bar Q^2 (Q^2(J(p)J(-p))}&=8p^2 (\wt C_0(p^2)-4\wt C_{1/2}(p^2)+3\wt C_1(p^2)).}}
 The relations \bmssx\ actually apply whether or not the $\CJ$ symmetry is spontaneously broken.  When the symmetry is unbroken, they imply the relations of  \BuicanWS
\eqn\bmssri{m_{gaugino}=-{1\over 4}g^2 \int d^4 x\ev{Q^2(J(x)J(0))} \qquad\hbox{(unbroken),}}
\eqn\bmssrii{m^2 _Q=g^2T^A_Q \ev{J^A}-{1\over 128\pi ^2}g^4 c_2(r_Q)\int d^4 x\ln (x^2 M^2)\ev{\bar Q^2(Q^2(J(x)J(0)))}   \qquad\hbox{(unbroken),}}
which can be obtained as special cases of our more general \mgauginoq\ and \sfermintq.

As another special case,  consider non-gauge messenger Higgsed gauge fields  (broken gauge fields with a supersymmetric tree-level mass spectrum).  This happens if the pole contributions are supersymmetric: $\wt C_a^{pole}(p^2)=C_{susy}^{pole}=m_V^2 g^{-2}/p^2$, coinciding for $a=0, \half , 1$, and $\wt B_{1/2}^{pole}=0$.  It then follows that 
\eqn\xiislim{\Xi ^{AB}=-g^2\delta ^{AB}\left({p^2\over p^2+m_V^2}\right)^2 \big[\wt C_0^{reg}(p^2)-4\wt C_{1/2}^{reg}(p^2)+3\wt C_1^{reg}(p^2)\big]\qquad\hbox{(no gauge messengers)}}
and
\eqn\sigislim{\Sigma ^{AB}=g^2\delta ^{AB}{p^2\over (p^2+m_V^2)^2}\wt B_{1/2}(p^2) \qquad\hbox{(no gauge messengers)}.}
In this case, our general expression \sfermint\ properly reduces to the result of \IntriligatorFR\ for Higgsed gauge mediation without gauge messengers. 

As seen in \sigmss,  massless gauge fields (or, more generally, non-gauge messengers) have $\Sigma =\CO(g^2)$, and then \aqloopo\ and \altlampropb\ gives $A_Q=\CO(g^4)$, a 2-loop effect.  In this case, \aqloopo\ essentially reduces to an expression given in \DistlerBT\ for $a_{ijk}$.

\subsec{One-loop sfermion mases and a-terms with gauge messengers}

When there are gauge messengers, the $\wt C_a^{pole}$ for $a=0,1/2$ can have their poles can be shifted away from $p^2=0$ by susy-breaking effects (and $\wt B_{1/2}^{pole}$ can be non-zero when susy-and R-symmetry are broken).  Then, as  seen from \sfermmass, the massive gauge fields  can have $\Xi\neq 0 $ already at tree-level, $\CO(g^0)$, 
\eqn\xiexp{\Xi ^{AB} (p^2) = \Xi ^{A'B' (0) }(p^2)+ \Xi ^{AB (1)}(p^2)+\CO(g^4).}
The $\CO(g^0)$ term $ \Xi ^{A'B' (0) }(p^2)$ is present only for the broken generators $A',B'$, while the $\CO(g^2)$ term $\Xi ^{AB(1)}(p^2)$ is present for both broken and unbroken generators.  Explicitly, the $\CO(g^0)$ term is given by using the pole contributions in \sfermmass,
\eqn\xiexpo{\Xi ^{(0)}(p^2)={1\over 1+g^2\wt C^{pole}_0}-{4\over 1+g^2\wt C^{pole}_{1/2}+g^4\wt B^{pole\ \dagger }_{1/2} (1+g^2\wt C^{pole}_{1/2})^{-1}\wt B^{pole}_{1/2}/p^2}+{3\over 1+g^2\wt C^{pole}_1}.}
Using this in the second term in \sfermint\ yields a non-zero one-loop contribution to the visible sector sfermion $m_Q^2$, which is unique to the massive gauge messenger's direct coupling to the visible sector sfermions.

Let's consider the one-loop contribution to $m_Q^2$ in more detail.  The general expression, 
obtained by using  \xiexpo\ in the momentum integral \sfermint\ will be rather complicated.  But for a certain class of theories, the result is readily evaluated: in cases where $\wt B^{pole}_{1/2}=0$ (e.g.\ if there is an unbroken $U(1)_R$), and where the $\wt C_a(p^2)$ each have just a single pole term, the result of integrating \xiexpo\ in \sfermint\ is simply 
\eqn\sfermionmone{(m_Q^2)^{(1)}={g^2\Delta c_Q\over 16\pi ^2} m_V^2 \ln \left({m_V^6 m_C^2\over m_\lambda ^8}\right),}
where $\Delta c_Q\equiv c_2(r_Q')-c_2(r_Q)$ accounts for the sum over massive generators.  We'll see that $m_V^6m_C^2\leq m_\lambda ^8$,  so the one-loop masses \sfermionmone\ are tachyonic.   

Another case where our general expressions simplify is in the limit of small susy-breaking (whether or not one restricts to the class of theories mentioned before \sfermionmone).   
In this limit, the one-loop contribution to sfermion $m_Q^2$ (whether or not one restricts to the class of theories mentioned before \sfermionmone) reduces to 
\eqn\mqoneexpp{(m_Q^2)^{(1)}=-{g^2\over 16\pi ^2} \Delta c_Q \left({(\bar \phi , \phi)( \bar F, F)-(\bar \phi , F)( \bar F, \phi) \over (\bar \phi, \phi )   (\bar \phi, \phi )}\right) +\CO(|F|^4).}
The inner product in \mqoneexpp\ is defined as e.g. 
\eqn\innerprod{(\bar \phi , \phi )^{AB}\equiv \sum _i \phi _i^\dagger \{T^A, T^B\}\phi _i = (m_V^2)^{AB}/g^2, \qquad (\bar F, \phi )^{AB}\equiv \sum _i\bar F_i \{ T^A, T^B\} \phi _i,}
etc.\ (and we simplified \mqoneexpp\ by assuming that the broken generator expectation values satisfy $(\bar \phi , \phi )^{A'B'}=(\bar \phi, \phi )\d^{A'B'}$,  $(\bar F, \phi )^{A'B'}=(\bar F, \phi)\delta ^{A'B'}$, and $(\bar F, F)^{A'B'}=(\bar F, F)\delta ^{A'B'}$).

The expression \mqoneexpp\ for the one-loop $m_Q^2$ is manifestly always tachyonic.  For large $m_V$, it exhibits decoupling, giving  $m_Q^2 \sim |F|^2/|m_V|^2$.  This parameteric dependence is similar to the usual 2-loop masses in theories without gauge messengers,  $m^2\sim |F| ^2/|M|^2$, with $M$ the mass of the non-gauge messenger hidden sector field.  Actually, the one-loop contribution \mqoneexpp\ has even faster decoupling if the large $m_V$ is associated with a large goldstino pseudomodulus expectation value (as in the inverted hierarchy \WittenKV\ type models).  As we'll show (using results from \PolchinskiQK), for large goldstino pseudomodulus the one-loop $m_Q^2$ contribution \mqoneexpp\ scales as $m^2 \sim \alpha |F|^3/|M|^4$.  This is special to the one-loop contribution, and the two-loop contribution exhibits normal $m^2\sim \alpha ^2|F|^2/|M|^4$.  So in this case, if the goldstino pseudomodulus is sufficiently large, the one loop contribution could end up being parameterically suppressed as compared with the two-loop contribution.

Now consider the a-terms, given by \aqloopo.  When there are gauge messengers,  our general expression \altlampropb\ gives a contribution $\Sigma ^{A'B'}$ for the massive gauge fields which is $\CO(m_V^2\sim g^0)$, obtained from using  $\wt C_a^{pole}$ and $\wt B_{1/2}^{pole}$ contributions in \altlampropb.  Using this in \aqloopo\ gives a 1-loop contribution, $A_Q\sim \CO(g^2)$, given by 
\eqn\aqloop{\eqalign{A_Q^{(1)}&=-2g^2 \sum _{A'B'} T^{A'}_{r_Q} T^{B'}_{r_Q}\int {d^4 p\over (2\pi )^4}{1\over p^2}\Sigma ^{pole\ A'B'}(p^2)\cr &=
-2g^2(c_2(r_Q')-c_2(r_Q))\int {d^4 p\over (2\pi )^4}{1\over p^2}\Sigma ^{pole}(p^2)}}
where $\Sigma ^{pole \ A'B'}$ is the $\CO(g^0)$ term in \altlampropb\ obtained by replacing $\wt C_a\to \wt C_a^{pole}$ and $\wt B_{1/2}\to \wt B_{1/2}^{pole}$, and in the last line we  simplified the sum by taking $\Sigma ^{pole\ A'B'}=\delta ^{A'B'}\Sigma ^{pole}$ for the broken generators).  

In the small susy-breaking limit, we will verify that the one-loop expressions \mqoneexpp\ for the sfermion $m_Q^2$, and \aqloop\ for $A_Q$, agree with results which can be obtained from the one-loop effective K\"ahler potential of \GrisaruVE\ and references cited therein.  More generally, for arbitrary susy-breaking, these one-loop effects can in principle be determined from the one-loop Coleman-Weinberg potential, computed as a function of non-zero $Q$ and $F_Q$ background expectation values.  Aspects of this connection will be given in an appendix.  

\subsec{Two loop contributions to $m_Q^2$}

Our general expression \sfermint\ also gives the $\CO(g^4)$ two-loop contribution to $m_Q^2$, upon using the 
$\CO(g^2)$ term of \xiexp.  This contribution is present for both broken and unbroken gauge fields coupling to the visible sector.  For unbroken gauge fields, \ximss\ applies.  For broken gauge fields, we need to expand \sfermmass\ using \cbpolereg, treating the regular contributions as a perturbation, e.g. 
 \eqn\cregpole{{1\over 1+g^2 \wt C_a(p^2)}={1\over 1+g^2 \wt C_a^{pole}(p^2)}-{1\over 1+g^2 \wt C_a^{pole}(p^2)}g^2 \wt C_a^{reg}(p^2){1\over 1+g^2 \wt C_a^{pole}(p^2)}+\CO(g^4),}
where the first term is $\CO(g^0)$ and the second is $\CO(g^2)$.  

There are tachyonic contributions to the two-loop $m_Q^2$ masses, mediated by both massless and massive gauge fields, as in Fig. 2a and 2b.   For the massless gauge fields the expansion \ximss\ applies, and the functions $\wt C_a^{unbr}(p^2)$ there get additional contributions from a loop of massive gauge field, as in Fig. 3, which  contributes to $\wt C_{a}^{unbr}$ with the opposite sign as that of massive matter.  Consider, in particular, the UV behavior, which is independent of any spontaneous symmetry breaking, and is determined by the 
leading $x\to 0$ term in the current-current OPE, much as in \mss,
\eqn\opeex{J^A(x)J^B(0)={c\delta ^{AB}\over 16\pi ^4 x^4}+{f^{ABC}J^C(0)\over x^2}+\dots.}
Our normalization is such that a weakly coupled theory with  matter fields $\Phi _i$ in representations $r_i$ of the symmetry group $G$ would contribute to \opeex\ as 
\eqn\cmatterex{c_{matter}=\sum _i T_2(r_i) \qquad\hbox{(weakly coupled)},}
where as usual we define $\Tr (T_{r_i}^AT^B_{r_i})=T_2(r_i)\delta ^{AB}$ and an $SU(N)$ fundamental has $T_2({\bf N})=\half$. 
Unitarity requires that matter contributes $c_{matter}>0$.    But, with gauge messengers, the massive gauge fields yield an additional {\it negative} contribution.   As seen in background field perturbation theory, the coefficient $c$ is equal to the contribution of the massive fields  to the one-loop beta function $b_1=3T_2(G)-T_2(matter)$ of the gauge coupling $g$:
\eqn\cbeta{c=-(b_1'-b_1)\equiv c^{gauge}+c^{matter}, \qquad c^{gauge}\equiv 3T_2(G)-3T_2(G')<0.}
The  total $c$ in \cbeta\ can be positive or negative.  

\subsec{One-loop gaugino masses, enhanced already at $\CO(F)$}

Our general expression \mgaugino\ for the susy-breaking contribution to the gaugino mass applies for both the broken $G'/G$ generators, as well as the unbroken $G$ generators.  Let's consider, in particular,  the masses of the gaugino partners of the massless $G$ gauge fields.  In this case, using \sigmss, the expression \mgaugino\ reduces to the expression of \mss
\eqn\mgauginoun{m_{gaugino}^{unbr}=g^2 \wt B_{1/2}^{unbr}(0)\qquad\hbox{(unbroken (sub) groups).}}
Although the expression is the same as in cases without gauge messengers, the gauge messengers can still have a dramatic effect; they can provide a beneficial enhancement of these gaugino masses as compared with the standard, non-gauge messenger case.  

Let's first recall the suppression of gaugino masses in the case of gauge mediation without gauge messengers.   The effect is evident at $\CO(F)$ in a small-susy breaking expansion, and the higher order in $F$ terms in the more general case do not alter the result much.   In this limit hidden sector matter contributes 
\eqn\mgauginof{\eqalign{\wt B_{1/2}^{unbr}(0)&=\sum _i c_i {F_i \over M_i}+\CO\bigg(F\Big|{F\over M^2}\Big|^2\bigg)\qquad\hbox{(no gauge messengers)}\cr &=0+\CO\bigg(F\Big|{F\over M^2}\Big|^2\bigg) 
\qquad\hbox{(weakly coupled)},}}
where $c_i$ is the contribution to the OPE coefficient in \opeex, given by $c_i=T_2(r_i)$ in weakly coupled theories.  The vanishing in \mgauginof, at least in the context of weakly coupled examples (e.g. taking all $c_i=1$), comes from $\sum _i c_iF_i/M_i\to \Tr (FM^{-1})={d\over dX}\ln \det M=0$, where $X$ is the goldstino superfield and the vanishing is related to the condition that susy is spontaneously broken \PolchinskiAN, see \KomargodskiJF\ for a recent and general discussion.  

In theories with gauge messengers, the cancellation in \mgauginof\ does not occur, as was noted in  \refs{\PolchinskiAN, \PolchinskiAJ}.  There are now contributions from both the matter, and also from  massive gauge multiplets running in the loop,  as in Fig. 3, so $\wt B_{1/2}^{unbr}=\wt B_{1/2, matter}^{unbr}+\wt B_{1/2, gauge}^{unbr}$.  Consider, for example, weakly coupled theories where the goldstino pseudomodulus gets a large expectation value, $\ev{X}$, where the group is Higgsed $G'\to G$, resulting in a low-energy theory with relatively small susy-breaking.  The low-energy theory has a loop correction to the $G$ gauge kinetic superpotential term
\eqn\wlowcont{W_{low}\supset {k\over 32\pi ^2} \ln X \ (W_\alpha W^\alpha)_G,}
which leads to a gaugino mass proportional to the coefficient $k$ 
\eqn\mgauginof{m_{gaugino}^{unbr}={k\over 16\pi ^2}{F_X\over X}.}
As usual, the coefficient $k$ in \wlowcont\ is related to the contribution of the massive matter to the beta function, $k=c^{unbr}=c^{unbr, gauge}+c^{unbr, matter}$. 
In these classes of theories, as in \PolchinskiAJ, $k$ can also be related to the contribution of the integrated-out matter to the $U(1)_R$ ABJ anomaly $\Tr R G^2$, and thus to the amplitude for the R-axion to decay to two $V_G$ gauge fields:
\eqn\kexis{\eqalign{k&=c^{unbr, gauge}+c^{unbr, matter}=3(T(G)-T(G'))+\sum _i T_2(r_i)\cr 
k&=-(T(G')-T(G))-\sum _i (R_i -1)T_2(r_i),}}
where the $3(T(G)-T(G'))$ is the contribution of the $G'/G$ vector mutiplets ($T(G)=T_2({\rm adj})$, e.g. $T_2(SU(N))=N$) to $\wt B^{unbr}_{1/2, gauge}(0)$ and $R_i=R(\Phi _i)$ are the R-charges of the hidden sector matter fields under a (generally anomalous and approximate) tree-level R-symmetry.   
We'll illustrate all this for some weakly coupled examples in sect. 7.

\subsec{Generalizing to visible sector pseudomodulus effective potentials}

Suppose that the visible sector has some classical D and F flat moduli, as is the case in the MSSM.  Then susy-breaking effects from the hidden sector generate an effective potential $V_{eff}(Q, \bar Q)$, which reduces to $V_{eff}(Q, \bar Q)\approx m_Q^2|Q|^2$ near the origin.  The full form of the effective potential, for $\ev{Q}$ away from the origin, could be useful for some applications, e.g. cosmology, if the fields happen to start away from the origin (see \dMM\ for discussion and analysis in the context of weakly coupled ordinary gauge mediation).  We here find the general expression from
\eqn\veffone{V_{eff}=\half \Tr \int {d^4p\over (2\pi )^4}\bigg\{\ln (1+g^2\widetilde C_0^{tot})-2\ln \left[ (1+g^2\widetilde C^{tot}_{1/2})^2+{g^4|\widetilde B^{tot}_{1/2}|^2\over p^2}\right]+3\ln (1+g^2 \widetilde C^{tot}_1)\bigg\},}
where the $\Tr$ is over the adjoint gauge indices of $\wt C_a^{AB}$ etc. 
The dependence on $\ev{Q}$ arises because here 
\eqn\vishid{\wt C_a^{AB, tot}=\wt C^{AB}_a+\wt C_a^{AB, vis}, \qquad \wt B_{1/2}^{AB, tot}=\wt B_{1/2}^{AB}+\wt B_{1/2}^{AB, vis},}
where $\wt C_a^{AB, vis}$ and $\wt B_{1/2}^{AB, vis}$ are additional visible sector contributions to the current correlators.  Since we're interested in a classically supersymmetric $\ev{Q}$, we have
\eqn\visbc{\wt C_a^{AB, vis}={\bar Q \{ T^A, T^B\}Q\over p^2}+\CO(g^2), \qquad \wt B_{1/2}^{AB, vis}=0+\CO(g^2),}
where we'll now just consider leading order in $g^2$.  Using the pole contributions for the hidden sector $\wt C_a$ and $\wt B_{1/2}$ then gives the one-loop effective potential; we could similarly use the regular contributions to the current correlators to 
compute the two-loop effective potential.

We can also recover the a-term \aqloop\ from the one-loop effective potential \veffone, by generalizing \visbc\ to include a background expectation value for $F_Q$.  It suffices to work to $\CO(F_Q)$,  and the visible sector contributes 
\eqn\visbcf{(\wt C_a)^{AB}_{vis}={(\bar Q, Q)^{AB}\over p^2} +\CO(|F_Q|^2), \qquad (\wt B_{1/2})_{vis}^{AB}=-{(\bar Q, F_Q)^{AB}\over p^2}+\CO(|F_Q|^2).}
Using these as additional contributions in \veffone, we obtain an effective potential that, upon extracting the $\CO(\bar F_Q)$ term, gives a term in the visible sector soft-breaking effective lagrangian\foot{Such a soft-breaking potential was considered in \NibbelinkSI, and their result can be compared with the special case where our $\wt B_{1/2} (p^2)$ is replaced with a momentum-independent constant $m_\lambda$.}
\eqn\leffk{{\cal L}_{vis}\supset \int d^4 \theta \theta ^2 \wt k(\bar Qe^{2V}, Q)\supset \bar F_Q \partial _{\bar Q}\wt k(\bar Q e^{2V},Q).}  This yields at one loop
\eqn\kqis{\bar F_Q\wt \partial _{\bar Q}k(|Q|^2)=2\Delta c_Q(\bar F_Q, Q)\int {d^4 p\over (2\pi)^4}{2g^4 \wt B_{1/2}^{pole}/p^4\over \big(1+{g^2 |Q|^2\over p^2}+g^2 \wt C_{1/2}^{pole}\big)^2+{g^4|\wt B_{1/2}^{pole}|^2\over p^2}}.}  Expanding $\wt k(|Q|^2)$ around the origin, $k(|Q|^2)\approx A_Q|Q|^2+\CO(|Q|^4)$,   gives the A-term \aqloop.

\newsec{Gauge field propagators and current algebra}

We here give a general, current-algebra based description of current correlators, allowing for the possibility that the symmetry is spontaneously broken, $G'\to G$, as well as spontaneous supersymmetry breaking. Because the symmetry $G'$ is to be gauged, we do not include explicit symmetry breaking.  So the associated current is necessarily conserved, and thus satisfies  $\bar D^2\CJ = D^2\CJ =0$ (the covariant version, once the symmetry is gauged).  So, even if the symmetry is spontaneously broken,   \eqn\superj{\J=J+i\theta j-i\bar\theta\bar j-\theta\sigma^\mu\bar\theta j_\mu+\half\theta\theta\bar\theta\bar\sigma^\mu\partial_\mu j-\half\bar\theta\bar\theta\theta\sigma^\mu\partial_\mu\bar j-\frac{1}{4}\theta\theta\bar\theta\bar\theta(\Box J +J_D),}
with $\partial ^\mu j_\mu=0$ and the other components unconstrained (here $\ev{J^A_D}=\bar F_0T^AF_0$, which vanishes in the case of unbroken symmetry).  Equivalently, supersymmetry relates the higher components of the multiplet to the $J$ component as 
\eqn\jcomprel{j_\alpha = -i[Q_\alpha, J], \qquad j_\mu =-{1\over 4}\bar \sigma ^{\dot \alpha \alpha} _\mu(\{\bar Q_{\dot \alpha }, [Q_\alpha, J]\}-\{Q_\alpha , [\bar Q_{\dot \alpha}, J]\}),}
where, again, this holds whether or not any of the symmetries are spontaneously broken.

The current 2-point functions are constrained by the symmetries to have the same general form as in \mss, regardless of whether or not the symmetry is spontaneously broken\foot{It can be shown, as in \ShoreTG, that $\ev{j_\mu ^A(p)J^B(-p)}=ik_\mu f^{ABC}\ev{J^C}/k^2$, where $f^{ABC}$ are the group structure constants.  We'll take $f^{ABC}\ev{J^C}=0$, so $\ev{j_\mu^A(p) J^B(-p)}$ will play no role here.}: 
\eqna\compjj{}
$$\eqalignno{
&\vev{J(p)J(-p)}=\wt C_0(p^2)&\compjj a\cr
&\vev{j_\alpha(p)\bar j_{\dot\alpha}(-p)}=-\sigma_{\alpha\dot\alpha}^\mu p_\mu\wt C_{1/2}(p^2)&\compjj b\cr
&\vev{j_\mu(p)j_\nu(-p)}=-(p^2\eta_{\mu\nu}-p_\mu p_\nu)\wt C_1(p^2)&\compjj c\cr
&\vev{j_\alpha(p)j_\beta(-p)}=\epsilon_{\alpha\beta}\wt B_{1/2}(p^2)&\compjj d
}$$
We suppress the adjoint indices -- if the currents have indices $A$ and $B$, the functions in \compjj{}\ have adjoint indices, $\wt C^{AB}_a$ and $\wt B^{AB}_{1/2}$.  If the symmetry is unbroken, $\wt C^{AB}_a=\delta ^{AB}\wt C_a$ and $\wt B^{AB}_{1/2}=\delta ^{AB}\wt B_{1/2}$.   Also, as usual, we drop a $(2\pi )^4\delta ^{(4)}(0)$.

At short distances, i.e.\ large $p$, it does not matter whether or not symmetries are spontaneously broken: the theory becomes effectively supersymmetric, and any spontaneous breaking of the $\CJ$ symmetry becomes irrelevant.  The UV behavior of the current correlators is thus substantially the same as was discussed in \mss; the $C_a$, for $a=0,1/2, 1$, all have a universal dependence on the UV cutoff $\Lambda$ 
\eqn\cac{\lim _{x\to 0} C_a(x^2)={c\over 16\pi ^4}, \qquad \widetilde C_a(p^2)={c\over 8\pi ^2}\ln (\Lambda/M)+\hbox{finite},}
and $\wt B_{1/2}(p^2/M^2)$ is finite.  The coefficient $c$ is given by the OPE \opeex.  As remarked there, and we'll discus further in the next section, the constant $c$ appearing in \opeex\ and \cac\ need not be positive, since massive gauge bosons contribute negatively to $c$.

The gauge field propagators can be expressed as follows in terms of the current correlators \compjj{}.  
In a superspace notation, the 1PI correction to $\ev{\CV(p)\CV(-p)}$ is $-g^2\ev{\CJ \CJ}$, and summing these 1PI corrections gives the full gauge field propagators.  Consider first the spin 1 gauge field.  The 1PI self energy is $\Pi _{\mu \nu}(p^2)=(g_{\mu \nu}-p_\mu p_\nu/p^2)(-g^2 \wt C_1(p^2))$.  Summing the series of these gives the full gauge propagator\foot{We generally work in Euclidean space, but it should be understood that, e.g in the numerator of $\Delta _{\mu \nu}$, we  use Minkowski $p^\mu$ without introducing new notation.  Note that $g^2\wt C_1^{pole}=m_V^2/p_{Euc}^2=-m_V^2/p_{Mink}^2$, so the ordinary Minkowski space, Landau gauge propagator is recovered.} 
\eqn\fullgauge{\ev{V^A_\mu (p)V^B_\nu(-p)}\equiv -i\Delta _{\mu \nu}(p^2)={-i\over p^2(1+g^2\widetilde C_1(p^2))^{AB}}\left(g_{\mu \nu}-{p^\mu p^\nu \over p^2}\right) +\CO (g^4). }
Similarly, the scalar component $D$  of the vector multiplet $V$ has full propagator 
\eqn\Dpropb{\ev{D^A(p)D^B(-p)}\equiv -i\Delta (p^2)={-i\over (1+g^2\wt C_0(p^2))^{AB}}+\CO (g^4),}
where in \fullgauge\ and \Dpropb\ we have explicitly written the $A$, $B$ adjoint indices.  
The gaugino has propagator (now suppressing the adjoint indices)
\eqn\lampropb{\ev{\lambda _\alpha (p)\bar\lambda _{\dot \alpha}(-p)}\equiv -i\Delta _{\alpha \dot \alpha}(p^2)=
\frac{-ip_{\a\ad}/p^2 }{1+g^2\wt C_{1/2}+g^4\wt B_{1/2}^\dagger (1+g^2\wt C_{1/2})^{-1}\wt B_{1/2}/p^2}+\CO(g^4),}
which comes from summing 1PI terms as \foot{For simplicity, take $\wt C_{1/2}^{AB}$ and $\wt B_{1/2}^{AB}$ to commute,  $[\wt C_{1/2}, \wt B_{1/2}]=0$, though \lampropb\ is general.}
\eqn\propsum{\sum_{n, m=0}^\infty{2n+m\choose m}\Big(-g^2\wt C_{1/2}\Big)^m\bigg(-\frac{g^4|\wt B_{1/2}|^2}{p^2}\bigg)^n=\frac{1+g^2\wt C_{1/2}}{(1+g^2\wt C_{1/2})^2+g^4|\wt B_{1/2}|^2/p^2},}
where the combinatoric factor is because, while the $n$ $\wt B_{1/2}$ and $n$ $\wt B_{1/2}^\dagger$ necessarily alternate along the propagator line, the $m$ $\widetilde C_{1/2}s$ can be sprinkled anywhere among them.  We can also form the chirality flipped gaugino propagator, which differs from the sum \propsum\ in that we have an extra $\wt B_{1/2}^\dagger$ factor (correspondingly, replace $2n+m\to 2n+m+1$ in the combinatoric factor) and the sum then yields 
\eqn\altlampropbb{\Sigma_{\a\b}^\dagger=-\frac i2\ev{\lambda _\alpha (p)\lambda _\beta (-p)}=\frac{g^2}2\e_{\a\b}\frac{\wt B_{1/2}^\dagger}{(1+g^2\wt C_{1/2})^2p^2+g^4|\wt B_{1/2}|^2}+\CO(g^4).}

In the following subsection, we discuss how spontaneous symmetry breaking of the $\CJ$ symmetry, and supersymmetry, show up in the current-algebra.  We'll next discuss the coupling to gauge fields, and the mass-spectrum of the Higgsed vector multiplet in this description.

\subsec{Pole contributions to current correlators when the symmetry is spontaneously broken}

At long distances, spontaneous breaking of the $\CJ$ symmetry has a dramatic effect on the current correlators \compjj{}.   When the symmetry is unbroken, the current $\CJ$ has vanishing 1-point function, and the functions $\wt C_a(p^2)$ and $\wt B_{1/2}(p^2)$ in \compjj{} are regular, without poles, at $p^2=0$ \BuicanWS.  On the other hand, when there is spontaneous symmetry breaking, the  $\CJ$ has non-zero 1-point functions between the vacuum and NG boson and superpartner states.  These current one-point functions lead to the additional ``pole" contributions \cbpolereg\ to $\wt C_a'$ and $\wt B_{1/2}'$, where the prime indicates that it's a broken generator of the UV theory.

Consider first  the pole contribution to $\wt C_1'$, which is a standard effect.  We'll first consider the broken symmetry as a global symmetry, and next discuss the effect of making it a gauge symmetry.   When the $j_\mu$ symmetry is spontaneously broken, the current does not annihilate the vacuum, but instead creates a NG boson state:
\eqn\jmupi{\langle \pi ^{A'}(p)| j^{\mu B'}(x)|0 \rangle =iv_{A'} \delta ^{A'B'}p^\mu  e^{-ip\cdot x},}
where $A'$ runs over the broken generators (if the symmetry breaking is $G\to G'$, then $A'\in G'/G$) and $v_{A'}$  (a.k.a. $f_\pi$, but not here) is a real order parameter for the spontaneous symmetry breaking.  Current conservation $\partial _\mu j^\mu=0$ implies that $\pi $ is massless. We can write the associated contribution to the current as  $j_\mu ^{A'}\supset - i v_{A'}\partial _\mu \pi ^{A'}$.  The $j_\mu^{A'}$ two-point function can then factorize on two one-point functions, leading to 
\eqn\conepole{\wt C_1^{pole\ A'B'}(p^2)={v_{A'}^2\delta ^{A'B'}\over p^2},}
where the $1/p^2$ comes from the intermediate propagator with the massless NG boson $\pi$ and we have arranged the numerator in \conepole\ to be a diagonal matrix by a choice of basis. In a general basis,  $\wt C_1^{pole}$ is written in terms of a matrix $(v^2)^{AB}$,
\eqn\conepolegen{\wt C_1^{pole\ AB}(p^2)={(v^2)^{AB}\over p^2}.}
We also allow $A$ and $B$ in \conepole\ to run over any unbroken generators, simply taking $v_A=0$ for them.

Supersymmetry, even if spontaneously broken,  implies that  each NG boson has superpartners, and resides in a chiral superfield
\eqn\ngcomps{\Pi ^{A'}= (\sigma +i\pi )^{A'}(y)+\sqrt{2}\theta \psi _\pi ^{A'}(y)+\theta \theta F _\pi^{A'}( y),}
(with $y^\mu =x^\mu +i\theta \sigma ^\mu \bar\theta $, as usual \WessCP), and $\sigma$ and $\pi$ are real fields, with $\pi $ the NG boson (we'll often suppress the index $A'\in G'/G$).   We are here considering the ``doubled" NG case, in the terminology of the large literature\foot{The ``non-fully doubled" case  refers to the possibility that some of the NG supermultiplets can have both $\pi$ and $\sigma$ in \ngcomps\ being true NG bosons.  The fully doubled case occurs e.g. if $G/H$ is a symmetric space;  see e.g. \LercheQA\ for a nice discussion.    Much of the large literature on (pseudo) NG bosons in supersymmetric theories considers ungauged symmetries, where both fully doubled and non-fully doubled are viable possibilities.  Gauging a case with non-doubled NG bosons would lead to a peculiarity:  there would not be enough bosons in the spectrum to complete the longitudinal polarization of a massive vector supermultiplet, so supersymmetry would be ``shattered" \refs{\ShoreBH, \ShoreTG}.  We believe that this phenomenon can not occur, that the NG bosons of gauged symmetries must be fully doubled, to avoid gauge anomalies (as seen in the example presented in \ShoreTF).}  on NG bosons in susy theories.    As exhibited with \jmupi, the broken currents $j^{\mu A'}(x)$ act as interpolating fields, to create the NG boson state $|\pi ^{A'}(p)\rangle$ from the vacuum.  We'll now use the supermultiplet structure of the chiral superfield \ngcomps\ to show that $J^{A'}$ and $j_\alpha ^{A'}$ likewise act as interpolating fields to create the states $|\sigma ^{A'}(p)\rangle $ and $|\psi _\pi ^{A'}(p)\rangle$, respectively, from the vacuum.

Consider first the case of unbroken susy, which was already considered in \refs{\ShoreRI, \ShoreRJ}, whose results we'll now recall.  Writing the NG boson fields as the chiral superfield \ngcomps, we have e.g. $[Q_\alpha, \phi]=-i\psi _\alpha$, $\{ Q_\alpha, \psi _\beta\}=2i\epsilon _{\alpha \beta}F$,  $\{\bar Q_{\dot \alpha}, \psi _\alpha\}=4 \partial _{\alpha \dot \alpha}\phi$, etc., where $\phi =\sigma +i\pi$.  The states, on the other hand,  transform like the functional derivative with respect to the field, so $Q_\alpha |\phi \rangle =0$, $Q_\alpha |\psi _\beta\rangle =-i\epsilon _{\alpha \beta}|\phi\rangle$, $Q_\alpha |\bar \phi \rangle = 4i p_{\alpha \dot \alpha}|\bar \psi ^{\dot \alpha}\rangle$ etc., which yields e.g. 
\eqn\qstates{Q_\alpha |\sigma \rangle =-iQ_\alpha |\pi \rangle = 2ip_{\alpha \dot \alpha}|\bar \psi ^{\dot \alpha}\rangle , \qquad Q_\alpha |\psi _\beta\rangle = -i \epsilon _{\alpha \beta}|\phi\rangle, \qquad Q_\alpha |\bar \psi _{\dot \alpha}\rangle = -2ip_{\alpha \dot \alpha}|F\rangle,}
where $|\phi \rangle = |\sigma\rangle +i|\pi \rangle$.  It follows from \qstates\ that e.g. $[\bar Q_{\dot \alpha}, Q_\alpha]|\pi\rangle =-4\partial _{\alpha \dot \alpha}|\sigma\rangle$.  Combining this with the susy-relation \jcomprel\ between $j_\mu$ and $J$, it follows that
\eqn\qqstaterel{\langle \pi ^{A'}(p)|j^{B'}_\mu(x)|0\rangle =-{1\over 4} \bar\sigma _\mu ^{\dot \alpha \alpha}\langle \pi ^{A'}(p)|[\bar Q_{\dot \alpha}, Q_\alpha ]J^{B'}(x)|0\rangle = \langle \sigma ^{A'}(p)|ip_\mu J^{B'}(x)|0\rangle.} 
Likewise, we can use the susy relation \jcomprel, $j_\alpha =-i[Q_\alpha, J]$, and \qstates\ to relate $\langle \psi _\alpha |j_\beta |0\rangle$ to $\langle \sigma |J |0\rangle$.  The upshot is that supersymmetry relates the one-point function  as
\eqn\jones{\eqalign{\langle \sigma ^{A'}(p)|J^{B'}(x)|0\rangle &=v_{A'}\delta ^{A'B'}e^{-ip\cdot x}, \qquad \langle \pi ^{A'}(p)|j^{B'}_\mu (x)|0\rangle =iv_{A'}\delta ^{A'B'}p_\mu e^{-ip\cdot x}, \cr \langle {\psi}^{A'} _{\pi,\alpha}(p)|j^{B'}_\beta(x) |0\rangle &= -v_{A'}\delta ^{A'B'}\epsilon _{\alpha \beta}e^{-ip\cdot x},}}
with susy implying the  equality of the coefficient $v_{A'}$ in each one-point function in \jones.  
As seen from \jones, the current supermultiplet fields act as interpolating fields for creating the NG supermultiplet fields:
\eqn\jcompng{J ^{A'}\supset v_{A'}\sigma, ^{A'} \quad j^{A'}_\alpha \supset v_{A'}\psi ^{A'}_{\pi,\alpha}, \quad j^{A'}_\mu \supset -iv_{A'}\partial _\mu \pi ^{A'};} 
Writing it in superspace,
\eqn\jngb{\CJ ^{A'}\supset \half v_{A'}\Pi ^{A'}+h.c..}

The one-point functions \jones\ lead to pole contributions, analogous to \conepolegen, for $\wt C_0(p^2)$ and $\wt C_{1/2}(p^2)$, with the poles displaced from $p^2=0$ when susy is spontaneously broken. 
Consider first the case of unbroken supersymmetry.  Then all $\wt C_a(p^2)$ are equal, and $\wt B_{1/2}=0$.  In particular, $\wt C_0(p^2)$ and $\wt C_{1/2}(p^2)$ must then also have a pole contribution, coinciding with \conepolegen:
\eqn\capolesusy{\wt C_{a=0, 1/2, 1}^{pole}=\wt C_{susy}^{pole}={(v^2)^{AB}\over p^2} \qquad\hbox{(unbroken susy)}.}
If susy is unbroken,  the NG superpartners $\sigma$ and $\psi _\pi$ are massless, and the pole in $\wt C_0^{pole}$ and $\wt C_{1/2}^{pole}$ come from factorizing using \jones\ on an intermediate $\sigma$ or $\psi _\pi$ state, respectively.

We now allow for spontaneous supersymmetry breaking.  Of course,  $\wt C_1^{pole}$ is unchanged and the massless NG boson $\pi$ still resides  in a chiral superfield \ngcomps, with partners $\sigma$ and $\psi _\pi$,
but now the F-component $F_\pi ^{A'}$ in \ngcomps\ can have a non-zero expectation value.  
The one point functions \jones\ are not altered by spontaneous susy breaking: in particular, the order parameter $v_{A'}$ remains  the same for the spin $a=0,1/2, 1$ 1-point functions\foot{Once susy is spontaneously broken, \qstates\ should be re-expressed in terms of the supercurrent acting on the states, since $Q_\alpha$ does not exist (it's associated with a zero momentum goldstino).}.   The effect of the susy breaking is to give the fields $\sigma$ and $\psi _\pi$ susy-breaking mass splittings from the massless NG bosons $\pi$.   There are still pole contributions $\wt C_a'^{pole}(p^2)$, for $a=0$ and $a=1/2$, from where they factorize on the $\sigma$ and $\psi _\pi$ states,  respectively.  But now the pole are shifted away from $p^2=0$, and are instead at the susy-breaking masses $\delta m_0^2\equiv m_\sigma ^2$, and $\delta m_{1/2}^2\equiv m_{\psi _\pi}^2$:
\eqn\clom{\widetilde  C'_a(p^2)^{pole}={v^2\over p^2+\delta m_a^2} \qquad\hbox{single pole case},}
with $\delta m_1^2\equiv 0$, of course, since the NG boson $\pi$ remains massless.  As we have noted, the coefficient $v$ in the one-point functions \jones\ are equal, even if susy is spontanesouly broken.  So supersymmetry, even if spontaneously broken, implies that the residues in \clom\ are all equal to $v^2$, for $a=0,1,2$.  
As we'll discuss shortly, \clom\ is a special case, and more generally there are several poles.

An example of how the partners $\sigma$ and $\psi _\pi$ can get susy-breaking masses is via 
\eqn\wsplit{W_{split}=\half (\delta m_{1/2}+|\delta m_{1/2}|^2\theta ^2)\Pi ^{A'}\Pi ^{A'},}
where the $\theta ^2$ term keeps $\pi$ massless.  It follows from \wsplit\ that the susy-breaking mass splittings of the scalar and fermion partners of the NG boson are related
\eqn\mbreakreln{\delta m_0^2=2\delta m_{1/2}^2 \qquad\hbox{(F-term breaking)} }
which is a characteristic relation of F-term breaking.   We'll briefly discuss D-type breaking later, and note that the relation \mbreakreln\ does not hold in that case.  The example \wsplit\  is incompatible with an R-symmetry, but it has a simple variant that is compatible with unbroken R-symmetry: 
\eqn\wsplitt{W_{split,R}=\half |\delta m_{1/2}|^2\theta ^2 \Pi ^{A'}\Pi ^{A'}+\delta m_{1/2}\Pi ^{A'}\Psi ^{A'},}
where $\Psi ^{A'}$ are some other fields, and \wsplitt\ respects an unbroken $U(1)_R$ symmetry under which $R(\Pi ^{A'})=0$ and $R(\Sigma ^{A'})=2$.    It gives susy breaking mass splittings in the NG boson supermultiplet which also satisfy \mbreakreln.

In \clom\ we have written a single pole term.  More generally (and as we'll illustrate in weakly coupled examples), $\wt C_0^{pole}(p^2)$ and $\wt C_{1/2}^{pole}(p^2)$ can have a sum of pole terms
\eqn\casum{\wt C_{0}^{pole}=\sum _i  {r_{0, i}\over p^2+\delta m_{0,i}^2}, \qquad \wt C_{1/2}^{pole}=\sum _i {r_{1/2, i}\over p^2+\delta m_{1/2, i}^2}.}
Supersymmetry, even if it's spontaneously broken, implies that the  sum of the residues of these poles are constrained to equal $v^2$.   The multi-pole situation occurs if the NG superfield $\Pi $ couples to some other fields $\Psi$, by a slightly more involved analog of the superpotential \wsplitt, e.g. \eqn\wsplittt{W=\big[M+\theta ^2 (|M|^2+|\delta m^2|)\big]\Pi ^{A'}\Pi^{A'} +\delta m\Pi ^{A'} \Psi ^{A'}.}  In this case, the superpartners $\sigma ^{A'}$ and $\psi _{\pi}^{A'}$ of the NG boson $\pi ^{A'}$ mix with the other fields of the susy-breaking sector.

Perhaps it's worth commenting on our terminology.  
Once susy is broken, and the poles of $\wt C_0^{pole}(p^2)$ and $\wt C_{1/2}^{pole}(p^2)$ are displaced away from $p^2=0$, the reader might  question the meaning of the distinction between the ``pole" and ``regular" contributions in \cbpolereg\ to the $\wt C_{a\neq 1}(p^2)$ and $\wt B_{1/2}(p^2)$.  Our view is that it does remain useful, even when susy is spontaneously broken, to make the distinction between the ``pole" contributions, which come from factorizing the current two-point functions onto one-point functions, and the ``regular" contributions, which do not (it might be better to refer to the terms as ``reducible" and ``irreducible," but we'll stick with ``pole" and ``reg," respectively). An important distinction between the pole and regular contributions is that they contribute at different orders of small $g$ perturbation theory.  Again, this is because we take $m_V=\CO(g^0)$, and hence $\wt C_a^{pole}$ and $\wt B_{1/2}^{pole}$ are $\propto v^2 =\CO(g^{-2})$, versus the regular contributions, which are $\CO(g^0)$.

We can give a general parameterization of the leading susy-breaking effects by considering  the general form of $\wt C_a^{pole}(p^2)$ and $\wt B_{1/2}^{pole}(p^2)$ in the limit $p^2\to \infty$.  Because the spontaneouus susy breaking becomes irrelevant in this limit, the $\wt C_a^{pole\ AB}$ all coincide with $\wt C_1^{pole\ AB}$, as given in \conepolegen, to $\CO(1/p^2)$.  Any susy-breaking (whether or not it is small) then has the leading effect for large $p^2$ which can be parameterized as follows
\eqn\celse{\eqalign{\wt C_0^{pole }&=\wt C_1^{pole}+{ v^2 \delta m_0^2\over p^4}+\CO\bigg({1\over p^6}\bigg)\cr  \wt C_{1/2}^{pole}&=\wt C_1^{pole} +{v^2 \delta m_{1/2}^2\over p^4}+\CO \bigg({1\over p^6}\bigg)\cr 
\wt B_{1/2}^{pole}&=-{v^2 m_\chi  \over p^2}+\CO\bigg({1\over p^4}\bigg),}}
where we'll take \celse\ as definitions of susy-breaking effects $(\delta m_0^2)^{AB}$, $(\delta m_{1/2}^2)^{AB}$, and $m_\chi ^{AB}$ (and we suppress the $(AB)$ gauge indices in \celse).  In the case where $\wt C_0^{pole}$ and $\wt C_{1/2}^{pole}$ have the simple, single pole form \clom, the parameters $\delta m_0^2$ and $\delta m_{1/2}^2$ in \celse\ simply give the pole displacement from $p^2=0$, as in \clom.  More generally, whether  or not there is a single pole, we will find that the parameters $\delta m_0^2$ and $\delta m_{1/2}^2$ in \celse\ are related by \mbreakreln\ when the susy-breaking is only by F-terms.    Since $\wt B_{1/2}$ breaks the R-symmetry, having $R(\widetilde B_{1/2})=-2$, the parameter $m_\chi$ in \celse\ is also, of course, only non-vanishing if there is no unbroken R-symmetry: in terms of R-breaking spurions, it has $R(m_\chi )=-2$.

Also note that, because all $\wt C_a$ coincide to $\CO(1/p^2)$ in the limit $p^2\to \infty$, the sums of the residues of the poles in \casum\ must coincide with that of $\wt C_1^{pole}$ in \conepole, and the sum of residues of $\wt B_{1/2}^{pole}$ is equal to the coefficient $m_\chi v^2$ in \celse:
\eqn\ressum{\sum \hbox{Residues} (\wt C_a (p^2))=v^2, \qquad \sum \hbox{Residues}(\wt B_{1/2}(p^2))=m_\chi v^2.}
We'll see that  \ressum\ implies that $Str M^2|_{vector multiplet}$ is independent of $m_V^2$.

Our discussion so far did {\it not} assume a weakly coupled description.  We now give more explicit, general expressions for {\it weakly coupled} realizations of  spontaneous symmetry and susy breaking. Take the fields of the hidden sector to be $\Phi$ (suppressing flavor and gauge indices), which contribute to the current as $\CJ ^{A'}=\bar \Phi  T^{A'}\Phi$.  Now expand around the non-zero $\Phi$ background as $\Phi =\Phi _0+\delta \Phi$, allowing for susy-breaking via vevs $\Phi _0=\phi _0+\theta ^2 F_0$.  Globally, $\Phi$ can vary away from $\Phi _0$ in the NG directions as $\Phi = e^{\Pi ^{A'}T^{A'}}\Phi _0$, and we'll do a linearized expansion.   The current supermultiplet is expanded for general, small $\delta \Phi$ as
\eqn\currexp{\CJ ^{A'}=\bar \Phi _0T^{A'}\Phi _0+\bar \Phi _0T^{A'}\delta \Phi +\delta \bar \Phi T^{A'}\Phi _0+ \delta \bar \Phi T^{A'}\delta \Phi.}
The linear terms in $\delta \Phi$ show how the broken generators create the NG bosons \ngcomps\ from the vacuum, and we can write these linear term contributions as in \jngb, with 
\eqn\ngbis{ \CJ ^{A'}\supset \half v_A'\Pi ^{A'}+h.c. \qquad\hbox{with}\qquad \half v_{A'}\Pi ^{A'}=\bar \Phi _0T^{A'}\delta \Phi,}
where $v_{A'}$ are normalization constants.

The pole contributions then come from the  linear terms in \currexp, with the two-point functions of these terms giving poles from the $\ev{\delta \bar \Phi (p)\delta \Phi (-p)}$ propagators.  This gives 
\eqn\cfunctionsapp{\eqalign{
&\wt C_0^{pole\ AB}=\bar\phiT^{A}\frac1{p^2+M_0^2}\phiT^{B},\qquad \wt C_{1/2}^{pole\ AB}=\bar\phiT^{A}\frac1{p^2+M_{1/2}^2}\phiT^{B},\qquad \wt C_1^{pole\ AB}=\frac{\bar \phiT ^A\phiT ^B}{p^2},\cr
&\wt B_{1/2}^{pole\ AB}=-2\bar\phi _0T^B\frac1{p^2+\bar\M\M}\bar\M(\bar\phi _0T^A)^T}}
where here we denote $\phiT^A\equiv \big(T^A\phi_0,\bar\phi_0 T^A\big)$, $\bar \phiT ^A=\phiT ^{A\dagger}$ (using bars to denote conjugation to stay out of the way of the indices) and the mass matrices are,
\eqn\massesapp{M_0^2=\pmatrix{\bar\M\M+\D&\bar\F\cr \noalign{\vskip-6pt}&\cr
\F&\M\bar\M+\widetilde \D},\quad M_{1/2}^2=\pmatrix{\bar\M\M&0\cr \noalign{\vskip-6pt}&\cr
0&\M\bar\M},}
where $\M_{ij}=W_{ij}$, and $\F_{ij}=W_{ijk}\bar W^k$, and 
\eqn\dare{ \D ^j_i\equiv g^2(T^A)^j_i\ev{J^A}\qquad \widetilde \D ^i_j\equiv g^2(T^A)^i_j\ev{J^A}.}
We use a compact notation, where we suppress indices when the meaning is unambiguous, but they can quickly be restored by noting that (1) we take $\bar\phi$ to be a row vector and $\phi$ a column vector, and (2) the position of the indices is an indication of chirality.  For example, we have $\phi^i$ and $\bar\phi_i$.   As another example, the condition for gauge invariance of the superpotential is $W_iT^{Ai}{}_j\phi^j=0$, and we write the first derivative of this equation as $\M T^A \phi _0=(\bar F T^A)^T$.  As is evident from \cfunctionsapp, $\wt C_0^{pole}$ and $\wt C_{1/2}^{pole}$ are generally given by sums of pole contributions, with the poles shifted away from $p^2=0$, as in \clom.

The   $\D$ term in \massesapp\ is associated with $D$-terms (see \djvev).  Note that the masses $M_0^2$ and $M_{1/2}^2$ in \massesapp\ are similar to the standard expressions for tree-level masses, but with $g\to 0$.  The $g\neq 0$ contributions to the masses will instead come from summing the gauge corrections to the propagator.  That said, the inclusion of the D-term contribution \dare\ perhaps looks strange, because of the $g^2$ factor.  Actually, \dare\ should be understood as a $g$-independent contribution, in the $g\to 0$ limit.
The point will be that, despite appearances, the $\D$ contribution to the masses  \massesapp\   can be independent of $g$, (much as in the example of \semidirect), as we'll illustrate with the FI model example.  For much of the following discussion, we will anyway focus on  $F$-term breaking, and set $\D=0$.

Summarizing, weakly coupled theories have $\wt C_1^{pole\ AB}(p^2)$ given by \conepolegen, with
\eqn\vsquaredeq{(v^2)^{AB}\equiv \bar \phiT ^A \phiT^B\equiv \bar\phi _0\{ T^A, T^B\} \phi _0\equiv (\bar \phi _0, \phi _0)^{AB}.}  
The susy-breaking mass shifts are parameterized as in \celse, and considering the functions \cfunctionsapp\ for $p^2\to \infty$  reveals that weakly coupled theories with only $F$-term breaking, $\D =0$ have (using inner product notation as in \vsquaredeq)
\eqn\wcsplittings{(\delta m_0^2)^{AB}={2(\bar F_0, F_0)^{AB} \over (\bar \phi _0, \phi _0)^{AB}}=2(\delta m_{1/2}^2)^{AB}, \qquad  m_\chi ^{AB}={(\bar \phi _0, F_0)^{AB}\over (\bar \phi _0, \phi _0)^{AB}}.}
In particular, if {\it any} charged matter field has non-zero F-term, it follows from \wcsplittings\ that $\delta m_0^2>0$ and $\delta m_{1/2}^2>0$ are strictly positive: the locations of the $\wt C_0$ and $\wt C_{1/2}$ poles are necessarily shifted away from $p^2=0$ in theories with only F-term breaking.  We'll illustrate how \wcsplittings\ are modified when there is $D$-term breaking, $\D\neq 0$, in the FI model example.

\subsec{Gauging the $G'$ symmetry}

We now couple the $G'$ currents to $G'$ gauge fields, with
\eqn\linti{\CL _{int}\supset -g \int d^4\theta \CJ ^AV^A\supset g(JD-\lambda  j -\bar \lambda \bar j -j^\mu V_\mu).}
 The NG boson chiral superfields $\Pi ^{A'}$ are then, of course, eaten and become the extra components of the massive $G'/G$ vector multiplets: $\pi$ becomes the longitudinal component of the gauge field, $\sigma$ becomes the real scalar $C$, and $\psi _\pi$ becomes the additional gaugino $\chi$.  This can be illustrated in various (super) gauge-choice, e.g. unitary gauge, setting $\Pi ^{A'}$ to zero in \jngb\ and \ngbis.  Instead, we'll employ Wess-Zumino gauge to set $C$ and $\chi$ to zero in the gauge multiplet, and their d.o.f. return as coming from the $\Pi^{A'}$.  We'll also use super-Landau gauge.  One could use a general gauge adding, and later decoupling, unphysical ghost degrees of freedom, but we'll mostly stick to super-Landau gauge.

Before getting into the details of the gauge multiplet propagators, let us note a few basic points.
The coupling of \jngb\ to the vector multiplet yields the terms
\eqn\ngbvterms{{\cal L}\supset \int d^4 \theta g\Big(\half v_{A'}\Pi ^{A'}+h.c.\Big)\CJ ^{A'}\supset gv_{A'}(\sigma ^{A'}D^{A'}+i\psi _{\pi}^{A'}\lambda ^{A'}+h.c. -iv^{A'}_\mu \partial ^\mu \pi ^{A'}).}
As usual, the $\pi ^A$ coupling in \ngbvterms\ is associated with the gauge field $v_\mu ^{A'}$ getting a mass
\eqn\mvsqis{(m_V^2)^{AB}=g^2 (v^2)^{AB}.}
Likewise, the $\sigma ^{A'} D^{A'}$ coupling in \ngbvterms\ corresponds to the $CD$ vector multiplet coupling in unitary gauge, with $\sigma ^{A'}\leftrightarrow C^{A'}$, and the interaction in \ngbvterms\ gives the propagating real scalar a supersymmetric mass equal to \mvsqis.  Likewise, \ngbvterms\ gives $\psi _{\pi}^{A'}\leftrightarrow \chi ^{A'}$, the additional gaugino of the massive vector multiplet, with \ngbvterms\ pairing the fermions with supersymmetric mass equal to \ngbvterms.    We can also consider at this stage the susy-breaking effects.  For example, if $\sigma ^{A'}$ gets an additional susy-breaking mass-squared $\delta m_0^2$, then the real scalar of the vector multiplet has 
\eqn\mvsqis{(m_C^2)^{AB}=g^2 (v^2)^{AB}+(\delta m_0^2)^{AB}.}
This mass shift is positive in unitary theories with only F-term breaking, $m_C^2\geq m_V^2$.  The fermions can get masses from both the coupling in \ngbvterms\ and also, for example, the susy-breaking coupling $\delta m_{1/2}$ appearing in \wsplitt:
\eqn\fermterms{\CL \supset ig v_{A'}\psi _{\pi}^{A'}\lambda ^{A'}+\delta m_{1/2}\psi _{\pi}^{A'}\psi _{\Psi}^{A'}\qquad\hbox{(R-symmetric case)},}
which respects a $U(1)_R$ symmetry under which $R(\lambda)=R(\psi _\Psi)=-R(\psi _{\pi})=1$.  This gives
\eqn\fermeigens{\eqalign{\hbox{two fermions of mass}=\sqrt{m_V^2+\delta m_{1/2}^2}:&\qquad\frac{\delta m_{1/2} \psi _\Psi +im_V\l}{\sqrt{m_V^2+\delta m_{1/2}^2}},\qquad\psi _\pi ,\cr
\hbox{one fermion of mass}=0:&\qquad \frac{\delta m_{1/2}\l+im_V\psi _\Psi }{\sqrt{m_V^2+\delta m_{1/2}^2}}.
}}
The fermions on the top line are  the two gaugino mass eigenstates of the massive vector multiplet, which are here degenerate with each other, but  not with the vector multiplet: $m_{\lambda _1}^2=m_{\lambda _2} ^2=m_V^2+\delta m_{1/2}^2>m_V^2$.  The degeneracy of the gauginos in this case follows from the R-symmetry, as they then marry with a Dirac mass. In theories without R-symmetry, there can be Majorana mass terms $m_\chi \neq 0$, and then $m_{\lambda _1}^2\neq m_{\lambda _2}^2$.  The massless fermion in \fermeigens\ is also a consequence of  the $U(1)_R$ symmetry, as it is then needed for $\delta m^2_{1/2}\neq 0$, to contribute a massless fermion with $R=1$ to the $\Tr R$ and $\Tr R^3$ 't Hooft anomalies. 

We'll now re-derive the above mass spectrum from the current-algebra based approach, directly from our expressions for the gauge multiplet propagators and the pole contributions to the current correlators.  
Consider first the full propagator \fullgauge\ for the gauge field, using $\wt C_1=\wt C_1^{pole}+\wt C _1^{reg}$, with $\wt C_1^{pole}$ given by \conepole.  The result is the   
the massive gauge field propagator 
\eqn\massiveprop{\Delta _{\mu \nu}=\frac{1}{p^2 -m_V^2}\left(g_{\mu \nu}-{p_\mu p_\nu\over p^2}\right)\left(1-g^2 \wt C _1^{reg}(p^2){1\over p^2 -m_V^2}\right)+\CO(g^4),}
where 
\eqn\mvsqis{(m_V^2)^{AB}=g^2 (v^2)^{AB}=g^2v^2_A\delta ^{AB}.}
The pole in  $\wt C_1(p^2)$ at $p^2=0$ eliminates the pole of the gauge field propagator \fullgauge\ at $p^2=0$; instead, the pole of \fullgauge\ is at $p^2=m_V^2$.
This is the ``Higgs" (Schwinger, actually \SchwingerTN) mechanism for giving a gauge boson a mass, re-deriving \mvsqis\ from the current-algebra based approach.
Dropping the $g^2 \wt C^{reg} _1$ term in \massiveprop, which is a one-loop correction, the leading, tree-level propagator for the gauge field is 
\eqn\massivepropo{\Delta _{\mu \nu}^{(0)}={(g_{\mu \nu}-{p_\mu p_\nu\over p^2})\over p^2 -m_V^2}={g_{\mu \nu}-{p_\mu p_\nu\over m_V^2}\over p^2-m_V^2}+{p^\mu p^\nu/m_V^2\over p^2}.}
This exhibits the unphysical massless ghost, which cancels with the eaten NG boson\foot{One could add gauge fixing terms for a supersymmetric $R_\xi$ gauge (see e.g. \FortinGX\ and references therein), where the unphysical ghost has mass $\xi m_V^2$, interpolating between Landau gauge, $\xi =0$,  to unitary gauge, $\xi \to \infty$, where the ghost and eaten NG boson decouple.}.

Similarly, for the propagator \Dpropb\ of the real scalar, using $\wt C_0=\wt C_0^{pole}+\wt C^{reg} _0$ yields 
\eqn\DDprop{\Delta ={1\over 1+g^2\widetilde C_0^{pole}(p^2)}\left(1-g^2 \wt C^{reg}_0(p^2){1\over 1+g^2\widetilde C_0^{pole}(p^2)}\right)+\CO(g^4).}
At tree-level, dropping the 1-loop correction $g^2\wt C_0^{reg}(p^2)$ for the moment, the propagator \DDprop\ has poles where $1+g^2\wt C_0^{pole}(p^2)=0$.  As we'll see, this is complicated in general.  In the simple case where 
$\wt C_0^{pole}$ has a single pole term, as in \clom, at    $p^2=-\delta m_a^2$, then \DDprop\ gives for the tree-level propagator of the real scalar of the vector multiplet
\eqn\DDprope{\Delta ^{(0)}={(p^2+\delta m_0^2) \over p^2+\delta m_0^2 +m_V^2},}
with pole at 
\eqn\mcis{m_C^2=m_V^2+\delta m_0^2,}
which is the mass of the propagating scalar field (it's the $C$ field in unitary gauge) of the massive vector multiplet\foot{We refer to the propagating real scalar of the vector multiplet as $C$, even though we here set $C=0$ in WZ gauge.  Then the two real scalars, $D$ and the component $\sigma$ of the NG chiral superfield $\Pi$, have a matrix of kinetic terms to diagonalize.  One linear combination of $D$ and $\sigma$ remains a non-propagating auxiliary field, and can thus be set to be a constant.  The other eigenvalue is a propagating scalar field, called $C$ above, can thus be equivalently thought of as $D$ or $\sigma$.}.   We have thus re-derived \mvsqis\ from the current-algebra approach, and see that \mvsqis\ applies in the simple, single pole case.  

Likewise, the tree-level $\Delta _{\alpha \dot \alpha}^{(0)}$ is obtained by replacing $\wt C_{1/2}(p^2)$ and $\wt B_{1/2}(p^2)$ in \lampropb\ with their pole contributions 
 $\wt C^{pole}_{1/2}(p^2)$ and $\wt B^{pole}_{1/2}(p^2)$.  The locations of the poles are complicated in general, but for theories with  $\wt B_{1/2}=0$,  and with $\wt C_0^{pole}$ and $\wt C_{1/2}^{pole}$ having a single pole as in \clom, the result is 
\eqn\lampropnob{\Delta _{\alpha \dot \alpha}={p_{\alpha \dot \alpha}(p^2+\delta m_{1/2}^2)\over p^2(p^2+\delta m_{1/2}^2+m_V^2)}\!\left(\!1-g^2\wt C^{reg} _{1/2}(p^2){(p^2+\delta m_{1/2}^2)\over p^2(p^2+\delta m_{1/2}^2+m_V^2)}\!\right)+\CO(g^4)\quad(\wt B_{1/2}=0).}
The poles here correspond to the mass eigenvalues discussed in \fermeigens: there is the massive gaugino pole at 
\eqn\mlamis{m_\lambda ^2=m_V^2+\delta m_{1/2}^2,}
and an additional pole at $p^2=0$, as in \fermeigens, coming from how the $\delta m_{1/2}$ susy-splitting between the (two degenerate) gauginos and the gauge field occurs, as in \wsplitt, in a theory with unbroken $U(1)_R$ symmetry.  

Let's now consider cases where $\wt C_0^{pole}$ and $\wt C_a^{pole}$ involve a sum of pole terms, as in \casum.  This happens if the NG superfield $\Pi$ couples to other hidden sector fields $\Psi$, e.g. as in \wsplittt.  In that case, the NG boson's partners $\sigma$ and $\psi _\pi$ have a mass mixing matrix with those other hidden sector fields.  Including also the couplings in \ngbvterms, results in the massive vector multiplet fields $C$, $\lambda$, $\chi$ getting masses given by $m_V^2$ plus appropriate eigenvalue of the mass mixing matrix.  We don't evaluate it explicitly because we'll instead use the current algebra approach.    Consider first the tree-level propagator $\Delta ^{(0)}(p^2)$ in \DDprop, 
  in the case \casum, where $C_0^{pole}$ is the sum of say $n$ terms.  Then $\Delta ^{(0)}(p^2)$ has poles at the $n$ solutions of  $1+g^2\wt C_0^{pole}(p^2)=0$, which is an $n$-th order polynomial equation in $p^2$.  One of these solutions is the $m_C^2$ mass, and the other $n-1$ are masses of hidden sector states that mix with the $C$ field.  Similar considerations apply to $\Delta _{\alpha \dot \alpha}^{(0)}(p^2)$, which has poles corresponding to the gauginos and states they mix with.  
  
 In the simplest case, where the  $\wt C_a$ have a single pole, as in \clom, the relation \mbreakreln\ implies that the massive vector multiplet has tree-level mass supertrace
\eqn\mvstr{Str M^2|_{\hbox{vector multiplet}}=\hskip.1in-\hskip-.2in\sum _{\hbox{broken}\ A, B}\hskip-.15in\delta m_0^{2\ AB}=\hskip.1in-\hskip-.2in\sum _{\hbox{broken}\ AB}{2(\bar F, F)^{AB}\over (\bar \phi, \phi)^{AB}}<0\quad \hbox{(single pole)},}
where we used \mbreakreln\ to write $\delta m_0^2-4\delta m_{1/2}^2=-\delta m_0^2$ and in the last expression we specialized to the case of a weakly-coupled hidden sector.  

Note that the mass $m_V$ of the gauge field has dropped out of the supertrace \mvstr, and there is an analog of this statement also for the multi-pole case.  As discussed above, the $n$ roots of $1+g^2\wt C_0^{pole}(p^2)=0$ yield $n$ physical mass-squareds, which we'll call $m^2_{C_i}$.  It follows from the $a=0$ case of \ressum\ that $\sum _{i=1}^n m_{C_i}^2=m_V^2+\dots$, where $\dots$ are determined by the $\wt C_0^{pole}(p^2)$ pole locations, independent of $g$ and $m_V^2$.  Likewise, it follows from the $a=1/2$ case of \ressum\ that $\sum _{i=1}^n m_{\lambda _i}^2=m_V^2+\dots$.  So if we take a super-trace over the massive vector multiplet, along with the hidden sector states that they mix with, we get $m_V^2(1-4+3)+\dots$ and the $m_V^2$ drops out, generalizing the observation about \mvstr.  

This observation that $m_V$ drops out of $Str M^2|_V$ can be applied as follows. 
Consider taking $m_V$ to be much larger than all other mass scales in the theory (say in the single-pole case, for simplicity).   One could then consider a low-energy description of the hidden sector, in which the ultra-massive vector multiplet was integrated out.  If the hidden sector has spontaneous susy-breaking at tree-level, with a weakly coupled description and canonical K\"ahler potential in the UV, then the full theory has $Str M^2=0$.  Since the massive vector multiplet has supertrace \mvstr, the remaining light fields of the hidden sector low-energy theory for $m_V$ large and integrated out must have opposite supertrace
\eqn\mvstrr{Str M^2|_{low}=+\hskip-.1in\sum _{\hbox{broken}\ A, B}\hskip-.15in\delta m_0^{2\ AB}=\sum _{\hbox{broken}\ A, B}{2(\bar F, F)^{AB}\over (\bar \phi,\phi )^{AB}}>0.}
This can be compared with the general formula \GrisaruSR\
\eqn\strlow{Str M^2=2R_{k\bar \ell}F^k \bar F^{\bar \ell},}
and the result of \BuicanWS\ for the curvature of the tree-level K\"ahler potential of the light fields when vector multiplets are integrated out:
\eqn\rvm{R_{k\bar \ell}=\sum _{\hbox{broken}\ A, B}{2((T^A)^j_k(T^B)_{\bar \ell j}+ (T^A)_{k\bar\ell} \Tr ' T^B)\over (\bar \phi , \phi )^{AB}}.}
Using this in \strlow, the first term in \rvm\  indeed agrees with \mvstrr, and the second term does not contribute here because we're taking $\ev{D^A}=0$ (and we see from \dfreln\ that we're here assuming that $\bar F T^A F=0$).

\newsec{Gauge mediation by unbroken $G$ gauge fields for $G'\to G$}

Consider the general situation with UV gauge group $G'$, spontaneously broken to subgroup $G\subset G'$ at some energy scale $m_{V}$. In this section, we consider the susy-breaking effects which are mediated by the massless $G$ gauge fields.  As a concrete example, we can consider $G'=SU(5)_{GUT}$ and $G=SU(3)\times SU(2)\times U(1)$, but our discussion will be completely general. 
Because unbroken gauge groups do not have tree-level susy-splittings (gauge messengers), the $G'$ mediated effects can be considered in the framework of \mss, without much modification.  The main modification is that, for non-Abelian groups, the massive  $G'/G$ gauge fields contribute in the loop to affect the $G$ current correlators, acting as additional messengers with unusual signs.  

The results for the soft masses are as follows.  
The gaugino superpartners of the massless gauge fields get soft mass
\eqn\gauginomass{M_{G\ gaugino}=g^2 \widetilde B^{unbr}_{1/2}(0).}
The visible sector sfermions $Q_f$ get diagonal soft masses
\eqn\sfermionmass{\eqalign{m_{Q}^2&=g^2 c_2(r_Q)\int {d^4 p\over (2\pi)^4}{1\over p^2} \Xi ^{unbr}(p^2)+g^2\hskip-.2in\sum _{broken\ A'B'}\hskip-.2inT_{r_Q}^{A'}T_{r_Q}^{B'}\int {d^4 p\over (2\pi)^4}{1\over p^2} \Xi ^{A'B'},\cr \Xi ^{unbr}(p^2)&=-g^2 (3\wt C^{unbr}_1(p^2)-4\wt C^{unbr}_{1/2}(p^2) +\wt C_0^{unbr}(p^2)),}}
where $c_2(r_Q)$ is the quadratic Casimir summing over only the massless gauge fields, restricting the adjoint indices $A,B$ to $G$ in \sfermint.  The unbroken gauge fields do not have the pole contributions to their propagators, so $\Xi ^{unbr}$ reduces as in \sfermionmass\ to the expression of \mss.  The broken contribution in \gauginomass\ comes from  the massive 
$G'/G$ gauge fields coupling directly to the visible sector fields, as in Fig. 2b, and we defer discussing these contributions to the following section.   Since we're  just discussing the unbroken contributions in this section, we'll henceforth drop the ``unbroken" superscript reminders.

The massive $G'/G$ gauge fields contribute to the $G$ currents, and thus  to the unbroken $\wt B_{1/2}$ and $\wt C_a$  correlators:
\eqn\bcgm{\eqalign{\wt C^{unbr}_a(p^2)&=\wt C_a^{gauge}(p^2)+\wt C_a^{matter}(p^2),\cr \wt B^{unbr}_{1/2}(p^2)&=\wt B_{1/2}^{gauge}(p^2)+\wt B_{1/2}^{matter}(p^2).}}
The ``matter" contributions are those from the hidden sector which may or may not be weakly coupled, much as in \mss.  The ``gauge" contribution can be computed in $g$ perturbation theory.  They account for the correction, from a loop of massive gauge fields, to the propagators of the unbroken gauge fields (the analog of the correction to the photon propagator from a $W^\pm$ loop), as in Fig. 3.  Note that the group theory implies that both of the gauge fields running in the loop, labeled $A'$ and $B'$ in Fig. 3,  are massive -- there is no diagram with one internal massive and one internal massless gauge field.  This is because  a broken and an unbroken generator can only combine to give a broken generator\foot{To see this, take all the  $G'$ generators to satisfy 
$(T^AT^B)=\half \delta ^{AB}$, $\Tr (T^{A'} T^{B'})=\half \delta ^{A'B'}$, $\Tr (T^AT^{B'})=0$, 
where $A, B$ run over the unbroken subgroup $G$, and $A', B'$ run over the broken $G'/G$.  
It follows from the last equation that $f^{AB'C} \sim \Tr (T^A[T^{B'}, T^{C}])=\Tr ([T^C, T^A], T^{B'})=0$.}, 
$[G'/G, G]\subset G'/G$, i.e. $ f^{AB'C}=0$.  Since diagrams with a gauge loop like Fig. 3 would have a factor of $f^{AB'C}$ from a diagram with one massive and one massless internal propagator, it can only get contributions if both internal propagators are for massive $G'/G$ components.

The UV behavior of the functions $\wt C_a(p^2)$ is controlled by a constant coefficient, $c$, as in \cac, $\wt C_a\sim c\log \Lambda$.  $c$ gives the contribution of the sector to the beta function of the gauge coupling when the symmetry is gauged, as is seen from the relation of $\wt C_1(p^2)$ to the gauge field self-energy, and the relation of that to the beta function in background field gauge.  So there are gauge and matter contributions to $c$, associated with their contributions in \bcgm, given by the contribution of these fields to the gauge beta function: 
\eqn\cbeta{c=-(b_1'-b_1)\equiv c^{gauge}+c^{matter}, \qquad c^{gauge}\equiv 3T_2(G)-3T_2(G').}
While unitarity implies that $c^{matter}\geq 0$, we see that $c^{gauge}<0$.  The total $c$ in \cbeta\ can be positive or negative.  The same coefficient $c^{gauge}$ in \cbeta\ gives the group theory dependence of the functions $\wt C_a^{gauge}(p^2)$ and $\wt B_{1/2}^{gauge}(p^2)$ when the Higgsing   $G'\to G$ is at a single scale,  $(m_V^2)^{A'B'}=m_V^2\delta ^{A'B'}$. In that case, the  sum 
over the internal $G'/G$ gauge fields in the loop in diagrams like Fig. 3, gives
(using $f_{AB'C'}=i(T_A)_{B'C'}$):
\eqn\groupfacts{ (\wt C_a^{gauge})^{AB}, \ (\wt B_a^{gauge} )^{AB}\sim \quad \Tr _{G'/G}(T^AT^B)=(T_2(G')-T_2(G))\delta ^{AB}\equiv -{1\over 3}c^{gauge}\delta ^{AB}.}
Aside from the overall coefficient in \groupfacts, the functions  $\wt C_a^{gauge}$ and $\wt B_{1/2}^{gauge}$  are otherwise independent of the  choice of groups $G'$ and $G$.

Let's first consider the function $C^{gauge}_{susy}(p^2)$ in the case where susy is unbroken.  The 
relevant loop diagrams with internal Higgsed vector multiplet components could then be computed in supersymmetric background gauge field formalism; as discussed in sect.~6.5 of \superspace, the loop correction to the gauge field propagator associated with internal gauge field loops is just a factor of $(-3)$ times that of a matter field, coming from three chiral ghosts.  (See also 
\refs{\FortinGX-\GrisaruWC\OvrutWA} for discussion of supersymmetric $R_\xi$ gauge.)  Using \groupfacts, the factor is $(-3)(-c^{gauge}/3)=c_{gauge}$.  So, if supersymmetry were unbroken, we would have $\wt B_{1/2}^{gauge}=0$ and all $\wt C_a(p^2)$ would equal
\eqn\cgaugesusy{\eqalign{\wt C_a^{gauge}(p^2)&=\wt C_{susy}^{gauge}(p^2)= c^{gauge} \int {d^4 k\over (2\pi)^4} {1\over (k^2+m_V^2)((p+k)^2+m_V^2)},\cr 
&= c^{gauge} \int {d^4 k\over (2\pi)^4} {1\over k^2[1+g^2\wt C_{susy}^{pole}(k)]}\frac1{(p+k^2)[1+g^2\wt C_{susy}^{pole}(p+k)]}.}}

With broken supersymmetry, there are analogous expressions for the various $\wt C_a^{gauge}$.  For example,  considering the diagram for $\ev{j_\alpha \bar j_{\dot \alpha}}$ with internal gaugino, with propagator $\Delta _{\beta \dot \beta}(k)$, and internal gauge field, with propagator $\Delta _{\mu \nu}(p+k)$, yields 
\eqn\altcgaugehalf{\wt C_{1/2}^{gauge}(p^2)= c^{gauge}  \int \frac{d^4 k}{ (2\pi)^4} \frac{1}{ k^2[1+g^2\wt C_{1/2}^{pole}(k)]}\frac1{(p+k)^2[1+g^2 \wt C_1^{pole}(p+k)]},}
where, for simplicity,  we wrote the expression for $\wt B_{1/2}=0$.    
The function $\wt C_1^{gauge}$ can also be explicitly determined, though it is rather lengthy (it gets contributions from each field in the massive vector supermultiplet running in the loop, and in background field gauge these are related to the contribution of each field to the one-loop beta function) so we will not write it out here.  Finally, the function $\wt B^{gauge}_{1/2}$ gets a contribution from the loop with gaugino correlation function $\Sigma (p^2)$ and $\Delta _{\mu \nu}(p+k)$ for the massive fields in the loop:
\eqn\bhalfg{\eqalign{\wt B_{1/2}^{gauge}(p^2)&=\sum _{A'B'} \int {d^4 k\over (2\pi)^4}\Sigma ^{A'B'}(p^2) \Delta^{A'B'} _1((p+k)^2)\cr 
&= 2c^{gauge}\int {d^4 k\over (2\pi)^4}\frac{ g^2 \wt B_{1/2}'(k)}{[1+g^2\wt C_{1/2}'(k)]^2k^2+g^4|\wt B_{1/2}'(k)|^2}\frac1{((p+k)^2 +m_V^2)},}}
where $\wt B_{1/2}'$ and $\wt C_{1/2}'$ denote the pole contributions of the massive gauge fields.  
In particular, 
\bhalfg\ contributes to the mass of the gaugino partners of unbroken, massless gauge fields as $m_\lambda = g^2\wt B^{matter}_{1/2}(0)+g^2\wt B^{gauge}_{1/2}(0)$.

\newsec{Massive gauge messengers' direct coupling to the visible sector}

The effects discussed in the previous section, are not too different from the general gauge mediation setup of \mss, aside from the massive $G'/G$ messenger contribution with $c_{gauge}<0$.   We now discuss the direct coupling of the massive $G'/G$ gauge messengers to the visible sector matter.  Because this is  direct coupling of susy-breaking messengers to the visible sector, it leads to more dramatic differences from the non-gauge messenger case, with visible sector soft terms generated at one lower loop order.

The contribution of the massive gauge fields to the sfermion $m_Q^2$ \sfermint\ is given by
\eqn\msfermiongm{m_Q^2\supset g^2 \sum _{A', B'}T_{r_Q}^{A'}T_{r_Q}^{B'}\int {d^4 p\over (2\pi )^4}{1\over p^2}\Xi ^{A'B'}(p^2),}
where the sum is over the broken generators $A', B'$, and $\Xi ^{A'B'}(p^2)$ is as defined in \xiis, and given by \sfermmass, with $\wt C_a^{A'B'}$ and $\wt B_{1/2}^{A'B'}$ given by the sum of pole and regular contributions, as in \cbpolereg.  The regular contributions don't know that the symmetry is spontaneously broken, so they're all proportional to $\delta ^{A'B'}$. If the pole contributions are also such that the $A'$ and $B'$ broken generators have a single Higgsing scale, then they're also proportional to $\delta ^{A'B'}$, so $\Xi ^{A'B'}(p^2)=\Xi (p^2)\delta ^{A'B'}$, and then the sum over broken generators in \msfermiongm\ gives 
\eqn\cdiff{\sum _{broken\ A'}T_{r'_Q}^{A'}T^{A'}_{r'_Q}=\Delta c_Q {\bf 1}_{|r_Q|}, \qquad \Delta c_Q\equiv c_2(r_Q')-c_2(r_Q),}
with $\Delta c_Q>0$.  For simplicity, we'll consider this single Higgsing scale case in what follows.  Our results can be straightforwardly adapted for a more general case by adding analogous contributions for each Higgsing scale.

The one loop contribution to $m_Q^2$ is given by \msfermiongm\ and the expression \sfermmass\ for $\Xi$, upon using the $\CO (g^0)$ term in $\Xi$, coming from the pole contributions to the 2-point functions
\eqn\Xio{\Xi ^{(0)}(p^2)=\left({1\over 1+g^2\wt C^{pole}_0(p^2)}-{4\over 1+g^2\wt C^{pole}_{1/2}+{g^4|\wt B^{pole}_{1/2}|^2\over p^2(1+g^2\wt C^{pole}_{1/2})}}+{3\over 1+g^2\wt C_1^{pole}}\right).}
The two loop contribution to $m_Q^2$ is also given by \msfermiongm, using the $\CO(g^2)$ contribution $\Xi ^{(1)}(p^2)$, obtained from expanding \sfermmass\ using \cbpolereg, with the ``regular" contributions giving the higher loop order.  Since it is a straightforward expansion, see e.g. \cregpole, we won't bother to write out explicitly the general, lengthy expression for $\Xi ^{(1)}(p^2)$.

To go farther, we will need to evaluate the integral in \msfermiongm\ using the explicit expressions for $\wt C_a$ and $\wt B_{1/2}$ in a given theory.  We now evaluate it explicitly in those theories where $\widetilde B_{1/2}=0$ and $\wt C_a^{pole}$ have single pole, as in \clom.  In these theories, the above expressions for the  one and two loop contributions to $m_Q^2$ from the massive gauge fields simplify to yield  
\eqn\msfermiongmm{\eqalign{m_Q^2&=-g^2 \Delta c_Qm_V^2 Str \int {d^4 p\over (2\pi )^4}{1\over p^2}{1\over p^2+M_a^2}\cr &+g^4\Delta c_QStr \int {d^4 p\over (2\pi )^4}{1\over p^2}\bigg(1-{m_V^2\over p^2+M_a^2}\bigg)^2\wt C _a.}}
where we used $(1+{m_V^2\over p^2+\delta m_a^2})^{-1}=1-{m_V^2\over p^2+M_a^2}$, where $M_a^2=m_V^2+\delta m_a^2$ are the masses of the components of the massive vector multiplet, and $Str$ means to sum over $a$, with $a={1/2}$ weighted by $(-4)$ and $a=1$ weighted by $3$.  The first term in \msfermiongmm\ are the one-loop contribution to the sfermion masses; doing the integral, this gives the one-loop contribution to be
\eqn\msfermiononel{(m_Q^2)^{(1)}={g^2\over 16\pi ^2}\Delta c_Qm_V^2 \ln \left({m_C^2 m_V^6\over m_\lambda ^8}\right),  \qquad (\hbox{single pole case})}
where $m_C^2=m_V^2+\delta m_0^2$ and $m_\lambda ^2=m_V^2+\delta m_{1/2}^2.$  
The terms $\ln m_C^2-4\ln m_{\lambda}^2+3\ln m_V^2$ in \msfermiononel\ come from the three diagrams in Fig. 4.   The expression \msfermiononel\ gives tachyonic $(m_Q^2)^{(1)}<0$ (using e.g. \mbreakreln), contributing to visible sector $Str M_Q^2<0$.  The expression \sfermionmone\ must exhibit decoupling, $m_Q^2\to 0$, for $m_V^2\to \infty$; it indeed does, since the susy-splittings remain fixed, $m_C^2/m_V^2\to 1$ and $m_\lambda ^2/m_V^2\to 1$, in this decoupling limit. 

\vskip-.1in
$$
	\matrix{\displaystyle
		\epsfxsize=1.8 truein\epsfbox{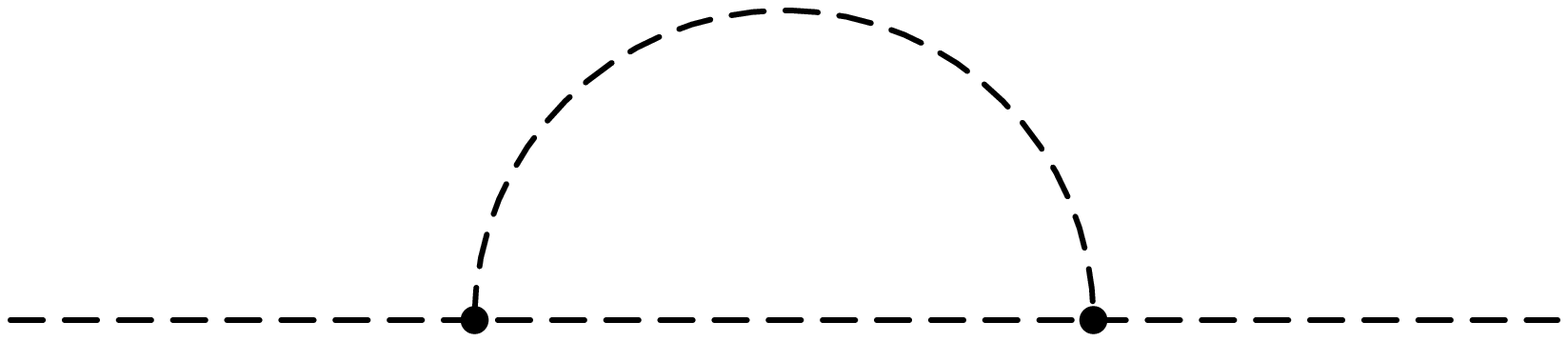}&\cr
		\noalign{\vskip-16pt}&}
		{\hskip10pt}
	\matrix{\displaystyle
		\epsfxsize=1.8 truein\epsfbox{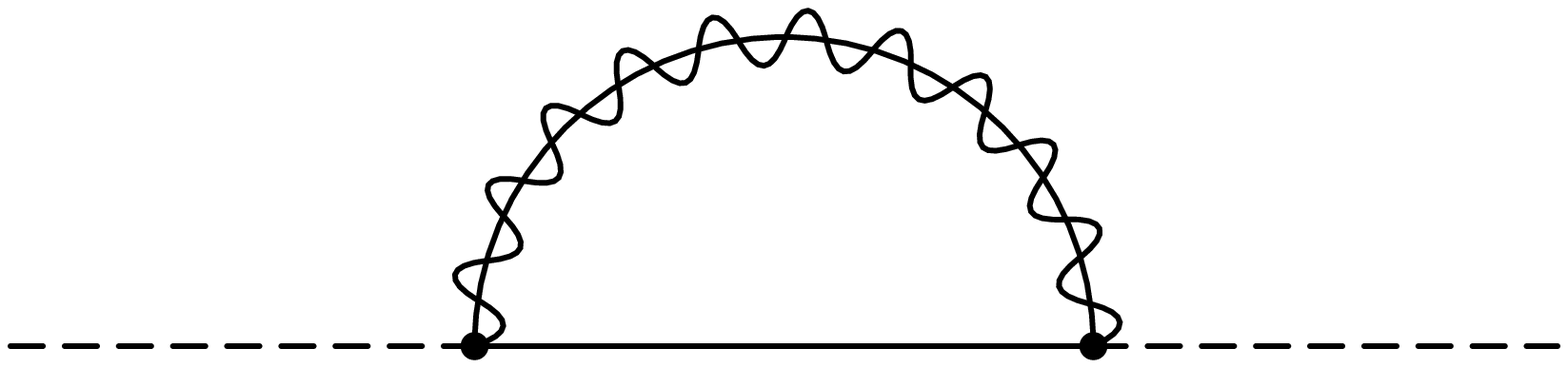}&\cr
		\noalign{\vskip-15pt}&}
		{\hskip10pt}
	\matrix{\displaystyle
		\epsfxsize=1.8 truein\epsfbox{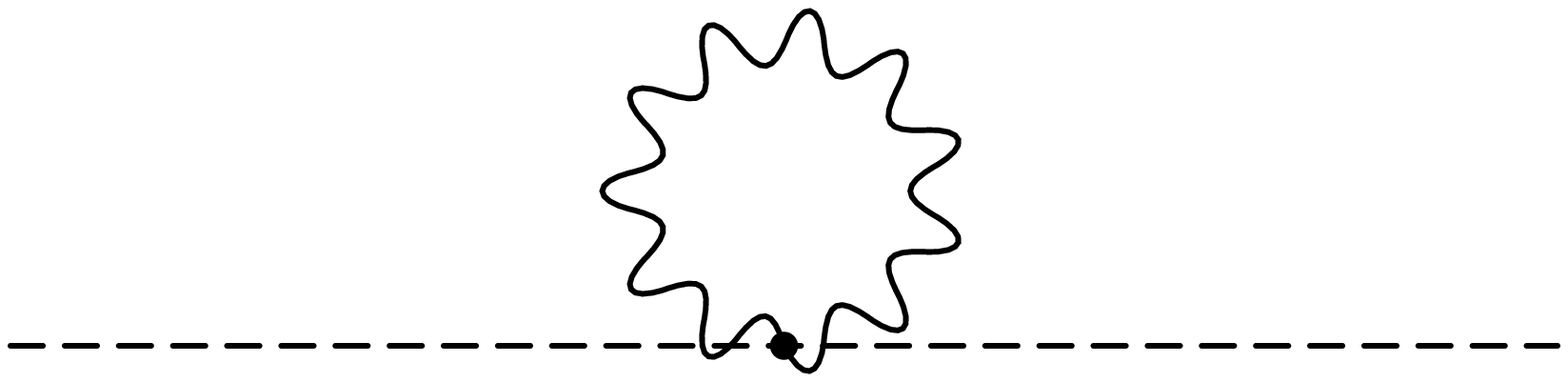}&\cr
		\noalign{\vskip-15pt}&}
$$
\vskip.1in
{\small\noindent{\bf Figure 4.}    One-loop diagrams contributing to the mass of a visible sector scalar.  In Landau gauge, the ``sunrise'' diagram for the gauge boson vanishes.
}\vskip.2in

The second term in  \msfermiongmm, is the gauge-messenger generalization of the familiar two-loop expression (again, specialized to this case of single poles and $\wt B^{pole}_{1/2}=0$):
\eqn\gaugemesstwol{(m_Q^2)^{(2)}\!\!\supset \!g^4\! \Delta c_Q \!\!\int\!\! {d^4 p\over (2\pi)^4}{1\over p^2}\!\bigg[\!\Big({p^2+\delta m_0^2\over p^2+m_C^2}\Big)^2 \!\wt C_0^{reg}\!-4\Big({p^2+\delta m_{1/2}^2\over p^2+m_\lambda ^2}\Big)^2 \!\wt C _{1/2}^{reg}\!+3\Big({p^2\over p^2+m_V^2}\Big)^2 \!\wt C _1^{reg}\bigg]\!,}
which should be added to the contribution from any unbroken gauge fields (discussed in the previous section).  This is similar in form to the non-gauge-messenger case (setting $M_a^2=m_W^2$ equal, \gaugemesstwol\ agrees with \IntriligatorFR, and setting all $m_a^2=0$ it agrees with \mss).     
Again, for the more general case of  multi-pole contributions to $\wt C^{pole}_{a=0, \half}(p^2)$, and $\wt B_{1/2}^{pole}(p^2)\neq 0$ the generalization of \gaugemesstwol\ is obtained by using \sfermint\ with the $\CO(g^2)$ contribution to $\Xi ^{AB}$ obtained by expanding \sfermmass\ to $\CO(g^2)$; it's straightforward, and the expression is lengthy.

$$\epsfxsize=2.7in\epsfbox{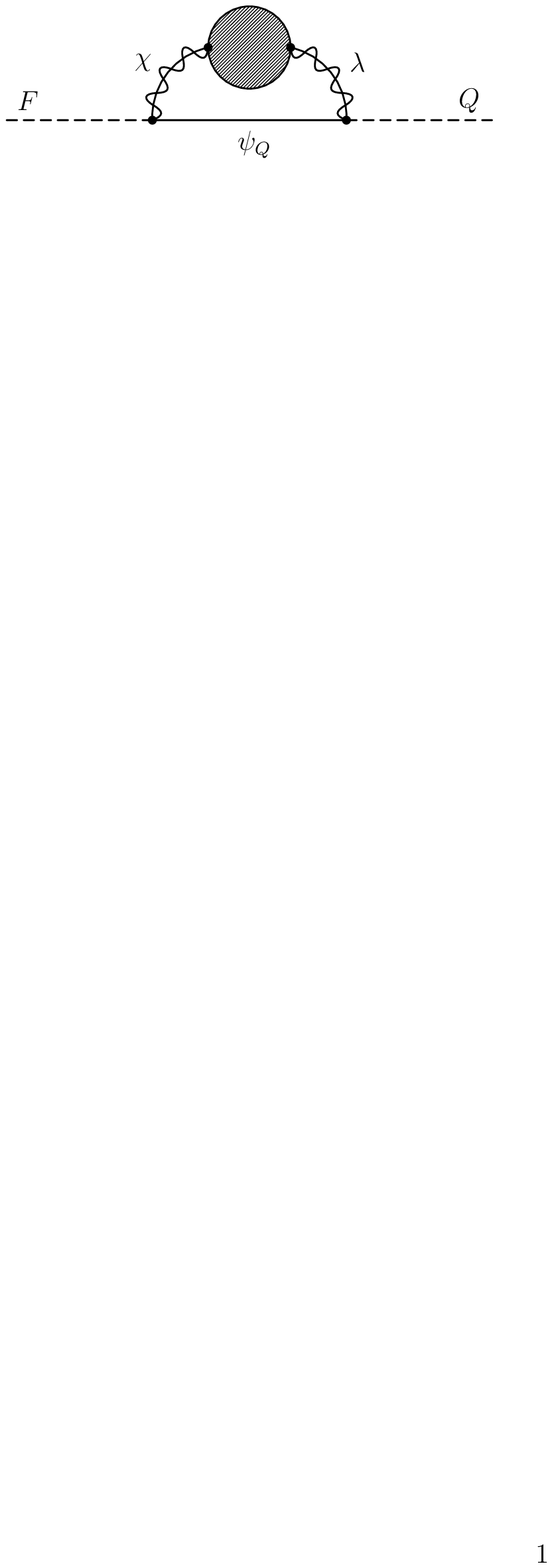}$$
\vskip- .1in
{\small\noindent{\bf Figure 5.}   The diagram giving $A_Q$ when $F_Q$ is not integrated out.}\vskip.2in

We now consider the $A$ terms.  Keeping the auxiliary components $F_Q$ of the visible sector matter fields $Q$, $A$-terms arise from a term 
\eqn\leffa{\CL _{eff}\supset \int d^4 \theta \ A_Q\theta ^2 \ \bar Q e^{2gV}Q\supset A_Q \bar F_Q Q.}
The first expression illustrates that  $A_Q$ multiplies a super-gauge invariant soft term.  
The coefficient $A_Q$ is associated with the diagram\foot{We choose to illustrate it in super-unitary gauge, where  $\chi \in V$ is kept.} in Fig.~5, which gives the general expression \aqloopo, whether or not there are gauge messengers.  Note that this diagrams 
looks rather different from the loop diagram which generates e.g. a trilinear coupling $
{\cal L}\supset -a_{ijk}Q_iQ_jQ_k$ but computing that loop is equivalent to \aqloopo, with $a_{ijk}=Y_{ijk}(A_{Q_i}+A_{Q_j}+A_{Q_k})$, where $Y_{ijk}$ is the supersymmetric Yukawa coupling in $W\supset Y_{ijk}Q_i Q_jQ_j$.
When there are no gauge messengers,  \aqloopo\ gives a 2-loop expression for $A_Q$.  On the other hand, when there are gauge messengers, the contribution of the massive gauge fields gives a one-loop contribution as in \aqloop, 
\eqn\aqloop{A_Q^{(1)}=-2g^2\Delta c_Q\int {d^4 p\over (2\pi )^4}{1\over p^2}\Sigma ^{pole}(p^2).}

Note the similarity of \aqloop\ and the 1-loop contribution to unbroken $G$ gaugino masses from broken $G'/G$ gauge messengers, $m_\lambda ^{unbr, gauge}=g^2 \wt B_{1/2}^{unbr, gauge}(0)$.  Using \bhalfg, 
\eqn\mlamcont{m_\lambda ^{unbr}\supset g^2\cdot 2 c^{gauge} \int {d^4k\over (2\pi)^4}{1\over p^2+m_V^2}\Sigma ^{pole}(p^2).}
The only difference (aside from $\Delta c_Q$ vs $c^{gauge}$) is the $1/p^2$, from the $Q$ propagator in \aqloop, vs the $(p^2+m_V^2)^{-1}$, from the massive vector propagator in \mlamcont .

We now consider the full sfermion effective potential, which can potentially be useful for cosmological applications, as in \dMM.  For sfermions near the origin, the effective potential reduces to the quadratic term associated with the sfermion mass \msfermiongm.  For sfermions far from the origin, the susy-breaking does not have as large of an effect, and the potential flattens, with the characteristic $\log$ behavior of a pseudomodulus.  The one-loop effective potential comes from diagrams with a
 $D$,  $\lambda _\alpha$, and $v_\mu$ loop.
Each is corrected by a $\sum _n$ terms, with $n$ current 2-point functions put in the loop (and a $1/n$ symmetry factor, hence the logs); this yields: 
\eqn\veffone{V_{eff}^{(1)}=\half \Tr _{G'}\int {d^4p\over (2\pi )^4}[\ln (1+g^2\widetilde C_0)-2\ln \left[ (1+g^2\widetilde C_{1/2})^2+{g^4|\widetilde B_{1/2}|^2\over p^2}\right]+3\ln (1+g^2 \widetilde C_1)],}
where $\Tr _{G'}$ runs over all generators of the UV group $G'$.  
This is the generalization of the effective potential given in \IntriligatorFR\ to the case of gauge messengers.    As we're now used to, we expand this expression to $\CO(g^2)$, with the pole pieces contributing to $g^2 \wt C_a$ and $g^2\wt B_{1/2}$ at $\CO(g^0)$, and the  regular terms  $\wt C _a$ and $\wt B _{1/2}$ contributing to \veffone\ at $\CO(g^2)$.

We want to compute the effective potential \veffone\ as a function of visible sector background expectation values $\ev{Q}$.  To do so, note that the the visible and hidden sectors are decoupled before the gauge interactions are turned on, so we have $\CJ =\CJ _{hidden}+\CJ _{visible}$ and in \veffone\ 
\eqn\vishid{\wt C_a=(\wt C_a)_{hidden}+(\wt C_a)_{visible}, \qquad \wt B_a=(\wt B_a)_{hidden}+(\wt B_a)_{visible}.}
Since we're interested in the effective potential with the visible sector fields $Q$ expectation values only along $D$ and $F$ flat directions,  the visible sector correlation functions are supersymmetric:
\eqn\visbc{(\wt C_a)^{AB}_{visible}={\bar Q \{T^{A}, T^{B}\}Q\over p^2}, \qquad (\wt B_{1/2})_{visible}=0,}
where we here only keep the visible sector pole term because we work to leading order in $g^2$.  Plugging \visbc\ and \vishid\ into \veffone\ then gives
\eqn\veffoneis{\eqalign{V_{eff}^{(1)}(|Q|^2)=\half \Tr _{G'}\int {d^4p\over (2\pi)^4}\Bigg\{&\ln \Big(1+{g^2(\bar Q, Q)\over p^2}+g^2 \wt C_0\Big)+3 \ln \Big(1+{g^2(\bar Q, Q)\over p^2}+{m_V^2\over p^2}\Big) \cr &-2\ln\Bigg[\Big(1+{g^2(\bar Q, Q)\over p^2}+g^2 \wt C_{1/2}(p^2)\Big)^2 +{g^4|\wt B_{1/2}|^2\over p^2}\Bigg]\Bigg\}.}}
Taking ${dV_{eff}\over d|Q|^2}$ of \veffoneis\ at $\ev{Q}\approx 0$ indeed reproduces the (one and two loop) sfermion mass $m_Q^2$ given above (with $\sum _{AB} (\bar Q, Q)^{AB} \to 2c_2(r_Q)|Q|^2$); in particular, using $\wt C^{pole}_a$ and $\wt B^{pole}_{1/2}$ in \veffoneis\ reproduces the 1-loop \msfermiongm.  The potential for large $|Q|^2$ should reduce to the expression found in \IntriligatorFE\ for Higgsing pseudomoduli far from the origin:
\eqn\vefflargeQ{V_{eff}^{(1)}(|Q|) \approx -{g^2\over 8\pi ^2} \sum _i|F_{\Phi _i}|^2 (c_2(r_{\phi _i}')-c_2(r_{\phi _i}))\ln |gQ|, \qquad \hbox{for}\qquad |\ev{Q}|\gg |\ev{\phi _i}|}
where $r_{\phi _i}'$ is the representation of hidden sector field $\Phi _i$ above the scale of the large $\ev{Q}$ Higgsing, and $r_{\phi _i}$ is that below. In this limit, the integral \veffoneis\ is dominated by the region of large $p^2\sim |Q|^2$, where 
we can use the expansions \celse\ to show that  \veffoneis\ yields \vefflargeQ\ upon using the weak coupling expressions \wcsplittings. 

We can also consider a susy-breaking effective potential which generalizes the a-term.  This potential also follows from the one-loop effective potential \veffone, by generalizing \vishid\ and \visbc\ to include a background expectation value for $F_Q$.  It suffices to work to $\CO(F_Q)$, dropping terms which are $\CO(|F_Q|^2)$ and higher in visible sector F-terms (keeping arbitrary order in hidden sector F-terms).  Using the small F expansion expressions in the appendix, the visible sector contributes 
\eqn\visbcf{(\wt C_a)^{AB}_{visible}={(\bar Q, Q)^{AB}\over p^2} +\CO(|F_Q|^2), \qquad (\wt B_{1/2})_{visible}^{AB}=-{(\bar Q, F_Q)^{AB}\over p^2}+\CO(|F_Q|^2).}
Using these as additional contributions in \veffone, we obtain an effective potential that is similar to \veffoneis, but with the hidden sector $\wt B^{pole\ AB}_{1/2}$ in \veffoneis\ replaced with $\wt B_{1/2}^{pole\ AB}=- (\bar Q, F_Q)^{AB}/p^2$.  Extracting the $\CO(\bar F_Q)$ term, this gives a term in the visible sector soft-breaking effective lagrangian\foot{Such a soft-breaking potential was considered in \NibbelinkSI, and their result can be compared with the special case where our $\wt B_{1/2} (p^2)$ is replaced with a momentum-independent constant $m_\lambda$.}
\eqn\leffk{{\cal L}_{vis}\supset \int d^4 \theta \theta ^2 \wt k(\bar Qe^{2V}, Q)\supset \bar F_Q \partial _{\bar Q}\wt k(\bar Q e^{2V},Q).}  This yields
\eqn\kqis{\bar F_Q\wt \partial _{\bar Q}k(|Q|^2)=2\Delta c_Q(\bar F_Q, Q)\int {d^4 p\over (2\pi)^4}{2g^4 \wt B_{1/2}^{pole}/p^4\over \big(1+{g^2 |Q|^2\over p^2}+g^2 \wt C_{1/2}^{pole}\big)^2+{g^4|\wt B_{1/2}^{pole}|^2\over p^2}}.}  Expanding $\wt k(|Q|^2)$ around the origin, $k(|Q|^2)\approx A_Q|Q|^2+\CO(|Q|^4)$,   gives the A-term \aqloop.

\newsec{Susy relations and small susy-breaking expansions}

\subsec{susy relations among current correlators, and gauge propagators}
As shown in  \BuicanWS , the current supermultiplet structure implies the relations \bmssx,
 \eqn\bmssx{\eqalign{\ev{Q^2(J(p)J(-p))}&=-4\wt B_{1/2}(p^2),\cr
 \ev{\bar Q^2 (Q^2(J(p)J(-p))}&=8p^2 (\wt C_0(p^2)-4\wt C_{1/2}(p^2)+3\wt C_1(p^2)).}}
 Here we instead note that there are analogous relations for the gauge field propagators
 \eqn\sigmaxisusy{\eqalign{{1\over 4p^2}\ev{Q^2(D(p)D(-p)}&=\ev{\bar \lambda ^{\dot \alpha}(p)\bar \lambda _{\dot \alpha}(-p)}\cr 
{1\over 8p^2}\ev{\bar Q^2(\bar Q^2(D(p)D(-p)))}&=\Xi (p^2),}}
where $\Xi (p^2)$ is as defined in \xiis.   These relations are  obtained from the susy-transformations of the fields in the vector multiplet, e.g. \WessCP:
\eqn\susygauge{\delta _\xi D=\bar \xi \bar \sigma ^m\partial _m \lambda -\xi \sigma ^m\partial _m \bar \lambda, \qquad \delta _\xi \lambda =i\xi D-\sigma ^{mn}\xi v_{mn}.}
The susy relations for the gauge multiplet are related to those of the current multiplet by Legendre transform, since $D\leftrightarrow \delta / \delta J$,  $\lambda _\alpha \leftrightarrow \delta /\delta j^\alpha$, $V_\mu \leftrightarrow \delta/\delta j^\mu$.

Both \bmssx\ and \sigmaxisusy\ apply whether or not the $\CJ$ symmetry is spontaneously broken.  In the case where the $\CJ$ symmetry is not spontaneously broken, the relations \bmssx\ and \sigmaxisusy\ coincide with each other (to the relevant $g^2$ order), and they imply  the relations \bmssri\ and \bmssrii\ of \BuicanWS.  When the $\CJ$ symmetry is broken, on the other hand, it is the relations \sigmaxisusy, and not the relations \bmssx, which are relevant for computing the soft susy-breaking terms, as in \mgauginoq\ and \sfermintq.

\subsec{connection with small susy-breaking limits}

The above results can be connected with spurion-based methods when susy-breaking effects are small.
The idea is that 
\eqn\qthetarel{\ev{Q^2(\star)}=4\ev{(\star)_{susy}}|_{\theta ^2} +\CO(F|F|^2),  \qquad \ev{\bar Q^2(Q^2(\star))}=16 \ev{(\star)_{susy}}|_{\theta ^2\bar \theta ^2} +\CO(|F|^4)}
where $(\star)_{susy}$ denotes the quantity computed first in the limit of unbroken susy, and then replacing the parameters with spurions, which can have  susy-breaking $\theta ^2$ and $\theta ^4$ components.  Applied to \bmssx\ this yields the following relations, which were originally obtained in 
 \refs{\DistlerBT, \IntriligatorFR}:
\eqn\gammathetaii{\wt C_{susy}^{AB}(p^2)|_{\theta ^2}=-\wt B_{1/2}^{AB}+\CO(F|F|^2), \qquad \wt C_{susy}^{AB}(p^2)|_{\theta ^2\bar \theta ^2}=\half  p^2(\wt C_0^{AB}-4\wt C_{1/2}^{AB}+3\wt C_1^{AB})+\CO(|F|^4).}

The relations  \gammathetaii\ remain valid in the gauge messenger case, where the symmetry is spontaneously broken, and then  the quantities $\wt C_a$ and $\wt B_{1/2}$ above should be understood as the sum of the regular and pole contributions.  
Consider, in particular, the pole contributions.  In the limit of unbroken susy, we have 
\eqn\csusyis{(\wt C_{susy}^{pole})^{AB}={\bar \Phi \{T^A, T^B\}\Phi \over p^2}\equiv {(\bar \Phi , \Phi)^{AB}\over p^2}.}
To leading order in small susy-breaking, the relation \csusyis\ is preserved, where we 
use a spurion analysis, with chiral superfields $\Phi =\phi +\theta ^2 F$.  We are considering multiple spurions, and could decorate $\Phi$ with flavor and gauge indices, but we'll suppress the indices.   We then have the superspace expansion
\eqn\phiiexp{(M_V^2)^{AB}=g^2(\bar \Phi ,\Phi )^{AB}=
 (m_V^2)^{AB}+\theta ^2 (m_V^2)^{AB}m_\chi ^{AB}+h.c. +\half \theta ^2\bar \theta ^2( m_V^2)^{AB}(\delta m_0 ^2)^{AB},}
where we define
\eqn\mvec{(m_V^2)^{AB}\equiv g^2(\bar \phi _0, \phi _0)^{AB}, \qquad m_\chi ^{AB}\equiv {(\bar \phi _0, F_0)^{AB}\over  (\bar \phi _0, \phi _0)^{AB}}, \qquad   (\delta m_0^2)^{AB}\equiv 2{(\bar F_0,F_0)^{AB}\over  (\bar \phi _0, \phi _0)^{AB}}.}
The relations \gammathetaii\  imply that 
\eqn\bpolea{\wt B_{1/2}^{pole\ AB}=-{(\bar \phi ,  F)^{AB}\over p^2}+\CO(F|F|^2),}
\eqn\cdiffp{\wt C_0^{pole\ AB}-4\wt C_{1/2}^{pole\ AB}+3\wt C_1^{pole\ AB}={2(\bar F, F)^{AB}\over p^4} +\CO(|F|^4).}
These relations can be explicitly  verified in the weakly coupled case, by expanding out the  pole contributions given in \cfunctionsapp.  We can also immediately verify them in the limit of large $p^2$, via the expansions \celse.

But it is the susy relations \sigmaxisusy\ among the propagators, rather than the 
susy relations \bmssx\ among the current correlators, which are relevant for finding the 
 soft breaking terms in the gauge messenger case.  Applying the relation \qthetarel\ in the small susy-breaking limit, we find the appropriate relations, analogous to \gammathetaii:
\eqn\sigreln{\Sigma ^{AB}(p^2) = {1\over p^2}\Delta _{susy}^{AB}|_{\theta^2}+\CO (F|F|^2),}
and
\eqn\xireln{\Xi ^{AB}(p)={2\over p^2}\Delta _{susy}^{AB}|_{\theta ^2\bar \theta ^2} +\CO(|F|^4),}
where we have for the  full $\ev{D^A(p)D^B(-p)}$ propagator \eqn\deltasusy{ \Delta _{susy}^{AB}=\bigg({1\over 1+g^2\wt C_{susy}}\bigg)^{AB},}
initially computed for unbroken susy, and then extended to depend on susy-breaking spurions.  We compute \deltasusy\ to $\CO(g^2)$, using 
\eqn\cpolereg{g^2\wt C_{susy}=g^2\wt C_{susy}^{pole}+g^2 \wt C_{susy}^{reg}={M_V^2\over p^2}+g^2 \wt C_{susy}^{reg}(p^2)}
with the first term $CO(g^0)$ and the second $\CO(g^2)$.  The propagator then has $\CO(g^0)$ and $\CO(g^2)$ terms:
\eqn\deltasexp{\Delta _{susy}(p^2)=\left({p^2\over p^2+M_V^2}\right)^{AB}+\left[\left({p^2\over p^2+M_V^2}\right)(-g^2 \wt C_{susy}^{reg})\left({p^2\over p^2+M_V^2}\right)\right]^{AB}+\CO(g^4).}

\subsec{hidden sector contributions to visible sector wavefunction $Z_Q$}

The above expressions motivate considering  the hidden sector contribution to the wavefunction $Z_Q$ factor, which can be computed from the diagrams of fig. 1, computed with non-zero external momentum $k$, taking $\partial /\partial k^2$ before setting $k\to 0$.  In the limit of unbroken susy, this gives
\eqn\zqis{Z_Q\supset 1-2g^2\sum _{A,B} T_{r_Q}^A T_{r_Q}^B\int {d^4p\over (2\pi)^4}{1\over p^4}\Delta^{AB} _{susy}(p^2).}
To leading order in the small susy-breaking limit, this relation is preserved as a relation in superspace, with both sides picking up $\theta ^2$ and $\bar \theta ^2$ components.  Comparing \zqis\ with \sfermintq\ and \aqloopoq, and using \qthetarel\ and \sigreln\ and \xireln, we then have in the small susy-breaking limit:
\eqn\amzq{A_Q=Z_Q|_{\theta ^2}+\CO(F|F|^2), \qquad \widetilde m_Q^2=-Z_Q|_{\theta ^2\bar \theta^2} +\CO(|F|^4).}
With these relations, we recover the expected relations from spurion methods.

Consider first the one-loop $\CO(g^2)$ contribution to the Z-factor \zqis, which is obtained by using the first term in \deltasexp.  Doing the $d^4p$ integral using
\eqn\pintis{\int {d^4p\over (2\pi)^4}{1\over p^2(p^2+m^2)}=\hbox{constant}+{1\over 16\pi^2}\ln {\Lambda ^2\over m^2},}
where $\Lambda$ is the UV momentum cutoff, this term is 
\eqn\zqiso{Z ^{(1)}_Q={\Delta c_Qg^2 \over 8\pi ^2}\ln {M_V^2\over \Lambda ^2},}
where for simplicity we take $(M_V^2)^{AB}=\delta ^{A'B'}M_V^2$.  As a check, note that this same result can be obtained from the one-loop effective K\"ahler potential \GrisaruVE
\eqn\keffis{K_{eff}^{(1)}=-{1\over 32\pi ^2}\left[\Tr\ M_c^\dagger M_c \ln \left( {M_c^\dagger M_c\over \Lambda ^2}\right)-2\Tr\ M_V^2\ln \left({M_V^2\over \Lambda ^2}\right)\right],}
associated with integrating out massive chiral and vector superfields, respectively.  Since visible sector expectation values also contribute to the vector multiplet mass, we take $M_V^2\to g^2(\bar Q, Q)+M_{V,hid}^2$ and, expanding around $Q=0$, 
\eqn\keffexp{\eqalign{K_{eff}^{(1)}&\supset {g^2\over 16\pi ^2} \sum _{A'B'}(\bar Q, Q)^{A'B'} \ln \left({g^2 (\bar \Phi, \Phi)^{A'B'}\over \Lambda ^2}\right),\cr &=Z_Q^{(1)}(\Phi , \bar \Phi )\bar Q Q\qquad Z_Q^{(1)}(\Phi , \bar \Phi)={g^2\over 8\pi ^2}\Delta c_Q \ln \left({g^2(\bar \Phi , \Phi)\over \Lambda ^2}\right),}}
where in the last line  we took the  broken contributions to satisfy $(\bar \Phi , \Phi)^{A'B'}=(\bar \Phi , \Phi)\delta ^{A'B'}.$  This indeed agrees with the expression \zqiso.

We can likewise consider \zqis\ to two-loops, $\CO(g^4)$, by  using the second term in \deltasexp.  Separating out the contributions from the massless $G$ generators, and the broken $G'/G$ generators, respectively, this gives (with $c_2(r_Q)\equiv c_Q$ and $c_2(r_Q')-c_2(r_Q)\equiv \Delta c_Q$)
\eqn\zqist{Z^{(2)}_Q=2g^4 \int {d^4 p\over (2\pi)^4}{1\over p^4}\left(c_Q\wt C_{susy}^{unbr}(p^2)+\Delta c_Q \left({p^2\over p^2+M_V^2}\right)^2\wt C_{susy}^{reg, brok}(p^2)\right).}

\subsec{One loop a-terms and sfermion masses, in the small susy-breaking limit}

Let's consider first the one-loop contribution to $A_Q$ and $m_Q^2$, which are obtained from \amzq\ and \zqiso. Using $(M_V^2)^{AB}=g^2(\bar \Phi , \Phi )^{AB}$ and the expansion \phiiexp, we obtain 
the one loop A-term:
\eqn\aqapprox{A_Q^{(1)}={g^2\Delta c_Q\over 8\pi ^2}\ln M_V^2|_{\theta ^2}={g^2\over 8\pi ^2}\Delta c_Qm_\chi \qquad \hbox{where}\qquad m_\chi ={(\bar \phi _0, F_0)\over (\bar \phi _0, \phi _0)}.}
Similarly, the one-loop contribution to $m_Q^2$ obtained from \amzq\ and \zqiso\ is 
\eqn\mqoneexp{\eqalign{(m_Q^2)^{(1)}&=-{g^2\over 16\pi ^2} \sum _{A', B'}T_{r_Q}^{A'}T_{r_Q}^{B'}\left({(\bar \phi , \phi)( \bar F, F)-(\bar \phi , F)( \bar F, \phi) \over (\bar \phi, \phi )   (\bar \phi, \phi )}\right) ^{A'B'} +\CO(|F|^4),\cr &=
-{g^2\over 16\pi ^2} \Delta c_Q \left({(\bar \phi , \phi)( \bar F, F)-(\bar \phi , F)( \bar F, \phi) \over (\bar \phi, \phi )   (\bar \phi, \phi )}\right) +\CO(|F|^4),}}
where in the second line, we simplified by taking the inner products of the expectation values appearing in \mqoneexp\ to all be proportional to $\delta ^{A'B'}$. The one-loop gauge messenger contribution to the sfermion mass-squared is tachyonic, as is manifest from \mqoneexp.

The result \mqoneexpp, in the special case when $(\bar \phi, F)=0$,  reduces to the small susy-breaking limit of the $\wt B_{1/2}=0$ result \msfermiononel.   
Writing $m_C^2=m_V^2+\delta m_0^2$ and $m_{\lambda}^2=m_V^2+\delta m_{1/2}^2$, taking $\delta m_0^2\ll m_V^2 $ and $\delta m_{1/2}^2\ll m_V^2$.  In this limit, \msfermiononel\ reduces to 
\eqn\msfermonela{(m_Q^2)^{(1)}\approx {g^2\Delta c_Q\over 16\pi ^2}(\delta m_0^2-4\delta m_{1/2}^2).}
Using the relation \mbreakreln, the result \msfermonela\ indeed agrees with \mqoneexp\ with $\wt B_{1/2}=0$.

As we have shown, the result \zqiso\ can also be obtained directly from the 1-loop effective K\"ahler potential \keffexp, so the result \mqoneexp\ for the one-loop $m_Q^2$ at $\CO(|F|^2)$ can be viewed as a direct consequence of the one-loop K\"ahler potential.  We see that the common view that there are not one-loop contributions to $m_Q^2$ applies to the $M_c^\dagger M_c$ term in \keffis, but generally not to the $M_V^2$ term. In the appendix,  we will verify more generally, to arbitrary order in $|F|^2$, that our results about the one-loop contributions to $m_Q^2$ can equivalently be re-derived directly from the one-loop Coleman-Weinberg effective potential.

\subsec{One loop sfermion masses are suppressed for large goldstino pseudomodulus vev}

The numerator in \mqoneexpp\ has an interesting property: it is constant along the goldstino pseudomodulus flat direction.  As proved in \refs{\PolchinskiQK,  \RayWK}, all models of tree-level spontaneous F-term supersymmetry breaking (with canonical K\"ahler potential) have a classical pseudomodulus -- that of the goldstino -- 
which can be parameterized as
\eqn\goldflat{\phi _i^x=\phi _i^0+x F_i^0,}
where $x$ is an arbitrary complex parameter, associated with the expectation value of the superpartner of the goldstino, and $\phi _i^0=\ev{\phi _i}$ is chosen as a zero of the D-term for any gauge interactions, with $F_i^0=\ev{F_{\phi _i}}$.  It was shown in \PolchinskiQK\ that the classical $D$ and $F$ terms are independent of the parameter $x$, in particular $F_i^x=F_i^0$, i.e. constant along the goldstino pseudomodulus.  In terms of the quantities appearing in \mqoneexp, we therefore have
\eqn\xquants{\eqalign{(\bar \phi ^x, \phi ^x)^{AB}&=(\bar \phi ^0, \phi ^0)^{AB}+ x(\bar \phi  ^0, F^0) ^{AB}+x^*(\bar F^0, \phi ^0)^{AB}+|x|^2 (\bar F^0, F^0)\cr (\bar F^x, \phi ^x)^{AB}&=(\bar F^0, \phi ^0)^{AB}+x(\bar F^0, F^0)^{AB}}}
from which it immediately follows that the numerator in \mqoneexp\ is constant along the goldstino pseudomodulus
\eqn\xquantd{\left((\bar \phi ^x, \phi ^x)( \bar F ^x, F^x)-(\bar \phi ^x, F ^x)( \bar F ^x, \phi ^x)  \right)^{AB}=\left((\bar \phi ^0, \phi ^0)( \bar F ^0, F^0)-(\bar \phi ^0, F ^0)( \bar F ^0, \phi ^0)  \right)^{AB}.}

If the goldstino pseudomodulus has a large expectation value, and if $(\bar F^0, F^0)\neq 0$, then the gauge group is (perhaps partially) Higgsed at some correspondingly large mass scale $M$.  This scale can happen to be parameterically much larger than the typical mass scale $\mu$ of the superpotential.  
This is what happens in the inverted hierarchy model \WittenKV, where the dynamically generated GUT scale is much larger than the mass scale in the superpotential, and the large expectation values there are indeed along the goldstino pseudomodulus.   It follows from \xquantd\ that the numerator in \mqoneexp\ does not scale with that large scale $M$, but is instead $\CO(|\mu ^6|)$.  In such cases, the one-loop contribution \mqoneexpp\ to sfermion masses is parameterically of the form 
\eqn\mqonesupp{(m_Q^2)^{(1)}\sim {g^2\over 16\pi ^2}{|\mu| ^6\over |M|^4}.}
This has a $|\mu /M|^2\ll 1$ parameteric suppression as compared with the two-loop contributions,
\eqn\mqtwos{(m_Q^2)^{(2)}\sim \left({g^2\over 16\pi ^2}\right)^2{|\mu |^4\over |M|^2}.}
So, if $|\mu /M|^2\ll g^2/16\pi ^2$, the one-loop contribution to sfermion $m_Q^2$ could be negligible compared with the two-loop contributions!  This was briefly noted already long ago in a comment appearing in \DimGM\ in the context of the inverted hierarchy type models, but no explanation was given there (it was stated without calculation that a certain diagram would be $\CO(\mu ^6/M^4)$, but other diagrams, which could have contributed at $\CO(\mu ^4/M^2)$, were not considered).  Here we have explained it as coming from the particular general property  of the goldstino pseudomodulus.

There can be pseudomoduli other than the goldstino (see e.g. \IntriligatorFE), and the cancellation in \xquantd\ and above noted 
suppression of the one-loop $m_Q^2$ would not occur for large expectation values of those pseudomoduli.

\subsec{One-loop gaugino masses and two-loop $m_Q^2$ contributions}

Consider the one-loop gaugino masses for the superpartners of the massless gauge fields, and the two-loop sfermion $m_Q^2$, to leading order in a small susy-breaking expansion.  We have seen that these soft terms are given by 
\eqn\mbreaks{m_{gaugino}^{unbr}\approx -g^2 \wt C^{unbr}(0)|_{\theta ^2}, \qquad (m_Q^2)^{(2)}\approx -Z_Q^{(2)}|_{\theta ^2\bar \theta ^2},}
where $Z_Q^{(2)}$ is given in \zqist.  In both expressions, $\wt C_{susy}(p^2)$ is promoted to superspace, via replacing $m_V^2\to M_V^2$, as given in \phiiexp.  We'll now show how the quantities \mbreaks\ can be evaluated in terms of the coefficient $c$ of \cac\ which enters into the leading UV singularity of the $JJ$ OPE \opeex, in close analogy with how it worked in the non-gauge messenger case discussed in \refs{\DistlerBT, \IntriligatorFR}.  This will allow us to similarly connect, now for theories with gauge messengers, with the expressions obtained in \gr\ using analytic continuation in superspace.

The relevant functions in \mbreaks\ and \zqist\ are of the general form
\eqn\cgenform{\wt C(p^2)={c\over 16\pi ^2}\ln {\Lambda ^2\over M_V^2}+f\Big({p^2\over M_V^2}\equiv y\Big)}
where $f(y)$ is a non-singular function, which is regular at $y=0$ and satisfies
\eqn\flim{ \lim _{y\to \infty}f(y)\to -{c \over 16\pi ^2}\ln y,}
so the $M_V^2$ dependence drops out for large $p^2$.  
To give a concrete example, the unbroken gauge group has both matter and gauge contributions, $C^{unbr}(p^2)=C^{unbr, matter}(p^2)+C^{unbr,gauge}(p^2)$, where the gauge contribution is given by \cgaugesusy\ with $m_V^2\to M_V^2$ as \phiiexp, 
\eqn\cgaugesusya{\wt C^{unbr, gauge}(p^2) = c^{gauge}  \int {d^4k\over (2\pi)^4}{1\over (k^2+M_V^2)((p+k)^2+M_V^2)},}
which is of the form \cgenform\ with $c=c^{gauge}$ given in \cbeta.  Our point here will be that  the quantities \mbreaks\ can be evaluated knowing only the coefficients $c$ in \cgenform\ and \flim, without needing to know the detailed expressions like \cgaugesusya.

Using the superspace expansion of $M_V^2$, see \phiiexp\ and \mvec, it follows from \cgenform\ that
\eqn\cgenformtt{\eqalign{\wt C|_{\theta ^2}(y)&=-\Big({c\over 16\pi ^2}+y {d\over dy}f(y)\Big)m_\chi  \cr 
\wt C|_{\theta ^2\bar \theta ^2}(y)&=-\Big({c\over 16\pi ^2}+y{df\over dy}\Big) \Big(\half \delta m_0^2 -m_\chi \bar m _\chi\Big)+y{d\over dy}\Big(y{d\over dy}f\Big)m_\chi \bar m_\chi.}}
The quantity $\half \delta m_0^2-m_\chi \bar m_\chi=((\bar F, F)(\bar \phi, \phi)-(\bar F, \phi)(\bar \phi, F))/(\bar \phi , \phi)^2$ is the same combination appearing in the one-loop $m_Q^2$ in \mqoneexp, whose numerator is constant along the goldsino pseudomodulus direction.

The one-loop gaugino masses for the superpartners of the unbroken gauge fields are thus given, to leading order in small susy-breaking by \mbreaks\ and \cgenformtt\ to be
\eqn\mglead{m_{gaugino}^{unbr}\approx -g^2\wt C^{unbr}|_{\theta ^2}(y=0)={g^2 \over 16\pi ^2}c^{unbr}m_\chi .}
As we have already discussed, this one-loop $\CO(F)$ term is generally non-zero, unlike theories without gauge messengers where the gaugino masses are suppressed  \PolchinskiAN.  The gaugino mass \mglead\ and the a-term $A_Q$ in \aqapprox\ are both given by a loop factor times $m_\chi$, and differ only in the group theory factor $\Delta c_Q$ vs $c^{unbr}$.  The more general difference appears in \aqloop\ vs \mlamcont, but the $1/p^2$ vs $(p^2+m_V^2)^{-1}$ propagator in the loop there has no effect in the $\CO(F)$ contribution here (as seen from \cgenformtt, since $y=0$ is determined from $c$, which itself can be determined from $y\to \infty$, where $m_V^2$ is insignificant).

Let's now use the second line of \cgenformtt\ to evaluate the two-loop contribution to $m_Q^2$ in \mbreaks, using \zqist,
\eqn\momint{(m_Q^2)^{(2)}\approx -2{g^4\over 16\pi ^2}\int _0^\infty {dy\over y}\left (c_Q\wt C^{unbr}|_{\theta ^2\bar \theta ^2}(y)+\Delta c_Q \left[\left(y\over y+1\right)^2 \wt C^{reg, brok}\right]_{\theta ^2\bar \theta ^2}(y)\right).}
The $\wt C^{unbr}$ integral  is easily computed upon using \cgenformtt\ and \flim.  The $\wt C^{reg, brok}$ integral is similar, upon replacing $\wt C \to \left({y\over y+1}\right)^2\wt C$  on both sides of \cgenformtt, i.e. 
\eqn\stepis{\eqalign{\left[\left(y\over y+1\right)^2 \wt C^{reg, brok}\right]_{\theta ^2\bar \theta ^2}&= -({c^{brok}\over 16\pi ^2}+y{d\widehat f \over dy})(\half \delta m_0^2-m_\chi \bar m_\chi)\cr &
+m_\chi \bar m_\chi y{d\over dy}\left[y{d\widehat f\over dy}+2{c^{brok}\over 16\pi ^2}\left({y\over y+1}\right)^2\right],}}
where $\widehat f\equiv \left({y\over y+1}\right)^2 f$ and the last term comes from $({c\over 16\pi ^2}\ln {\Lambda ^2\over M_V^2})_{\theta ^2}\left({y\over y+1}\right)^2_{\bar \theta ^2} +(\theta ^2\leftrightarrow \bar\theta ^2)$.  So the integrals in \momint\ are all of total derivatives, and yield 
\eqn\momintapp{(m_Q^2)^{(2)}\approx 2\left({g^2\over 16\pi ^2}\right)^2\left[\left( c_Q c^{unbr}-\Delta c_Q c^{brok}\right)m_\chi \bar m_\chi-(IR)\cdot (\half \delta m_0^2-m_\chi \bar m_\chi)\right],}
(the sign flip of the $c^{brok}$ term relative to the $c^{unbr}$ term is thanks to the extra contribution from the last term in \stepis).  Here $(IR)$ is given by contributions coming from the IR region of the integral involving the term 
proportional to  $\half \delta m_0^2-m_\chi \bar m_\chi$ in \cgenformtt,
\eqn\iris{(IR)\approx  (c_Qc^{unbr} +\Delta c_Q c^{brok})\ln y_{IR}+16\pi ^2 c_Q f^{unbr}(y=0).}
 The apparent IR divergence  for $y_{IR}\to 0$ includes the renormalization of the one-loop $m_Q^2$, see also (A.31), and again  $\half \delta m_0^2-m_\chi \bar m_\chi\to 0$ far along the goldstino pseudomodulus direction. The coefficient of the $m_\chi m_{\bar \chi}$ term in \momintapp, upon using the relations $c^{brok}=-b'$ and $c^{unbr}=b-b'=c^{matter}-3T(G')+3T(G)$, is $c_Qc^{unbr}-\Delta c_Q c^{brok}=c_Q b + c_Q'b'-2c_Q b'$.  This result  connects with that obtained by the method of \gr, which is reviewed and extended in appendix A.

\newsec{Gauge messenger examples}

\subsec{A simple class of F-term breaking models}

In this section, we will illustrate our general results for a simple class of models.  Suppose that the susy-breaking hidden sector has a symmetry group $G'$, with chiral superfields $X$, ${\bf \phi _1}$, and ${\bf \phi _2}$, in the singlet, ${\bf r_1}$, and ${\bf r_2}$ representations of $G'$, respectively.  We take the reps ${\bf r_1}$, and ${\bf r_2}$ to be real, with ${\bf r_1}\otimes {\bf r_1}\supset 1$ and ${\bf r_1}\otimes  {\bf r_2}\supset 1$.  To break susy, we take an O'Raifeartaigh-type tree-level superpotential 
\eqn\super{W=\half h X {\bf \phi _1}\cdot {\bf \phi _1}+m{\bf \phi _1}\cdot {\bf \phi _2}+fX,}
where ${\bf \phi _1}\cdot {\bf \phi _1}$ denotes the $G'$ singlet.  The $G'$ symmetry group can be weakly gauged, and used to mediate susy-breaking to a visible sector.

Such examples have been much studied in the old and recent literature.  For example we could take $G=SU(N)$ and ${\bf r_1}$ and ${\bf r_2}$ to be adjoints, with the $SU(5)$ case similar to the 
inverted hierarchy model of \WittenKV, and the $SU(2)$ case was considered in toy models for the inverted hierarchy model  in 
\refs{\GinspargTQ, \PolchinskiAN, \PolchinskiAJ}. The general class of models \super\ was considered recently \IntriligatorPY\ as toy models for spontaneous R-symmetry violation: the theory has a $U(1)_R$ symmetry with $R(X)=R({\bf \phi _2})=2$, and $R({\bf \phi _1})=0$, which is spontaneously broken by the goldstino pseudomodulus $\ev{X}\neq 0$ if $\eta \equiv g^2/h^2$ is sufficiently large  \refs{\DineXT, \IntriligatorPY}.

The $G'$ symmetry is either unbroken, if $y\equiv | hf/m^2|$ satisfies $y\leq 1$, or is broken to a subgroup, $G'\to G$, if $y\geq 1$.   Upon gauging $G'$, the theory with $y<1$ does not have gauge messengers, while the theory with $y>1$ does.  So we'll here be interested in the case $y>1$, and we'll take $m$ and $f$ to be real, and set $h=1$.  The vacua have $\ev{X}=X_0$ arbitrary at tree level and 
 \eqn\vevsg{\eqalign{\ev{{\bf\phi _1}}\cdot\ev{{\bf\phi _1}}&=-v^2, \qquad \ev{{\bf\phi _2}}=-{X_0\over m}\ev{{\bf\phi _1}}, \qquad v\equiv \sqrt{2(f-m^2)}\cr -\ev{\bar F_{\phi _2}}&=m\ev{{\bf \phi _1}}, \qquad -\ev{\bar F_X}=m^2, \qquad \ev{F_{\phi _1}}=0.}}
The expectation value of the pseudomodulus $X_0$ in the quantum theory depends on the value of $\eta \equiv g^2/h^2$.  If $\eta$ is sufficiently small, then $X_0=0$, and if $\eta$ is sufficiently large then $X_0$ is non-zero, and determined in terms of $\eta$.  So we'll consider both the $X_0=0$ and $X_0\neq 0$ cases, where the R-symmetry is either unbroken, or spontaneously broken by $X_0\neq 0$, respectively.    In either case, $\ev{{\bf \phi _1}}\neq 0$ breaks the $G'$ symmetry to some subgroup, $G$, and $\ev{F_{\phi _2}}\neq 0$ means that there are gauge messengers.

Since these examples are weakly coupled, it's straighforward to compute the $G'$ currents $\CJ$ and their correlators.  In particular, the short distance OPE \opeex\ has 
\eqn\cmatter{c_{matter}=\sum _{i=1}^2 T_2({\bf r_i}).} 
Let's now comment on gaugino masses, considering the case where ${\bf r_1}={\bf r_2}$.  In the theory without gauge messengers, taking $y<1$,  the one-loop $G'$ gauginos remain massless to $\CO(F)$ (even if the R-symmetry is spontaneously broken by $X_0\neq 0$), as in \mgauginof, because $\tau \sim \ln \det \pmatrix{hX&m\cr m&0}$ is independent of $X$.  But in our $y>1$ case the gaugino partners of the unbroken $G$ gauge fields {\it do} get a one-loop mass at $\CO(F)$, given by \kexis
\eqn\mgaugino{m_{gaugino}=-{g^2\over 16\pi^2}(T(G')-T(G)){(\bar \phi, F)\over (\bar \phi , \phi)},}
where we take $T(r_1)=T(r_2)=T(G')-T(G)$, as in the example of \PolchinskiAJ.  This small-susy-breaking result is a good approximation when $X_0$ is large.

In the following two subsections, we'll focus on the case $G'=SO(N)$, with the two matter fields in the representations ${\bf r_1}={\bf r_2}={\bf N}$.  For $y>1$, the expectation values \vevsg\ break $SO(N)\to SO(N-1)$.  We'll first discuss in detail the case $N=2$, where the group is completely broken.  Next, we'll discuss $N>2$, where there are massless $SO(N-1)$ gauge fields.

\subsec{The example with $G'=SO(2)\cong U(1)$, a massive gauge field}

In this section, we consider the simplest model of spontaneous susy and gauge symmetry breaking.  We could equivalently write the hidden sector theory in this case in a $U(1)$ notation as
\eqn\superui{W=fX+ X\phi _{1+}\phi _{1-} +m\phi _{1+}\phi _{2-}+m\phi _{2+}\phi _{1-},}
where the $\pm$ subscripts are the $U(1)$ charges, and $X$ is neutral.  There is a messenger parity symmetry, $\phi _{i+}\leftrightarrow \phi _{i-}$ (this symmetry is ensured by  $\Tr\ T=0$ in $SO(2)$), so no $\ev{J}$ is generated at one-loop.  
To facilitate our later generalization to $SO(N)$, we'll stick with the $SO(2)$ notation, with superpotential \super, with fields $\phi _1^{(a)}$ and $\phi _2^{(a)}$, with $SO(2)$ index $a=1,2$, and $SO(2)$ generator $T=\frac1{\sqrt2}\pmatrix{0&i\cr -i&0}$.  
The vacua can be taken to be $\ev{X}=X_0$ (the classically undetermined goldstino pseudomodulus) and as in \vevsg, with in particular 
\eqn\vevs{\vev{\phi_1}= \pmatrix{iv\cr0},\qquad\vev{\phi _2}=-{X_0\over m}\vev{\phi _1},  \qquad -\vev{\bar F_{\phi _2}}=\pmatrix{imv\cr0},}
which are such that  the $SO(2)$ D-term vanishes, which minimizes the energy.

Since $\ev{\phi _1}\neq 0$, the $SO(2)$ is broken, and since $\ev{\bar F_{\phi_2}}\neq 0$, there are gauge messengers.  The breaking order-parameters are 
\eqn\orparms{(\bar \phi _0, \phi _0)=v^2\bigg(1+\Big|{X_0\over m}\Big|^2\bigg), \qquad (\bar F_0, F _0)=m^2 v^2, \qquad (\bar\phi, F_0 )=-X_0v^2.}  Once we gauge $SO(2)$, the mass of the $SO(2)\cong U(1)$ gauge field is 
\eqn\mworis{m_V^2=g^2(\bar \phi , \phi )=g^2v^2\bigg(1+\Big|{X_0\over m}\Big|^2\bigg).}
The tree-level undetermined expectation value $\ev{X}=X_0$ is associated with the goldstino pseudomodulus and, as we have noted on general grounds, $X_0$ indeed cancels in 
\eqn\phifcomb{(\bar \phi _0, \phi _0)(\bar F_0, F_0)-(\bar F_0, \phi _0)(\bar \phi _0, F_0)=v^4 m^2.}

It's straightforward to work out the spectrum of the theory in components, using Wess-Zumino gauge.  We'll discuss this spectrum first, before getting into the description in terms of current-correlators.    We consider arbitrary values of $m_V/m$, even with $g\ll 1$, by taking sufficiently large $v$, via taking $f$ sufficiently large.

The expectation value of the pseudomodulus  $X_0=\ev{X}$  in the quantum theory depends on the value of $\eta =g^2/h^2$ \refs{\DineXT, \IntriligatorPY}, where $g$ is the gauge coupling and $h$ in the  Yukawa coupling of the cubic term in \superui\ (which we set to 1 here): if $\eta$ is small, $\ev{X}=0$, and for larger $\eta$ we can have $\ev{X}\neq 0$.  For $\ev{X}=0$, the $U(1)_R$ symmetry of \super\ is unbroken, whereas for $\ev{X}\neq 0$ it is spontaneously broken.  We'll now first consider the $\ev{X}=0$ case, and next the $\ev{X}\neq 0$ case.

This hidden sector theory has 5 chiral superfields, including the NG chiral superfield $\Pi$, which is eaten by the vector multiplet.  The unbroken $U(1)_R$ symmetry for $\ev{X}=0$ implies that one chiral superfield plays the role of the field $\Psi$ in \wsplitt.  Using \ngbis\ and \vevs, and comparing \wsplitt\ with \super, we identify $\Pi$ and $\Psi$ as the fluctuation of the second component of $\phi _1$ and $\phi _2$ respectively, 
\eqn\exphsigma{\Pi = \phi _1^{(2)} \qquad \Psi= \phi _2^{(2)}, \qquad \hbox{with}\qquad \delta m_{1/2}=m} (where the superscripts indicate the second component under $SO(2)$).   In particular,  $\hbox{Im}( \Pi)$  is the (eaten) massless NG boson and $\hbox{Re}( \Pi)$, gives the real scalar with  $m_C^2=m_V^2+2m^2$. The complex scalar component of $\Psi$ has complex mass-squared equal to $\delta m_{1/2}^2=m^2$.  The fermion components $\psi _\pi$ and $\psi _\Psi$ and the gaugino $\lambda$ all mix, as in \fermterms, and the mass eigenstates are as in \fermeigens : there are two gauginos, with mass $\sqrt{m_V^2+\delta m_{1/2}^2}$, and one massless fermion.  (This massless fermion, together with the massless goldstino, contribute to saturate the 't Hooft anomaly $\Tr ~U(1)_R=2$ of the theory with gauged $SO(2)$; without gauging $SO(2)$, the 't Hooft anomaly is $\Tr~U(1)_R=1$ and the only massless fermion is the goldstino.)   In summary, the spectrum of the massive vector multiplet is 
given by 
\eqn\vecmassex{{\centertable{
\vrule height3.5ex depth2.5ex width 0.8pt #\tabskip=.6em 
&  \hfil#\hfil  &  #\vrule  &  \hfil$#$\hfil  &  #\vrule  &    \hfil$#$\hfil        &   #\vrule  &    \hfil$#$\hfil&#\vrule
width 0.6pt \tabskip=0pt\cr
\noalign{\hrule height 0.6pt}
&     field   &                &        V_\mu           &                 &    \hbox{Re}( \Pi)    &                 &      \delta m_{1/2} \psi _\Psi +i m_V\lambda , \quad \psi _\Pi      &\cr
\noalign{\hrule}  
&  mass${}^2$  &           &         m_V^2=2g^2(f-m^2)          &                   &     m_C^2=m_V^2+ 2m^2                  &                 &         m_\lambda ^2 = m_\chi ^2=m_V^2+m^2  & \cr
\noalign{\hrule height 0.6pt}
}}}
where $V_\mu$ contains the NG boson $\approx \hbox{Im} (\Pi)$ as the longitudinal component.  
The vector multiplet thus has (at tree-level)
\eqn\vecmassexstr{Str M^2|_{\rm vector multiplet}=-2m^2.}
Of course the full theory has $Str M^2=0$, and indeed the complex scalar $\Psi |=\phi _2^{(2)}|$ has mass-squared $m^2$, and thus contributes $Str M_\Psi^2=2m^2$, and so zeros out the contribution \vecmassexstr\ of the vector multiplet.    If we consider the theory in the limit $m_V\gg m$, the massive vector can be integrated out to obtain a low energy theory satisfying \mvstrr.

The spectrum of the remaining three chiral superfield degrees of freedom of the hidden sector have net $Str M^2=0$ and are less interesting for our considerations here, but we nevertheless record them for completness: 
\eqn\somestates{{\centertable{
\vrule height3.5ex depth2.5ex width 0.8pt #\tabskip=.6em 
&  \hfil#\hfil  &  #\vrule  &  \hfil$#$\hfil  &  #\vrule  &    \hfil$#$\hfil        &   #\vrule  &  \hfil$#$\hfil  &#&\hfil$#$\hfil&#&\hfil$#$\hfil&#\vrule
width 0.6pt \tabskip=0pt\cr
\noalign{\hrule height 0.6pt}
&     field   &                &         {mX+iv\phi _2^{(1)}\over \sqrt{2f-m^2}}         &                 &    {m\phi _2^{(1)}+ivX\over \sqrt{2f-m^2}}   &                 &  &&        \phi _1^{(1)}  &&   &\cr
\noalign{\hrule}  
&  mass${}^2$  &           &         0_S         &                   &     (2f-m^2)_S                  &                 &         (2f-m^2)_F&\vrule&(2f)_B&\vrule&   (2f-2m^2)_B    &\cr
\noalign{\hrule height 0.6pt}
}}}
where $0_S$ and $(2f-m^2)_S$ refer to a supersymmetric masses for the complete supermultiplet, and $(2f-m^2)_F$ is the mass of the fermion component, and $(2f)_B$ and $(2f-2m^2)_B$ are the masses of the real scalars $\hbox{Re} (\phi _1^{(1)})$ and $\hbox{Im} (\phi _1^{(1)})$ respectively.  The $(0)_S$ multiplet is the goldstino and pseudomodulus supermultiplet.  The other fields in \somestates\ are very heavy,
 with masses much bigger than all other fields in the spectrum;  they have masses $\sim g^{-2}m^2$ if we take $f \sim m^2 (1+g^{-2})$ (to get  $m_V\sim m$) in the $g\ll 1$ limit.

Let's now re-derive the tree-level spectrum found above in terms of our general, current-algebra based analysis.  To do so, we consider the contributions of the current one-point functions to $\wt C_a^{pole}$.  We again first consider the case where $\ev{X}=0$, where there is an unbroken $U(1)_R$ symmetry and consequently $\wt B_{1/2}=0$.  Using 
\eqn\orjones{J\supset\sqrt2v \hbox{Re}( \phi _1^{(2)}), \qquad j\supset -iv\psi _1^{(2)}, \qquad j_\mu \supset \sqrt2v\hbox{Im}( \phi _1^{(2)})} we can directly compute the associated pole contributions, which are given by the general expressions \cfunctionsapp.  The result is
\eqn\orcpole{g^2\wt C_0^{pole}={m_V^2\over p^2+2m^2}, \qquad g^2\wt C_{1/2}^{pole}={m_V^2\over p^2+m^2}, \qquad g^2\wt C_1^{pole}={m_V^2\over p^2},}
up to $\CO(g^2)$ loop corrections.  In short, there are simple poles, at $p^2=\delta m_a^2$, with $\delta m_{1/2}^2=\half \delta m_0^2=m^2$.  (In particular, and as is always the case, $\wt C_{1/2}$ does not have a pole at $p^2=0$, despite the presence of the massless fermion in \fermeigens.)  Using \orcpole\ we immediately obtain the (Landau gauge, Minkowski space) gauge field propagators 
\eqn\sotwoprop{\eqalign{
i\vev{DD}&=\frac{p^2-2m^2}{p^2-m_V^2-2m^2}+\O(g^2),\cr
i\vev{\l_\a\lb_\ad}&=\frac{p^2-m^2}{p^2-m_V^2-m^2}\left(\frac{p_{\a\ad}}{p^2}\right)+\O(g^2),\cr
i\vev{V^\mu V^\nu}&=\frac{1}{p^2-m_V^2}\Big(g^{\mu\nu}-\frac{p^\mu p^\nu}{p^2}\Big)+\O(g^2),\cr}}
with the pole locations as expected based on our earlier discussion \vecmassex\ of the gauge multiplet spectrum.

We consider coupling the above  theory to a visible sector, with $SO(2)$ charged matter, $Q$.   For simplicity, consider the case of a single $SO(2)$ doublet $Q$, with tree-level superpotential\foot{We assume that $|\mu|\ll |m|$, to justify using the massless $1/p^2$ propagator in the loop integral for $m_Q^2$ and $A_Q$, though it's straightforward to generalize the expressions to include the $Q$ masses.} $W_{vis}=\mu Q Q$, where $\mu$ is a supersymmetric mass term.   Because of the unbroken $U(1)_R$ symmetry, there is no induced $A_Q$ term, i.e. no off-diagonal $B_\mu$ mass term for the scalars $Q$.  The one-loop soft mass-squared for the $Q$ scalars is given by the expression \msfermiononel, which here gives
\eqn\msfermor{m_{\widetilde Q}^{2\ (1)}= {g^2\over 32\pi ^2}m_V^2 \ln \left({m_C^2m_V^6\over m_\lambda ^8}\right)=-{g^2\over 32\pi ^2}m_V^2\ln \left(
 {(m_V^2+m^2)^4 \over m_V^6(m_V^2+2m^2)}\right),}
which is always negative, tachyonic; as a function of $x\equiv m^2/m_V^2$, which characterizes the relative susy-breaking of the gauge messengers, \msfermor\ starts at zero for $x=0$ (the unbroken susy limit), and decreases with $x$.  For small susy breaking, $m^2/m_V^2\ll 1$, \msfermor\ can be approximated as in 
\msfermonela, 
\eqn\msfermorlimm{m_{\widetilde Q}^{2\ (1)}\approx {g^2\over 32\pi ^2}(\delta m_0^2-4\delta m_{1/2}^2)=-{g^2\over 16\pi ^2}m^2 \qquad (m_V\gg m).}
In the large susy-breaking limit, $m^2/m_V^2\gg 1$, \msfermor\ becomes 
\eqn\msfermorlim{m_{\widetilde Q}^{2\ (1)}\approx -{g^2\over 32\pi ^2}m_V^2 \ln \left({m^6\over 2m_V^6}\right)=-{g^2\over 32\pi ^2}g^22(f-m^2)\ln \left({m^6\over 16 g^6(f-m^2)^3}\right), \quad (m_V\ll m.) }
In particular, taking $g\ll 1$, we obtain $m_{\widetilde Q}^2 \approx -3(g^2/32 \pi ^2)m_V^2 \ln g^{-2}$.  (The result in this limit was emphasized in the recent, independent work \BuicanVV).

Let's now reconsider all of the above in the theory with  $\ev{X}\neq 0$ (which, as mentioned above, is realized if $g^2/h^2$ is sufficiently large).   This has the interesting new effect of  spontaneously breaking the $U(1)_R$ symmetry, and indeed we find that it leads to $\wt B_{1/2}^{pole}\neq 0$.  We redo our calculations with 
\eqn\neworvevs{\vev X\equiv X_0,\qquad\vev{\phi_2}= -\frac {X_0}m\pmatrix{iv\cr0},}
where we still define $v\equiv \sqrt{2(f-m^2)}$. The mass of the $SO(2)$ gauge field is then given by
\eqn\mvorexii{m_V^2=g^2 \bar \phi _0 \{ T, T\}\phi_0 = g^2 v^2 \left(1+\bigg|{X_0\over m}\bigg|^2\right).}
The NG chiral superfield $\Pi$ is read off  from \ngbis\ and the vevs \vevs, and there is also an analog of the field $\Psi$ in \exphsigma, generalized now to $X_0\neq 0$:
\eqn\exphsigmaa{\Pi \sim \phi _1^{(2)}-{X_0\over m}\phi _2^{(2)}, \qquad \Psi \sim \phi _2^{(2)}+{X_0\over m}\phi _1^{(2)}.}
Taking $X_0$ to be real for simplicity, we find upon using the general expressions \cfunctionsapp\ 
\eqn\neworcs{\eqalign{
g^2\wt C_0^{pole}&=\frac{g^2 v^2\big[(p^2+m^2)+\frac{X_0^2}{m^2}(p^2+X_0^2+4m^2)\big]}{(p^2+m^2)(p^2+2m^2)+X_0^2p^2},\cr
g^2\wt C_{1/2}^{pole }&=\frac{g^2 v^2\big[(p^2+m^2)+\frac{X_0^2}{m^2}(p^2+X_0^2+3m^2)\big]}{(p^2+m^2)^2+X_0^2p^2},\cr
g^2\wt C_1^{pole}&=\frac{m_V^2}{p^2},\cr
g^2\wt B_{1/2}^{pole}&=\frac{ g^2 v^2 X_0(p^2+X_0^2+2m^2)}{(p^2+m^2)^2+X_0^2 p^2}.
}}
Note that $\wt C_0^{pole}$ and $\wt C_{1/2}^{pole}$ are not of the simple, single-pole form \clom, but each can be written as a sum of {\it two} simple pole terms.  The poles of  \neworcs\ are at $p^2=-m_{state}^2$, where $m_{state}$ is the mass eigenvalue of the appropriate state in the O'Raifeartaigh model with $X_0\neq 0$ (and $f>m^2$).  For example, the poles of both $\wt C_{1/2}^{pole}$ and $\wt B_{1/2}^{pole}$ in \neworcs\ are at the masses of the fermionic components of $\phi _1$ and $\phi _2$,  given by $m_{1/2}^2=\frac{1}{4}(|X_0|^2\pm \sqrt{ |X_0^2|+4|m|^2})^2$.  
It is also easily verified that these satisfy \celse\ in the large $p^2$ limit, with susy-breaking parameters given by \wcsplittings, which we have already evaluated in \orparms.

It is, in principle, straightforward to use the above expressions in our general formulae, to compute the sfermion masses and the $A_Q$ term.  But the required momentum integrals are complicated to evaluate, even with a computer, and the resulting expressions would not be very illuminating in any case.  So we will just consider the results in the limit where $X_0\gg m \sim gv$.  In this limit, the supersymmetric contribution to the vector multiplet mass is much larger than the susy-breaking mass splittings, so the spectrum is approximately supersymmetric.  Indeed, in this limit, \neworcs\ gives low-momentum $(p^2\ll m^2$) pole behavior which is approximately supersymmetric, with  $\wt C_0^{pole}\approx \wt C_{1/2}^{pole}\approx \wt C_{1}^{pole}\gg \wt B_{1/2}^{pole}$ :
\eqn\capprox{g^2\wt C_0^{pole}\approx {m_V^2\over p^2+(2m^4/X_0^2)}, \qquad \wt C_{1/2}^{pole}\approx {m_V^2\over p^2+(m^4/X_0^2)}, \qquad \wt B_{1/2}^{pole}\approx {m^2\over X_0}\wt C_{1/2}^{pole}.}
Since the theory is approximately supersymmetric in this limit, we can use the small-susy breaking approximations of section 6.  The one-loop contribution to the visible sector sfermion $m_Q^2$ is approximately given by \mqoneexpp,
\eqn\mqorapproxf{ \eqalign{(m_Q^2)^{(1)} &\approx -{g^2\over 32\pi ^2}\left({(\bar \phi ,\phi)(\bar F, F)-(\bar\phi, F)(\bar F, \phi)\over (\bar \phi, \phi )^2}\right)\cr 
&\approx -{g^2\over 32\pi ^2}{m^6\over |X_0|^4}.}}
The one-loop A-term is given by \aqapprox, which gives
\eqn\aqorapprox{A_Q\approx {g^2\over 16\pi ^2}{(\bar \phi , F)\over (\bar \phi , \phi)}\approx -{g^2\over 16\pi ^2}{X_0|m|^2 \over |X_0|^2}.}
The two-loop $m_Q^2$ in this limit is given by the approximation of \momintapp\ (neglecting the term $\half \delta m_0^2-m_\chi\bar m_\chi \to 0$, $c^{unbr}=0$ ($SO(2)$ is completely broken), $\Delta c_Q=\half$, and $c^{brok}=-b'=\sum _{r=1}^2 T_2(\phi _r)=2$) 
to be again tachyonic and given by 
\eqn\mqtwoapprox{(m_Q^2)^{(2)}\approx -2\left({g^2\over 16\pi ^2}\right)^2 m_\chi \bar m_\chi \approx -2\left({g^2\over 16\pi ^2}\right)^2 {|m|^4\over |X_0|^2}.}
This and \mqorapproxf\ illustrate the general point that, for sufficiently large goldstino pseudomodulus $X_0$, the one-loop $(m_Q^2)^{(1)}$ can be insignificant compared with  $(m_Q^2)^{(2)}$.

\subsec{The $SO(N)\to SO(N-1)$ gauge messenger models}

We again consider the model \super, now taking $\phi _1$ and $\phi _2$ in the $N$ dimensional representation of an $SO(N)$ gauge symmetry.  For $y>1$ the vacua are given by \vevsg, with 
\eqn\vevvs{\vev{\phi_1}= \pmatrix{iv\cr0\cr\vdots},\qquad \ev{\phi _2}=-{X_0\over m}\pmatrix{iv\cr0\cr\vdots}, \qquad-\vev{F_2^*}=\pmatrix{imv\cr0\cr\vdots}}
where we take $D^A=0$ to minimize the energy.  This is compatible with \dfreln\ since $\bar F T^A T=0$; again, this is general for O'R models. The $G'=SO(N)$ symmetry is broken to $G=SO(N-1)$ by $\vev{\phi_1}\ne0$, and $\vev{F_{\phi_2}}\neq 0$ implies that there are gauge messengers: the broken $SO(N)/SO(N-1)$ gauge fields have a susy-split spectrum.  The unbroken $SO(N-1)$ gauge multiplets are massless at tree-level.  The $SO(N-1)$ gauginos get a one-loop mass in the theory with $\ev{X}\equiv X_0\neq 0$. When we couple this hidden sector to a visible sector, say a chiral superfield $Q$ in the $N$ of $SO(N)$, we'll obtain sfermion contributions coming from both the unbroken $SO(N-1)$ gauge fields, and also the broken gauge fields.   We'll initially consider the $X_0=0$ case, and next $X_0\neq 0$.

We choose a basis in which the first $N-1$ generators are broken, $(T^{A'})_{bc}=(i/\sqrt2)(\delta _{b, 1}\delta _{A'+1, c} - \delta _{c, 1}\delta _{A'+1, b})$, for $A'=1\dots N-1$.  The gauge boson mass matrix is then 
\eq{(m_V^2)^{AB}=\cases{g^2 v^2 \delta ^{AB}& if $A\leq N-1$\cr 0 & otherwise}.}
Briefly, the tree-level mass spectrum of the other states are as follows.  There are three superfield's worth of degrees of freedom which are exactly as in \somestates, one being the massless goldstino and pseudomodulus superfield.  The other fields are associated with the broken generators, and give a spectrum which is $N-1$ copies of that discussed in the $SO(2)$ case,  with $N-1$ fields, $A'=1\dots N-1$, 
\eqn\exphsigmaa{\Pi ^{A'}= \phi _1^{(A'+1)} \qquad \Psi ^{A'}= \phi _2^{(A'+1)}, \qquad \hbox{with}\qquad \delta m_{1/2}=m,}
which are the NG supermultiplets,  and the $\Psi ^{A'}$ supermultiplets which couple to them as in \wsplitt.  The real scalar partners of the broken $SO(N)/SO(N-1)$ gauge fields come from 
the $N-1$ scalars  $\sigma ^{A'}=\hbox{Re} \Pi ^{A'}$, with $m_C^2=m_V^2+2m^2$.  The gaugino partners of the massive $SO(N)/SO(N-1)$ gauge fields have $m_\lambda ^2=m_V^2+m^2$, and come from  $N-1$ copies of the spectrum \fermeigens.  (This also gives $N-1$ massless fermions which are related to the unbroken $U(1)_R$ symmetry of the theory with $\ev{X}=0$: together with the goldstino and the massless $SO(N-1)$ gauginos, saturate the 't Hooft anomaly $\Tr ~ U(1)_R=\Tr  ~ U(1)_R^3=|SO(N)|+1$ of the theory with $SO(N)$ gauged.)

Now consider our current-algebra based approach,  using  the $\CJ$ components
\eq{
J^{A'}\supset \sqrt2v\hbox{Re}(\phi_1^{(A'+1)}),\qquad
j^{A'}\supset -iv\psi_1^{(A'+1)},\qquad
j_\mu^{A'}\supset\sqrt2v\p_\mu\hbox{Im}(\phi_1^{(A'+1)}),
}
from which one can calculate the following tree-level current correlation functions:
\eqn\orcs{
\wt C_a^{AB\ pole}=\cases{\wt C_a^{pole}\delta ^{AB} & $A\leq N-1$\cr 
0& otherwise,}}
where $\wt C_a^{pole}$ are the same as in the $SO(2)$ case above:
\eqn\orcpoles{ g^2\wt C_0^{pole}={m_V^2\over p^2+2m^2}, \qquad g^2 \wt C_{1/2}^{pole}={m_V^2\over p^2+m^2}, \qquad g^2 \wt C_1^{pole}={m_V^2\over p^2}.}
From \orcs, one immediately obtains the broken gauge field propagators,
\eqn\sothreeprop{\eqalign{
i\vev{DD}^{A'B'}&=\frac{(p^2-2m^2)}{p^2-m_V^2-2m^2}\delta ^{A'B'}+\O(g^2),\cr
i\vev{\l_\a\lb_\ad}^{A'B'}&=\frac{p^2-m^2}{p^2-m_V^2-m^2}\frac{p_{\a\ad}}{p^2}\delta ^{A'B'}+\O(g^2),\cr
i\vev{V^\mu V^\nu}^{A'B'}&=\frac{1}{p^2-m_V^2}\Big(g^{\mu\nu}-\frac{p^\mu p^\nu}{p^2}\Big)\delta ^{A'B'}+\O(g^2),\cr}}
where $A'$, $B'\leq N-1$ run over the broken generators.  The poles of these propagators agree with the spectrum described in the previous paragraph.

Now consider the susy-breaking effects for visible sector matter, considering the case of a field $Q$ in the fundamental of $SO(N)$.  For $\ev{X}=0$, the unbroken $U(1)_R$ symmetry ensures that the gauginos of the unbroken $SO(N-1)$ gauge group remain massless.  Also, for $\ev{X}=0$, 
this is in the class of models with single-poles, where we can simply apply \msfermiononel\ to give the one-loop contribution to the sfermion soft-breaking mass 
\eq{(m_Q^2)^{(1)}=\frac{g^2}{16\pi^2}\Delta c_Qm_V^2\left[\ln(m_V^2+2m^2)-4\ln(m_V^2+m^2)+3\ln(m_V^2)\right],}
which is tachyonic.  
The fundamental of $SO(N)$ has $c_2({\bf N})=\half (N-1)$. When we break $SO(N)\to SO(N-1)$, $Q$ decomposes as ${\bf N}\to {\bf (N-1)}+{\bf 1}$, with $c_2({\bf N-1})=\half (N-2)$ and $c_2({\bf 1})=0$, so $\Delta c_Q=\half , \ \half (N-1)$ for the $SO(N-1)$ fundamental and singlet, respectively. (The sum on broken generators  is $T^{A'}T^{A'}=\half \hbox{diag}(N-1,1,1,\dots)$.)

The two-loop contributions to the sfermion $m_Q^2$ can be computed from \gaugemesstwol, using expressions for $\wt C_a^{reg}(p^2)$; the expressions are similar to those in the appendix of \mss.  In the small susy-breaking limit $m_V^2\gg m^2$, this result can be approximated as in \momintapp, with $m_\chi =0$ for $X_0=0$ thanks to the unbroken $U(1)_R$ symmetry.

Let's now consider the theory when  $\vev X\equiv X_0\ne0$ in \vevvs, which spontaneously breaks  the $U(1)_R$ symmetry, yielding $\wt B_{1/2}\neq 0$.  The broken generators have $\wt B_{1/2}^{A'B'\ pole}=B_{1/2}^{pole}(p^2)\delta ^{A'B'}$, with $B_{1/2}^{pole}(p^2)$ given by the same expression as in \neworcs\ for the $SO(2)$ case.  Likewise, $\wt C_a^{A'B'\ pole}= \wt C_a^{pole}\delta ^{A'B'}$, with $\wt C_a^{pole}$ given by same expressions \neworcs.  As in the $SO(2)$ case, one could in principle (numerically) evaluate the $p$ integral to compute the visible sector sfermion $m_Q^2$ and $A_Q$, but here we will again simply comment about the limit where the pseudomodulus vev $X_0$ is very large, $|X_0|\gg |m|$.     In this case the $SO(N)/SO(N-1)$ gauge fields are ultra-massive, with relatively small mass splittings.

The gaugino masses in this limit are given approximately by \mgaugino\ to be
\eqn\msonm{m_{gaugino}^{SO(N-1)}\approx -{g^2\over 16\pi ^2} {(\bar \phi , F)\over (\bar \phi , \phi)}\approx {g^2\over 16\pi ^2} {X_0|m|^2\over |X_0|^2}.}
The one-loop A-term is given approximately by \aqapprox\ to be 
\eqn\asonm{A_Q\approx {g^2\over 8\pi ^2} \Delta c_Q{(\bar \phi , F)\over (\bar \phi , \phi)}\approx - {g^2\over 8\pi ^2} \Delta c_Q{X_0|m|^2\over |X_0|^2}.}
The one-loop $m_Q^2$ is approximately given by \mqoneexp\ to be (as in \mqorapproxf)
\eqn\moneson{(m_Q^2)^{(1)}\approx -{g^2\over 16\pi ^2}\Delta c_Q {m^6\over |X_0|^4}.}
The two-loop $m_Q^2$ is given by \momintapp\ (with $c^{unbr}=b-b'=-1$, $c^{brok}=-b'=2-3(N-2)$): 
\eqn\mtwoson{(m_Q^2)^{(2)}\approx 2\left({g^2\over 16\pi ^2}\right)^2 (-c_2(r_Q) +(3N-8)\Delta c_Q) {|m|^4\over |X_0|^2},}
which is non-tachyonic for large enough $N$. 
For $r_Q={\bf N}$ then $c_2(r_Q)=\half (N-1)$ and $\Delta c_Q=\half, \ \half (N-1)$ for the $SO(N-1)$ fundamental and singlet, respectively.

\subsec{Example: the Fayet Iliopoulos model}

We now consider a $U(1)$ gauge theory with hidden sector matter fields $\Phi _\pm$ of charge $\pm 1$, and visible sector matter fields $Q_\pm$, of charge $\pm 1$.  The superpotential is $W=W_{hid}+W_{vis}$, with $W_{hid}=m\Phi _+\Phi _-$ and $W_{vis}=MQ_+Q_-$.   Clearly, the separation of the fields into hidden and visible sectors is artificial -- they're just two flavors of a whole model --  but we'll nevertheless consider first the hidden sector, and next its effect on the visible sector.  We'll consider the model with $g^2\xi >m^2$ and $M^2>m^2$ (where we take $\xi$, $m$, and $M$ to be real and positive for simplicity). The condition  $g^2\xi >m^2$ ensures that the $U(1)$ gauge group is Higgsed, resulting in gauge messengers.  The condition $M^2>m^2$ ensures that it's a hidden sector $\Phi $ field, and not a visible sector $Q$ field, which gets an expectation value.  The need to take $M>m$ is a peculiarity of this example, making it contrary to the usual setup of heavy hidden sector and light visible sector fields.  
Also, because the gauge field $D$ term is crucial for supersymmetry breaking, we can not reliably analyze this example by first setting the gauge coupling to zero.

For $g^2\xi >m^2$ the vacuum has 
\eqn\fimodelvevs{\ev{\Phi _+}=0, \quad \ev{\Phi _-}=v; \quad \hbox{hence}\quad \ev{\bar F_{\Phi _+}}=-mv, \quad \ev{F_{\Phi _-}}=0,}
where the EOM is satisfied for $v$ given by
\eqn\vfi{g^2 v^2=g^2\xi - m^2\equiv \half m_V^2}
where we  gauge rotate $v$ to be real.  The hidden sector contribution to the $U(1)$ gauge current $g\CJ =\delta \Gamma/\delta V$ is  $\CJ =\bar\Phi _+ \Phi _+-\bar\Phi _- \Phi _-+\xi$.  The $D$ equation of motion implies that
\eqn\dvevfi{\ev{D}=-g\ev{J}=- g\xi + g v^2=-\frac{m^2}g,}
which satisfies the relation \dfreln, relating $\ev{D}$ to the F-terms.   In discussing the 2-point functions, such as $\wt C_0$, we need to redefine with supercurrent to have vanishing expectation value, $\CJ ' =\CJ +\ev{\CJ}$.

The $U(1)$ gauge group is Higgsed and, since charged matter has a non-zero F-term in the vacuum \fimodelvevs, there are gauge messengers.  The  $\CJ ' $ current correlation functions can be computed using the \cfunctionsapp, where now there is a D-term contribution in \massesapp.   As noted after \dare, 
the D-term contribution  is $g$-independent, and this is seen explicitly from \dvevfi, $\D _-^-=g^2\ev{J}=-m^2$.  Using \cfunctionsapp\ then yields for the hidden-sector correlator functions 
\eqn\ficone{g^2 \wt C^{pole}_0= \frac{m_V^2}{p^2}, \qquad g^2\wt C_{1}^{pole}=\frac{m_V^2}{p^2},\qquad g^2\wt C_{1/2}^{pole}=\frac{m_V^2}{p^2+m^2},}
where $\wt C_0^{pole}(p^2)$'s pole location is at $p^2=0$ thanks to a cancellation between the non-zero $F$ and $D$ term contributions in \massesapp.    It follows from the fact that there is an unbroken tree-level $U(1)_R$ symmetry under which $R(\Phi _+)=2$, $R(\Phi _-)=0$ (it is anomalous, but preserved perturbatively) that
\eqn\bzero{\wt B_{1/2}=0.}
In particular, the susy-breaking parameters \celse\ are 
\eqn\fimbs{\delta m_0^2 =0, \qquad \delta m_{1/2}^2={(\bar F_0, F_0)\over (\phi _0, \phi _0)}=m^2, \qquad m_\chi = {(\bar F_0, \phi _0)\over (\phi _0, \phi _0)}=0.}
These do not satisfy the F-term specific relation \mbreakreln, because of the D-term contribution.  

Using \ficone\ in \fullgauge\ etc. yields the vector multiplet propagators, with
\eqn\fiprops{\Delta _0(p^2)=\Delta _1(p^2) ={p^2\over p^2+m_V^2}, \qquad \Delta _{1/2}(p^2)={p^2+m^2\over p^2+m_V^2+m^2}.}
The poles give  the vector multiplet spectrum (which of course can also be obtained by the standard analysis of the tree-level lagrangian in components, as in e.g. \WessCP): $m_C^2=m_V^2$ (because $\delta m_0^2=0$ in \fimbs) and the two gauginos are degenerate with each other (since $m_\chi =0$), with $m^2_{\lambda _1}=m^2_{\lambda _2}=m_V^2+m^2$. The $\CO (g^2v^2)$ contributions to the masses are supersymmetric, and the  vector multiplet has $Str M^2=-4m^2$.    (The remaining fields consists of the goldstino, the would-be NG boson, and a complex scalar with squared mass $2m^2$, so the entire spectrum has $Str M^2=0$.)

We now couple this hidden sector to the visible sector fields $Q_\pm$. At tree-level, the fermion components of $Q_\pm$ have supersymmetric mass $M$, and the sfermions have mass-squareds $M^2\pm m^2$, where the latter includes the tree-level sfermion susy-breaking mass 
\eqn\mfitree{(m_{Q_\pm}^2)^{(0)}= \pm m^2} coming from the D-term \dvevfi.  So we need $M^2>m^2$ if we want $\ev{Q_\pm}=0$ (this is just the statement that it's the lighter of the two $U(1)$ flavors which gets the expectation values as in \fimodelvevs).   At one-loop, the visible sector sfermion masses are given by \xiis\ and \sfermint, appropriately modifying the $1/p^2$ in \sfermint\  to account for the tree-level $Q$ masses (which are here not negligible) in the internal propagators in Fig. 1: 
\eqn\mfivis{(m_{Q_{\pm}}^2)^{(1)}\supset g^2\int\frac{d^4p}{(2\pi)^4}\left(\frac{p^2}{p^2+m_V^2}\frac1{p^2+M^2\pm m^2}-\frac{4(p^2+m^2)}{p^2+m_V^2+ m^2}\frac1{p^2+M^2}+\frac3{p^2+m_V^2}\right).}
Since we use Landau gauge, there is no contribution from diagram D4 in Fig. 1, only from diagram D5, so $M$ doesn't enter in the last term.  There is also a one-loop correction to the tree-level result \mfitree, from the running of the FI D-term between the mass scales $\sqrt{M^2+m^2}$ and $\sqrt{M^2-m^2}$.  

We can consider this model for arbitrary values of the ratio $m_V^2/m^2 \in [0, \infty]$.  In the limit where $m_V$ is the largest mass scale in the problem, we can integrate out the massive vector.  Since it has a nearly susy spectrum in this limit, there is an approximately supersymmetric low-energy effective theory, with $F$-term susy breaking.   (In terms of the gauge invariant $X=\Phi _+\Phi _-$, $K_{low}\approx \xi ^{-1}\bar X X-\half \xi ^{-2}(\bar X X)^2$, $W_{low}\approx mX$.) In this limit, we can compare \mfivis\ with the result obtained  from the 1-loop effective  K\"ahler potential
\eqn\konefi{K^{(1)}={1\over 16\pi ^2}\Tr M_V^2\ln ({M_V^2\over \Lambda ^2})\supset Z^{(1)}_{Q_\pm}(|\Phi |^2) (|Q_+|^2+|Q_-|^2),}
where $M_V^2=2g^2(|\Phi _+|^2+|\Phi _-|^2+|Q_+|^2+|Q_-|^2)$  and 
\eqn\zfiis{Z_{Q_{\pm}}=1+Z_{Q_\pm}^{(1)}=1+ {g^2\over 8\pi ^2} \ln ({|g\Phi _+|^2+|g\Phi _-|^2\over \Lambda ^2}).}
Using the $F$-term in \fimodelvevs\ gives 
\eqn\fisoftsq{\widetilde m_{Q^\pm}^{2, (1)}=-\ln Z_{Q_\pm}|_{\theta ^2\bar\t^2}=-{g^2\over 8\pi ^2} {|F_{\Phi _+}|^2\over v^2}=-{g^2\over 8\pi ^2}m^2.}

\newsec{Conclusions and outlook}

Looking to possible model building applications, we have seen that gauge messengers have some potentially useful, and also potentially problematic, differences from non-gauge messenger models.  Some of the useful differences have already been discussed in the literature.  Tachyonic two-loop contribution to sfermion $m_Q^2$s at the messenger scale sounds suspicious, but need not be fatal
if the gaugino masses are sufficiently large to drive the RG running of $m_Q^2$ to positive values at low energy (via the running in \maderivs).  (One might worry about a stripe phase at short distances; we will not consider this issue further here.) This is what happens in the scenario envisioned in \DermisekQJ, where the extra gauge fields of $SU(5)_{GUT}$ are gauge messengers.  There the squarks start off with tachyonic 2-loop $m_Q^2$s, which are driven positive in the IR by the gaugino masses.  The sleptons, on the other hand, start off with non-tachyonic 2-loop $m_Q^2$ at the messenger scale, which is fortunate as their weaker RG running could have been insufficient to drive them positive in the IR.  The result is a compressed spectrum of superpartners in the low-energy spectrum, which can alleviate little hierarchy tunings.  

These earlier studies, however,  did not include  the one-loop tachyonic contributions to $m_Q^2$.  The helpful RG running effect  from \maderivs\ is of 2-loop order (since the gaugino masses are 1-loop), which can be insufficient to  compensate for an initially tachyonic one-loop tachyonic contribution to $m_Q^2$, even for the squarks.  Sleptons also get tachyonic one-loop $m_Q^2$, and can again end up tachyonic in the IR.  Fortunately, as we have discussed, the one-loop $m_Q^2$ contribution can be suppressed, in models analogous to the inverted hierarchy \WittenKV, where the group is Higgsed at a scale $m_V\gg \sqrt{F}$, by the large expectation value of the goldstino pseudomodulus.  The one-loop result, $m_Q^2\sim -(g^2/16\pi ^2)|\mu |^6/m_V^4$,  is then potentially small compared with the  two-loop RG running effect, $m_Q^2\sim (g^2/16\pi ^2)^2|\mu |^4/m_V^2$, and the sfermions could end up non-tachyonic in the IR.  

From a bottom-up perspective, the bottom line of general gauge mediation \mss\ is that all visible sector soft terms have a fixed dependence in terms of six parameters: the gaugino masses are given by the three $\wt B_a(0)$ where $a=3,2,1$ for $SU(3)\times SU(2)\times U(1)$ (the $\wt B_a(0)$ are complex, but a hidden sector symmetry, e.g. CP, is imposed to avoid CP violating phases), and the sfermion masses are given by three real parameters, denoted in \mss\ by $A_a$, which we'll here rename 
(reserving $A$ for A-terms), and define them as 
\eqn\eais{E_a\equiv \int {d^4p\over (2\pi)^4}{1\over p^2}\Xi _a(p^2),}
with $E_a=\CO(g_a^2)$. With this notation, the general gauge mediation result \mss\ is 
\eqn\munbrc{m_Q^2 \supset \sum _{a=1}^3 g_a^2 c_2(r_Q; a) E_a.}
Because the 5 sfermion masses depend on the three parameters $E_a$, they obey the two  sum rules $\Tr \ m^2 U(1)_Y=\Tr~ m_Q^2 U(1)_{B-L}=0$ at the messenger scale \mss.  

In BSM theories with additional gauge fields, there are additional contributions to \munbrc, involving additional parameters, see e.g. \LuoKF\ for discussion.  For example, if the SM fields are charged under an additional massive $U(1)'$, the sfermion masses are modified from \munbrc\ by an additive shift, 
\eqn\mqdelta{\delta m_Q^2=q_Q^2 g_{new}^2 E_{new}, \qquad \hbox{where}\qquad E_{new}\equiv \int {d^4p\over (2\pi)^4}{1\over p^2}\Xi _{new}(p^2),}
where $q(Q)$ is the charge of $Q$ under $U(1)'$.  This is the case whether or not the $U(1)'$ is a gauge messenger, the only quantitative difference is the $g_{new}^2$ order of $E_{new}$, and the fact that gauge messengers tend to have $E_{new}<0$.  The addition of the new parameter $E_{new}$ would violate the above sum rules -- with a single new parameter $E_{new}$, one sum rule would remain.  Since gauge messenger models involve new BSM gauge fields, they introduce additional parameters like $E_{new}$, and thus violate the sum rules of \mss.  In examples where the gauge messengers  are like the $SU(5)_{GUT}/SU(3)\times SU(2)\times U(1)$ gauge fields, the analog of \mqdelta\ is 
\eqn\mqdeltagut{\delta m_Q^2 =g^2 \Delta c(r_Q)E_{new}.}

As we have discussed, gauge messengers also introduce A-terms already at one-loop order, which we'll parameterize in bottom-up fashion as
\eqn\aqparm{A_Q= g^2\Delta c(r_Q) A_{new}, \qquad A_{new}\equiv -2\int {d^4p\over (2\pi)^4}{1\over p^2}\Sigma _{new}^{brok}(p^2).}
Each broken gauge group factor having gauge messengers thus introduces an additional parameter $A_{new}$.  These parameters are generally complex, which could introduce problematically large CP violating phases.  But the same mechanism which eliminates the phases of the gauginos,  e.g. CP symmetry in the hidden sector, can ensure that the $A_{new}$ are real.

As we have mentioned, we can also apply our methods to models where the susy-breaking sector is coupled to an intermediate messenger sector, by some additional gauge interactions.  An example of this is the model of \semidirect, where the $SU(2)$ gauge fields of the 3-2 model act as gauge messengers to the intermediate sector of the doublets $\ell _i$.   (And the one-loop tachyonic masses that we have discussed generally here correspond to the  one-loop, negative $Str M^2$ found in the example of \semidirect).

\noindent {\bf Acknowledgments:}

This research was supported in part by UCSD grant DOE-FG03-97ER40546.   MCS was also supported by World Premier International Research Center Initiative (WPI Initiative), MEXT, Japan in the final stage of this work.

\appendix{A}{Susy breaking, spurions}

\subsec{Soft terms; the Higgs mechanism}

The supersymmetric effective action, to two derivatives, is
\eqn\ssusy{S_{susy}=\int d^8z K(\bar \Phi e^{2V},\Phi) +\int d^6z (W(\Phi)+{1\over 4}\tau (\Phi) W_\alpha  W^\alpha )+c.c.,}
where $d^8z=d^4x d^4\theta$, $d^6z=d^4x d^2\theta$ and $``+c.c."$ always refers only to the complex terms.  The soft terms can be written (using the notation of  \NibbelinkSI) as
\eqn\ssoft{\eqalign{S_{soft}&=\int d^8z\left[\theta ^2\bar \theta ^2 \widetilde K(\bar \Phi e^{2V},\Phi)+ \theta ^2 \widetilde k (\bar \Phi e^{2V}, \Phi)+c.c.)\right]\cr &+\int d^6z\theta ^2(\widetilde W(\phi)+{1\over 4}\widetilde \tau (\Phi)W_\alpha W^\alpha )+c.c..}}
There is a redundancy in this description \refs{\YamadaID, \NibbelinkSI}, and we'll absorb $\widetilde W(\phi)$ into a holomorphic addition to $\widetilde K$.  The soft term $\widetilde k$ affects the solution of the auxiliary field $F$ equations of motion: 
\eqn\Fsol{\widehat F ^i=-K^{i\bar j}(\bar W_{\bar j}+\widetilde k_{\bar j}),}
where the hat is a reminder that $F^i$ has been solved for in terms of the other fields.  The matter contribution to the scalar potential is then
\eqn\vscalar{V_F=-\widetilde K+K^{i\bar j}(\bar{\widetilde k}_i+W_i)(\widetilde k_{\bar j}+\bar W_{\bar j})\equiv V_F^{susy}+V_F^{soft}.}
The $A$ terms can be regarded as coming from $\widetilde k$, as $\widetilde k\approx A_Q Q^\dagger e^{2V}Q+\dots$.

Now consider the quadratic terms for a massive vector multiplet:
\eqn\massvec{\CL _{susy}\supset \int d^4 \theta\ m_V^2  V^2+{1\over 4}\int d^2 \theta W_\alpha W^\alpha  +c.c.,}
where the components of the vector multiplet are \WessCP
\eqn\vis{V=C+i\theta \chi + \theta ^2 N-\theta \sigma ^m\bar \theta v_m+ i\theta ^2\bar \theta [\bar \lambda +{i\over 2}\bar \sigma ^m \partial _m \chi ] 
+\half \theta ^2\bar \theta ^2(D+{1\over 2}\Box C)+c.c..}
For $m_V\neq 0$ in \massvec, the dynamical fields consist of the massive spin 1 field $v_m$, the real scalar $C$, and two fermions $\lambda _\alpha$ and $\chi _\alpha$.  
(The sign of the $m_V^2$ term in \massvec\ is correct, because $V^2|_{\theta ^2\bar \theta ^2}=-\half v_m v^m +\dots$ \WessCP.)  $C$ and $\chi$ are rescaled by a factor of $m_V$ to have canonical kinetic terms and mass dimension, and $C$, $v_m$, $\lambda$, and $\chi$ all have supersymmetric mass $m_V$ (with the $\chi$ and $\lambda$ fermions getting mass by paring up).

We can add the soft breaking mass terms, $\delta m_0, m_\chi, m_\lambda$, for the gauge multiplet
\eqn\massvecsoft{\CL _{soft}\supset \int d^4\theta (-{1\over 2}m_V^2 \delta m_0^2 \theta ^2\bar\theta ^2  V^2+m_\chi m_V^2 \theta ^2 V^2)+\int d^2 \theta \ m_\lambda  \theta ^2  W_\alpha ^2+c.c..}
The mass eigenvalue $m_C$ of the propagating real scalar and the mass matrix $M_{\lambda \chi}$ of the fermions $(\lambda _\alpha, \chi _\alpha)$
are
\eqn\mcis{m_C^2=m_V^2+\delta m_0^2 , \qquad M_{\lambda \chi}=\pmatrix{m_\lambda & m_V\cr m_V&m_\chi},}
and the  minus sign of the $\delta m_0^2$ term in \massvecsoft\ yields a positive contribution to $m_C^2$ in \mcis.   The diagonal components $m_\lambda$ and $m_\chi$ in \mcis\ break the $U(1)_R$ symmetry under which $R(\lambda)=-R(\chi)=1$.  We can also have mixing with fermions from the matter sector, which in the simplest case we can represent as a single additional fermion $\psi$, writing the mass matrix as e.g. 
$$\pmatrix{0&m_V&0\cr m_V&m_\chi &\delta m_{1/2}\cr 0&\delta m_{1/2} &0},$$
where there is an unbroken R-symmetry if $m_\chi=0$.

In the Higgs mechanism, the mass terms proportional to $m_V^2$ in \massvec\ and \massvecsoft\  arise from spontaneous symmetry breaking. We start with the gauge invariant matter kinetic terms
\eqn\mattkin{\CL \supset \int d^4\theta \bar \Phi e^{2gV}\Phi,}
and consider the effect when we take constant non-zero $\ev{\Phi}=\phi _0+\theta ^2 F_0$. 
We expand $e^{2gV}$ in the grassmann components \vis, and expand in $C$ (assuming  $\ev{C}=0$ for simplicity).
Then \mattkin\ yields 
\eqn\mvecab{\CL \supset \int d^4\theta (M_V^2)^{AB}V^AV^B,}
where $A$ and $B$ are adjoint indices and  $M_V^2$ is given by the superspace expansion \phiiexp\ and \mvec.  
For $F$-term breaking, and non-zero F-terms, $\delta m_0^2> 0$. As we'll soon explain, the sign flip of the $\delta m_0^2$ term between \mvec, where $M_V^2\supset +\half m_V^2\delta m_0^2 \theta ^2\bar \theta ^2$, and \massvecsoft, comes from integrating out the auxiliary fields; it's essentially the same sign flip as how $\CL \supset +|F|^2$ becomes $V\supset +|F_0|^2$ upon solving the $F$ EOM for $F_0$.

Let's first note that the auxiliary field $N^A$ plays no interesting role: the  term linear in $N^A$ vanishes by gauge invariance, because $V^A$ couples to the conserved current $\CJ ^A$, with $D^2\CJ ^A=0$.  Indeed, 
\eqn\linn{\CL\supset (\bar F_0 \frac{\partial e^{2gC}}{\partial C^A}\phi _0)N^A+c.c.}
and $\ev{\bar F_0T^Ae^{2gC}\phi _0}=0$ when $F_0$ is determined from a superpotential,  since $\bar F_0T^Ae^{2g\ev{C}}\phi _0=\sum _iW_i (T^A)_{r_i} \phi _{0,i}=0$ by gauge invariance of $W(\Phi _i)$.  So, regardless of whether or not one chooses Wess-Zumino gauge, the $N^A$ equations of motion always give $\ev{N^A}=0$.

Now let's consider the $D^A$ auxiliary fields.  
Upon adding the term $\CL \supset \half D^AD^A$ coming from the gauge kinetic terms in \massvec,
the $D^A$ equation of motion gives
\eqn\Dhatis{\widehat D^A=-\half\bar \phi _0\frac{\partial e^{2gC}}{\partial C^A}\phi _0=-g\bar \phi _0T^A\phi _0-2g^2\bar \phi _0\{T^A, T^B\}\phi _0C^B+\CO(C^2)}
where the hat is a reminder that $D$ is solved for in terms of the other fields.  We  solve for $\ev{C}$ by setting $\ev{\partial \CL /\partial C}=0$, and we'll suppose that $\ev{C}=0$ for simplicity.  The variation of the terms in ${\cal L}$ linear in $C$ yields 
\eqn\CDeqn{2g \bar F_0T^A F_0+(m_V^2)^{AB} \ev{D^B}=0,}
as found long ago \PolchinskiAN.   See \semidirect\ for discussion and an example with $\bar F_0T^AF_0\neq 0$.  
Upon replacing $D^A\to \widehat D^A$, $\CL \supset -\half \widehat D^A\widehat D^A$, with the correct sign to correspond to $V_D\geq 0$; expanding this term to quadratic order in $C$ gives $C^A$ its supersymmetric mass-squared terms $m_V^2$, as in \mvec, with correct (positive, non-tachyonic) sign.

The susy-breaking $C$ mass component $\delta m_0 ^2$ in \massvecsoft\ coming from expanding the kinetic terms \mattkin\ initially looks like a tachyonic contribution: $\CL _{kin}\supset (\bar Fe^{2gC}F)$.  
But this is an artifact of the fact that we still need to include $F_i W_i$ superpotential terms to account for how  $\ev{F}=F_{0}\neq 0$, as is familiar from how it happens that $V_F=+|W'|^2$.  Doing this, 
\eqn\clfs{\CL \supset -(\bar F_0e^{2gC}F_0) =-(\bar F_0 F_0)-4g^2 (\bar F_0\{T^A, T^B\} F_0) C^AC^B+\dots.}
The susy-breaking contribution to $m_C^2$ is thus non-tachyonic :
\eqn\mcex{(m_C^2)^{AB}=(m_V^2)^{AB}+(\delta m_0^2)^{AB}, \qquad (\delta m_0 ^2)^{AB}=
{\bar F_0\{T^A, T^B\}F_0\over \bar \phi _0\{ T^A, T^B\}\phi _0},}
with $m_V^2$ given by \mvec.

Finally,  the fermions $(\lambda _\alpha, \chi _\alpha)$ have the mass matrix as in \mcis, with $m_\chi ^{AB}$ given by \mvec, 
which is R-symmetry breaking if non-zero.   This interaction is needed to generate a one-loop A-term contribution.   Including the hidden sector fermions, there are also the terms
\eqn\morefermterm{\CL \supset \sqrt{2}i g\bar\phi _i T^A\lambda ^A\psi _i-\sqrt{2}g\bar F_i T^A\chi ^A\psi _i+W_{ij}\psi _i\psi _j+h.c..}
So the massive gaugino mass eigenstates are given by linear combinations of the original fields $\lambda$ and $\chi$, mixed with the $\psi _i$.

\subsec{Analytic continuation in superspace}

We denote the hidden sector fields by $\Phi$, and the visible sector fields by $Q$.  Analytic continuation in superspace  \refs{\Vadim, \gr, \AGLR}\ yields leading order susy-breaking effects, in the small susy-breaking limit, by starting with the supersymmetric low-energy effective theory
\eqn\svis{S_{vis}=\int d^8z Z_Q (\Phi, \bar \Phi)\bar Qe^{2V}Q+\int d^6 z (W(Q)+{1\over 4}\tau (\Phi) W_\alpha  W^\alpha)+h.c.}
and then allowing $\Phi$ and $\bar \Phi$ to pick up susy-breaking $\theta ^2$ and $\bar \theta ^2$ components.  This yields the  soft terms:
\eqn\softs{A_Q=(\ln Z_Q)|_{\theta ^2}, \qquad m_Q^2=-(\ln Z_Q)|_{\theta ^2\bar \theta ^2}, \qquad m_{gaugino}=(\ln \tau)|_{\theta ^2},}
where $(\ln Z_Q)|_{\theta ^2}=Z_Q|_{\theta ^2}/Z_Q|$ denotes the $\theta ^2$ component, and $Z_Q|$ denotes the component without $\theta ^2$ or $\bar \theta ^2$.

The standard result in the literature are obtained by assuming that the hidden sector fields decouple at a scale set by the expectation value of a single superfield $\ev{X}=M+\theta ^2F$ (where $X$ can be a spuion or actual field).  As an example, the hidden sector can have $W=\sum _{i=1}^N (\lambda_ i X +m_i) \Phi _i \widetilde \Phi _i$.  Then the susy-breaking affects the visible sector by way of the RG running to low-energy, down from the scale $M$.  Using primes to denote quantities for $\mu >M$,  the 
gauge coupling beta function is affected as ($t\equiv \ln \mu$ and $\alpha \equiv g^2/4\pi$ )
\eqn\betafns{{d\over dt}\alpha ^{-1}=\cases{{b'\over 2\pi}&for $\mu >M$\cr {b\over 2\pi}&for $\mu <M$},\qquad b'=3T_2(G')-T_2(r'), \quad b=3T_2(G)-T_2(r),}
where $G'$ is the gauge group and $r'$ is the matter content above the scale $M$, and $G$ and $r$ are those below.   Matching the running  $\tau $ at the scale $X$  gives a low energy $\tau =\tau (X)$ and 
\eqn\mgaugino{m_\lambda = g^2 \tau |_{\theta ^2}={\alpha \over 4\pi }(b-b')\ln X|_{\theta ^2}=-{\alpha \over 4\pi }\Delta b{F\over M},}
where, for any quantity $\Omega$, we define $\Delta \Omega \equiv \Omega '-\Omega$ to be the change below the scale where the hidden sector fields are integrated out.

The other standard expressions \refs{\Vadim, \gr, \AGLR, \LutySN}\ are one-loop A-terms given by 
\eqn\aterms{A_Q=\ln Z_Q(|X|^2)|_{\theta ^2}=\half \Delta {d\ln Z_Q\over dt} {F\over M}=- \Delta \gamma _Q{F\over M},}
where we define $d\ln Z_Q/dt\equiv -2\gamma _Q$ (our definition of $\gamma _Q$, with this conventional $-2$ factor, introduces some superficial differences with the above references, but they all drop out in the end) and the factor of $\half$ in \aterms\ comes from $(\ln |X|)|_{\theta ^2}=\half {F\over M}$.  Similarly, the two-loop (diagonal) sfermion soft masses are
\eqn\acsofti{\widetilde m^2_Q=-\ln Z_Q(|X|)|_{\theta ^2\bar \theta ^2}=-{1\over 2}\big| {F\over M}\big |^2 \left(-{\partial \gamma _Q'\over \partial g_i'}\beta _i'+2{\partial \gamma _Q\over \partial g_i}\beta _i'-{\partial \gamma _Q\over \partial g_i}\beta _i\right)}
which is two-loop because $\gamma$ and $\beta$ are each one-loop.

When there is no direct coupling of the messengers to the $Q$ fields, in particular when there are not gauge messengers, then $\gamma _Q'=\gamma _Q$, so the one-loop A-terms \aterms\ vanish, $A_Q=0$ at the messenger scale $M$ (and RG run to small, non-zero values at lower scales).  The two-loop sfermion mass  \acsofti\ simplifies in this case to 
\eqn\sfermacc{\widetilde m_Q^2=-{1\over 2}\big| {F\over M}\big |^2 {\partial \gamma _Q\over \partial g_i}\Delta \beta _i=2c_2(r_Q){g^4\over (16\pi ^2)^2}(b-b')\big|{F\over M}\big|^2,}
where in the last expression we used $\gamma _Q=-2c_2(r_Q)g^2/16\pi ^2$.  For this case, without gauge messengers, $b-b'=c>0$.

The literature also treats the gauge messenger case using the above formulae \refs{\Vadim, \gr, \AGLR}, 
assuming that there is a single susy-breaking chiral superfield spurion $X$. 
 As we'll discuss shortly, this latter assumption is generally invalid, and leads to results that are generally imprecise/incorrect.  In particular, that assumption leads to the standard, but incorrect, claim that $m_Q^2$ vanishes at one-loop, even in the gauge messenger case. In any event, the standard expression for the one-loop A-term at the messenger scale 
in the gauge messenger case comes from using \aterms, with $\gamma _Q\neq \gamma '_Q$, since $c_2(r_Q)\neq c_2(r'_Q)$, so
\eqn\atermsacgm{A_Q=2{g^2\over 16\pi ^2}\Delta c_Q{F\over M}.}  
The standard expression \acsofti\ in the case of gauge messengers gives 
 the two-loop sfermion masses at the messenger scale to be
\eqn\masssqsac{\widetilde m_Q^2=2 {g^4\over (16\pi ^2)^2}(c_2(r_Q')b'+c_2(r_Q)b-2c_2(r_Q)b')\big|{F\over M}\big|^2,}
which is typically  tachyonic \gr.

We can also consider using these methods to obtain the leading effective potential in the limit where some visible sector pseudomodulus has a large expectation value compared with the susy-breaking scale.  The leading order in small $|F|$ one-loop effective potential in this limit is well-known from the time of \WittenKV, coming from  the wavefunction renormalization of the tree-level vacuum energy of the hidden sector:
\eqn\veffQapprox{V_{eff}(|Q|)\approx \sum _i Z_i(|Q|)^{-1}|F_i|^2\approx  \sum _i \Delta \gamma _i ^{(1)}\ln |Q| |F_i|^2=-{g^2\over 8\pi ^2}\sum _i \Delta c_{Q_i} \ln |Q| |F_i|^2,}
where the last expression is the one-loop result for the case of gauge messengers, and $\Delta \gamma _i^{(1)}$ is the  change in the anomalous dimension of $\Phi _i$ at the scale $|Q|$ where the gauge group is partially Higgsed.  The sign of the potential is such that $Q$ has a runaway to large expectation values, which is how the hierarchy is generated in  \WittenKV.  Without gauge messengers, the Higgsing pseudomoduli are instead first lifted at two-loops to leading order; see \IntriligatorFE\ for discussion of how, more generally, gauge messengers lead to pseudomoduli being lifted at one-lower loop order than would happen without gauge messengers.   The result  \veffQapprox\ suggests that there must be a corresponding $\CO(|F|^2)$, one-loop mass term $(m_Q^2)^{(1)}$ for the $Q$ field near the origin.  As we discuss, that is indeed true.

Let us now revisit the above results, correcting the literature by allowing for multiple susy-breaking fields.   We'll refer to them all as $\Phi=\phi +\theta ^2 F$ (without writing explicitly the flavor or gauge indices), which plays the role of $X$ above.  Generalizing the discussion in \refs{\Vadim, \gr}, the  $\ell $ loop contribution to $Z_Q$ is given by $\ln Z_Q(\bar\Phi \Phi )=\sum _\ell \alpha ^{\ell -1}P_\ell(\alpha \ln (\bar\Phi , \Phi))$, where $(\bar \Phi, \Phi)$ is some inner product.  We now note that
\eqn\logtheta{\left[\ln (\bar \Phi, \Phi)\right]|_{\theta ^2}={(\bar \phi, F)\over (\bar \phi, \phi)}, \qquad \left[\ln (\bar \Phi, \Phi)\right]|_{\theta ^2\bar\theta ^2}={(\bar\phi, \phi )(\bar F, F)-(\bar \phi, F)(\bar F, \phi)\over (\bar\phi , \phi )^2},}
and in particular $[\ln (\bar \Phi, \Phi)]|_{\theta ^2\bar\theta ^2}\geq 0 $ can be non-zero in general, unlike the case with a single field $X=M+\theta ^2F$.  So the soft terms \softs\ get contributions 
\eqn\agen{A_Q=\sum _\ell \alpha ^\ell P'_{\ell }(\alpha \ln (\bar \phi, \phi)){(\bar \phi, F)\over (\bar \phi, \phi)},}
\eqn\mqgen{\eqalign{m_Q^2&=\sum _\ell \alpha ^\ell  P'_{\ell }(\alpha \ln (\bar \phi, \phi)) {(\bar\phi, \phi )(\bar F, F)-(\bar \phi, F)(\bar F, \phi)\over (\bar\phi , \phi )^2} \cr &+ \sum _\ell \alpha ^{\ell +1}P''_{\ell}(\alpha \ln (\bar \phi, \phi)){(\bar \phi, F)(\bar F, \phi)\over (\bar\phi , \phi )^2}.}}
The standard expressions in the literature are obtained upon specializing the inner products to the case of a single field, e.g. $(\bar \phi, \phi)\to |\phi |^2$ etc., in which case the top line of \mqgen\ vanishes.

So the top line of \mqgen\ are qualitatively new contributions, while the second line of \mqgen\ are the appropriate generalizations of standard expressions.  In particular, taking $\ell =1$, the top line of \mqgen\ yields the one-loop contribution to $m_Q^2$, 
\eqn\mqonex{(m_Q^2)^{(1)}=-{1\over 2} \Delta \gamma _Q^{(1)}{(\bar\phi, \phi )(\bar F, F)-(\bar \phi, F)(\bar F, \phi)\over (\bar\phi , \phi )^2},}
and the second line of $\mqgen$ yields the appropriate generalization of the standard expression \acsofti\ for the two-loop $m_Q^2$, in terms of one-loop RG quantities.  The top line of \mqgen, with $\ell =2$, generally yields a new,  additional contribution to the two-loop $m_Q^2$, similar to \mqonex\ with 
$\Delta \gamma _Q^{(1)}$ replaced with $\Delta \gamma _Q^{(2)}$. 
  
The  one-loop results can be seen directly from the  one-loop effective K\"ahler potential \GrisaruVE, see \keffis.  The term from integrating out massive chiral multiplets does not contribute to $(m_Q^2)^{(1)}$, because $\ln M_c^\dagger M_c=\ln M_c^\dagger +\ln M_c$ is the sum of a purely holomorphic and purely anti-holomorphic contribution.  But the $M_V^2$ term, from integrating out massive vector multiplets, does contribute.  Using \eqn\keffii{K_{eff}^{(1)}\supset {1\over 16\pi ^2}\sum _{AB}(M_V^2)^{AB}\ln {(M_V^2)^{AB}\over \Lambda ^2}\approx \bar Q_I \{T^A, T^B\}Q^I\ \cdot {g^2\over 16\pi ^2}\ln g^2{\bar\Phi \{ T^A, T^B \} \Phi\over |\Lambda |^2},}
this leads to the one-loop $m_Q^2$ contributions as in \mqonex,   
\eqn\mone{(m_I^{2 (1)})_{cd}=-{g^2\over 16\pi ^2}\sum _{A,B}C(r_I)^{AB}_{cd}{(\bar F ,F)^{(AB)}(\bar\phi, \phi)^{(A,B)}-(\bar F, \phi)^{(A,B)}(\bar\phi , F)^{(A,B)} \over (m_G^{2(AB)})^2},}
where $C(r_i)^{AB}_{cd}=(\{T_{r_I}^A,T_{r_I}^B\})_{cd}$ (recall that $A, B$ are $G$ adjoint indices, whereas $c,d$ are in the representation $r_I$) and the inner product notation is as in \innerprod. 
The result \mqoneexp\ is obtained upon taking the inner products $(\cdot, \cdot )^{AB}=(\cdot, \cdot)\delta ^{A'B'}$.

\appendix{B}{Relations among results}

In this appendix, we demonstrate the equivalence of three methods of computing the effective potential where the methods are simultaneously valid.  We first use the tree-level current correlator coefficients to compute the one-loop potential, and show that this is equivalent to the Coleman-Weinberg \ColemanJX\ potential.  We then expand the one-loop potential to leading order in susy breaking and show that the effective K\"ahler potential yields the same result.  For simplicity, we take $D=0=[\wt B_{1/2},\wt C_{1/2}]$ throughout this appendix.

\subsec{Small $g$: Coleman-Weinberg and Current Correlators}

The one-loop effective potential for a visible-sector sfermion, $Q$, is given in \veffone\ and reproduced here:
\eqn\veffcurrentapp{V_{eff}^{(1)}=\frac{\Tr}2\!\int\!\frac{d^4p}{(2\pi)^4}\!\left(\!\ln(1+g^2\wt C_0^{(0)})\!-\!2\ln\!\!\left[(1+g^2\wt C_{1/2}^{(0)})^2\!+\frac{g^4|\wt B_{1/2}^{(0)}|^2}{p^2}\right]\!\!+\!3\ln(1+g^2\wt C_1^{(0)})\!\right)\!.}
For weak coupling, the $\wt C_a^{(0)}\equiv \wt C_a^{pole}$ are given in \cfunctionsapp.  To compute this as a function of $Q$ (along a D-flat background), we use \vishid\ and \visbc:
\eqn\qdep{\wt C_a^{(0)AB}=\frac{(\bar Q, Q)^{(A,B)}}{p^2}+Q\hbox{-}independent.}
This makes it simple to compute the following scalar object: 
\eqn\veffcurrent{\eqalign{
Q^i\frac{\p V_{eff}^{(1)}}{\p Q^i}\!=\!\frac\Tr2(\bar Q,Q)\!\!\int\!\!\frac{d^4p}{(2\pi)^4}\frac1{p^2}\Bigg[\!\frac1{1+g^2\wt C_0^{(0)}}-\frac{4(1+g^2\wt C_{1/2}^{(0)})p^2}{(1+g^2\wt C_{1/2}^{(0)})^2p^2\!+g^4|\wt B_{1/2}^{(0)}|^2}+\frac3{1+g^2\wt C_1^{(0)}}\!\Bigg]
}}
We will similarly differentiate the Coleman-Weinberg potential and thereby avoid the difficulty of extracting meaningless constants.

The Coleman-Weinberg potential can be written as
\eqn\cw{V_{eff}^{(1)}=\frac12\hbox{Tr}\int\frac{d^4p}{(2\pi)^4}\bigg[\ln\Big(p^2+\wt M_0^2\Big)-\ln\Big(p^2+\wt M_{1/2}^2\Big)+3\ln\Big(p^2+m_V^2\Big)\bigg].}
Note that the scalar and fermion mass matrices here are not the same as those in \massesapp.  The scalar mass matrix is now (recall that we are taking $D=0$) 
\eqn\mzerog{\wt M_0^2=M_0^2+g^2\phiT^A\bar\phiT^A.}
The fermion mass matrix has to be expanded to include the gauginos.  The mass-squared matrix is 
\eqn\mhalfg{\wt M_{1/2}^2=\pmatrix{m_{1/2}^2&0\cr0&(m_{1/2}^2)^T},\qquad m_{1/2}^2=\pmatrix{\bar\M\M+2g^2T^A\phi_0\bar\phi_0T^A&-i\sqrt2gTF\cr i\sqrt2g\bar FT&g^2(\bar\phi_0,\phi_{0})}}

Now let's considering differentiating \cw.  The last term trivially agrees with the corresponding term in \veffcurrent\ because $g^2\wt C_1^{(0)}=m_V^2/p^2$.  The derivative of the scalar contribution in \cw\ involves the following.
\eqn\dqvcwzero{\Tr\ \frac1{p^2+\wt M_0^2}\frac\p{\p Q^i}\phiT^A\bar\phiT^A=\bar\phiT^A\frac1{p^2+\wt M_0^2}\frac{\p\phiT^A}{\p Q^i}+\frac{\p\bar\phiT^A}{\p Q^i}\frac1{p^2+\wt M_0^2}\phiT^A}
Expanding this in $g$, we get terms such as
\eqn\dqseries{\bar\phiT^A\bigg(\frac1{p^2+M_0^2}-\frac1{p^2+M_0^2}g^2\phiT^B\bar\phiT^B\frac1{p^2+M_0^2}+\dots\bigg)\frac{\p\phiT^A}{\p Q^i}.}
Using the identities, 
\eqn\dqids{M_0^2\frac{\p\phiT}{\p Q}=0\qquad\hbox{and}\qquad\bar\phiT^B\frac{\p\phiT^A}{\p Q^i}=(\bar QT^BT^A)_i,}
one finds that the series can be expressed in powers of $\wt C_0^{(0)AB}$, giving 
\eqn\dtrln{\eqalign{
Q^i\frac{\p}{\p Q^i}&\Tr\ln(p^2+\wt M_0^2)=\frac{(\bar Q,Q)}{p^2}\Tr\frac1{1+g^2\wt C_0^{(0)}.}
}}
This shows that the scalar contribution of the Coleman-Weiberg potential agrees with that in \veffcurrent.  We omit the similar but more tedious calculation for the fermionic contribution.

\subsec{Small $F$: The Effective K\"ahler Potential and Current Correlators}

To leading order in susy breaking, the one-loop effective potential can be extracted from the one-loop effective K\"ahler potential.  We will now demonstrate that the result obtained in this way agrees with that from \veffcurrentapp.  First we reorganize the terms in $\wt C_{1/2}$.  Starting with the simplifying observation that the two blocks of the matrix contribute equally, and then using the gauge invariance of the superpotential,
\eqn\gaugeinv{\M T^A\phi=(\bar F T^A)^T,}
we find 
\eqn\chalfff{\eqalign{\wt C_{1/2}^{AB}&=2\bar\phi_0T^{A}\frac1{p^2+\bar\M\M}T^B\phi_0\cr
&=2\bar\phi_0T^{A}\left(\frac1{p^2}-\frac{\bar\M\M}{p^4}+\frac{(\bar\M\M)^2}{p^6}-\dots\right)T^B\phi_0\cr
&=\wt C_1^{AB}-\frac2{p^2}\bar F T^{A}\Delta_{1/2}T^{B}F,
}}
where $\Delta_{1/2}=(p^2+\bar\M\M)^{-1}$.  $\wt B_{1/2}$ has a similar expansion.  We use gauge invariance again to find, 
\eqn\bhalff{\eqalign{\wt B_{1/2}^{AB}&=-2\bar \phi_0T^{A}\frac1{p^2+\bar\M\M}T^{B}F\cr
&=-\frac{2}{p^2}\bar\phi_0 T^AT^BF+\O(|F|^2)
}}
This is as much as we need for the effective potential to leading order in $F$.  Finally, a similar set of manipulations reveals 
\eqn\czeroff{\wt C_0^{AB}=\wt C_1^{AB}+\frac{4}{p^4} \bar FT^AT^BF-\frac8{p^2}\bar FT^{A}\Delta_{1/2}^{-1}T^{B}F+\O(|F|^4).}
These relations are compatible with the large $p^2$ limit expansions \celse\ (which don't rely on small $F$ terms), and the relation \mbreakreln\ that $\delta m_0=2\delta m_{1/2}$ with F-term breaking.

If one now takes \chalfff, \bhalff, and \czeroff\ and uses them to expand the potential \veffcurrentapp, one finds that the terms involving $\Delta_{1/2}^{-1}$ cancel.  We can write the result in terms of $\Delta_1=(p^2+m_V^2)^{-1}$ as 
\eqn\veffint{V_{eff}^{(1)}=2g^2\int\frac{d^4p}{(2\pi)^4}\frac{1}{p^2}\Delta_1^{AB}\left(FT^BT^AF+2g^2 \bar FT^AT^B\phi_0\Delta_1^{BC}\bar\phi_0T^CT^AF\right)+\dots,}
where the ellipsis includes constant terms and terms higher order in $|F|$.

Now consider the one-loop effective K\"ahler potential, which can be written as
\eqn\keff{K_{eff}^{(1)}=\Tr\int\frac{d^4p}{(2\pi)^4}\frac1{p^2+M_V^2}-\frac12\Tr\int\frac{d^4p}{(2\pi)^4}\frac1{p^2+\bar\CM\CM}+constant}
Recall that $M_V^2$ is the superfield version of $m_V^2$:
\eqn\mhalfwf{M_V^2= g^2\left[(\bar\phi,\phi)+\tb^2(\bar F,\phi)+\t^2(\bar\phi,F)+\t^2\tb^2(\bar F,F)\right],}
The second term in \keff\ is irrelevant in the case we are considering because our visible-sector sfermions interact with susy-breaking only through the gauge interactions.  In manipulating the first term, it's useful to define $\varphi^A\equiv T^A\phi$, $f^{A}\equiv T^AF$.  The $D$-term of $K_{eff}$ can then be written as
\eqn\dofkeff{\eqalign{
\int d^4\t K_{eff}^{(1)}=2g^2\int\frac{d^4p}{(2\pi)^4}\Big[-(\Delta_1^{2})^{AB}\bar f^{B}f^{A}&+2g^2(\Delta_1^{2})^{AB}\bar f^{B}\varphi^{C}\Delta_1^{CD}\bar\varphi^{D} f ^{A}\cr
&+2g^2(\Delta_1^{2})^{AB}\bar\varphi^{B} f ^{C}\Delta_1^{CD}\bar f ^{D}\varphi^{A}\Big]
}}
Now we can simplify this expression with the identity,
\eqn\phideltacom{\varphi^A\Delta_1^{AB}=\wt\Delta_1\varphi^B,\qquad(\wt\Delta_1^{-1})^i{}_j\equiv p^2\d_j^i+2 g^2\varphi^{Ai}\bar\varphi_j^A,}
and a few other manipulations that mirror those in the appendix of \ISS.  In particular, we use \phideltacom\ in the first two lines, rearrange to arrive at the third, then integrate by parts, and finally we expand out $\wt\Delta_1$ and resum in terms of $\Delta_1$.
\eqn\keffsteps{\eqalign{
\int d^4\t K_{eff}&=2g^2\int\frac{d^4p}{(2\pi)^4}\Big[-(\Delta_1^{2})^{AB}\bar f ^{B} f ^{A}+2g^2(\Delta_1^{2})^{AB}\bar f ^{B}\wt\Delta_1\varphi^{D}\bar\varphi^{D} f ^{A}\cr
&\hskip 1.94in+2g^2\bar f ^{D}\wt\Delta_1^{2}\varphi^{B}\bar\varphi^{B} f ^{C}\Delta_1^{CD}\Big]\cr
&=2g^2\int\frac{d^4p}{(2\pi)^4}\Big[-(\Delta_1^{2})^{AB}\bar f ^{B} f ^{A}+(\Delta_1^{2})^{AB}\bar f ^{B}\wt\Delta_1(\wt\Delta_1^{-1}-p^2) f ^{A}\cr
&\hskip 1.94in+\bar f ^{D}\wt\Delta_1^{2}(\wt\Delta_1^{-1}-p^2) f ^{C}\Delta_1^{CD}\Big]\cr
&=2g^2\int\frac{d^4p}{(2\pi)^4}\Big[\Delta_1^{AB}\bar f ^{B}\wt\Delta_1 f ^{A}+p^2\frac{d}{dp^2}\Delta_1^{AB}\bar f ^{B}\wt\Delta_1 f ^{A}\Big]\cr
&=-2g^2\int\frac{d^4p}{(2\pi)^4}\Delta_1^{AB}\bar f ^{B}\wt\Delta_1 f ^{A}\cr
&=-2g^2\int\frac{d^4p}{(2\pi)^4}\frac{1}{p^2}\Delta_1^{AB}\left(\bar FT^BT^AF+2g^2 \bar FT^AT^B\phi_0\Delta_1^{BC}\bar\phi_0T^CT^AF\right),
}}
which is identical to \veffint.

\listrefs\end